\documentclass[aps,pra,reprint]{revtex4-2} 

\usepackage{fullpage}
\usepackage{graphicx}\usepackage{multirow}\usepackage{amsmath,amssymb,amsfonts}\usepackage{amsthm}\usepackage{mathrsfs}\usepackage[title]{appendix}\usepackage[dvipsnames]{xcolor}

\usepackage{bm}

\usepackage{physics}
\usepackage{comment}

\usepackage{etoolbox}
\patchcmd{\thebibliography}
  {\advance\leftmargin\labelsep}
  {\advance\leftmargin\labelsep\selectlanguage{english}}
  {}{}

\AtBeginDocument{\renewcommand{\selectlanguage}[1]{}}
\usepackage{float} \usepackage{algorithm}
\usepackage{algpseudocode}
\usepackage[colorlinks = true,
            linkcolor = blue,
            urlcolor  = blue,
            citecolor = blue,
            anchorcolor = blue]{hyperref}

\newcommand{\eg}{{\it e.g.}}
\newcommand{\ie}{{\it i.e.}}

\let\oldsim\sim 
\renewcommand{\sim}{{\oldsim}}
\newcommand{\Cov}{C}
\newcommand{\PCov}{\mathrm{PCov}}

\newcommand{\E}[1]{\mathbb{E}\left[#1\right]}
\newcommand{\OpD}{{\mathcal{D}}}

\newcommand{\avg}[1]{\langle #1\rangle}

\RequirePackage[normalem]{ulem} \RequirePackage{color}\definecolor{RED}{rgb}{1,0,0}\definecolor{BLUE}{rgb}{0,0,1}                        \providecommand{\DIFaddbegin}{} \providecommand{\DIFaddend}{} \providecommand{\DIFdelbegin}{} \providecommand{\DIFdelend}{}     \providecommand{\DIFaddbeginFL}{} \providecommand{\DIFaddendFL}{} \providecommand{\DIFdelbeginFL}{} \providecommand{\DIFdelendFL}{}   \newcommand{\DIFscaledelfig}{0.5}
\RequirePackage{settobox} \RequirePackage{letltxmacro} \newsavebox{\DIFdelgraphicsbox} \newlength{\DIFdelgraphicswidth} \newlength{\DIFdelgraphicsheight} \LetLtxMacro{\DIFOincludegraphics}{\includegraphics} \newcommand{\DIFaddincludegraphics}[2][]{{\color{blue}\fbox{\DIFOincludegraphics[#1]{#2}}}} \newcommand{\DIFdelincludegraphics}[2][]{\sbox{\DIFdelgraphicsbox}{\DIFOincludegraphics[#1]{#2}}\settoboxwidth{\DIFdelgraphicswidth}{\DIFdelgraphicsbox} \settoboxtotalheight{\DIFdelgraphicsheight}{\DIFdelgraphicsbox} \scalebox{\DIFscaledelfig}{\parbox[b]{\DIFdelgraphicswidth}{\usebox{\DIFdelgraphicsbox}\\[-\baselineskip] \rule{\DIFdelgraphicswidth}{0em}}\llap{\resizebox{\DIFdelgraphicswidth}{\DIFdelgraphicsheight}{\setlength{\unitlength}{\DIFdelgraphicswidth}\begin{picture}(1,1)\thicklines\linethickness{2pt} {\color[rgb]{1,0,0}\put(0,0){\framebox(1,1){}}}{\color[rgb]{1,0,0}\put(0,0){\line( 1,1){1}}}{\color[rgb]{1,0,0}\put(0,1){\line(1,-1){1}}}\end{picture}}\hspace*{3pt}}} } \LetLtxMacro{\DIFOaddbegin}{\DIFaddbegin} \LetLtxMacro{\DIFOaddend}{\DIFaddend} \LetLtxMacro{\DIFOdelbegin}{\DIFdelbegin} \LetLtxMacro{\DIFOdelend}{\DIFdelend} \DeclareRobustCommand{\DIFaddbegin}{\DIFOaddbegin \let\includegraphics\DIFaddincludegraphics} \DeclareRobustCommand{\DIFaddend}{\DIFOaddend \let\includegraphics\DIFOincludegraphics} \DeclareRobustCommand{\DIFdelbegin}{\DIFOdelbegin \let\includegraphics\DIFdelincludegraphics} \DeclareRobustCommand{\DIFdelend}{\DIFOaddend \let\includegraphics\DIFOincludegraphics} \LetLtxMacro{\DIFOaddbeginFL}{\DIFaddbeginFL} \LetLtxMacro{\DIFOaddendFL}{\DIFaddendFL} \LetLtxMacro{\DIFOdelbeginFL}{\DIFdelbeginFL} \LetLtxMacro{\DIFOdelendFL}{\DIFdelendFL} \DeclareRobustCommand{\DIFaddbeginFL}{\DIFOaddbeginFL \let\includegraphics\DIFaddincludegraphics} \DeclareRobustCommand{\DIFaddendFL}{\DIFOaddendFL \let\includegraphics\DIFOincludegraphics} \DeclareRobustCommand{\DIFdelbeginFL}{\DIFOdelbeginFL \let\includegraphics\DIFdelincludegraphics} \DeclareRobustCommand{\DIFdelendFL}{\DIFOaddendFL \let\includegraphics\DIFOincludegraphics} \RequirePackage{listings} \RequirePackage{color} \lstdefinelanguage{DIFcode}{ moredelim=[il][\color{red}\sout]{\%DIF\ <\ }, moredelim=[il][\color{blue}\uwave]{\%DIF\ >\ } } \lstdefinestyle{DIFverbatimstyle}{ language=DIFcode, basicstyle=\ttfamily, columns=fullflexible, keepspaces=true } \lstnewenvironment{DIFverbatim}{\lstset{style=DIFverbatimstyle}}{} \lstnewenvironment{DIFverbatim*}{\lstset{style=DIFverbatimstyle,showspaces=true}}{} 

\begin{document}

\title{Covariance Analysis of Impulsive Streaking}
\author{Jun Wang$^{1,2,3}$}
\thanks{These two authors contributed equally.}
\author{Zhaoheng Guo$^{2,3,4}$}
\thanks{These two authors contributed equally.}
\author{Erik Isele$^{1,2,3}$}
\author{Philip H. Bucksbaum$^{1,2,3}$}
\author{Agostino Marinelli$^{1,3}$}
\author{James P. Cryan$^{1,3}$}
\email{jcryan@slac.stanford.edu}
\author{Taran Driver$^{1,3}$}
\email{tdriver@stanford.edu}
\affiliation{$^{1}${Stanford PULSE Institute}, {SLAC National Accelerator Laboratory}, {{Menlo Park}, {CA} {94025}, {USA}}}
\affiliation{$^{2}${Department of Applied Physics}, {Stanford University}, {
{Stanford}, {CA} {94305}, {USA}}}
\affiliation{$^{3}${SLAC National Accelerator Laboratory}, {{Menlo Park}, {CA} {94025}, {USA}}}
\affiliation{$^{4}${Paul Scherrer Institute}, {Villigen}, {Switzerland}}

\date{\today}

\begin{abstract}
We present a comprehensive framework of modeling covariance in angular streaking experiments. 
Within the impulsive streaking regime, the displacement of electron momentum distribution~(MD) provides a tight connection between the dressing-free MD and the dressed MD. 
Such connection establishes universal structures in the composition of streaking covariance that are common across different MDs, regardless of their exact shape. 
Building on this robust framework, we have developed methods for retrieving temporal information from angular streaking measurements. 
By providing a detailed understanding of the covariance structure in angular streaking experiments, our work enables more accurate and robust temporal measurements in a wide range of experimental scenarios.
\end{abstract}
\maketitle

\section{Introduction}\label{sec:introduction}

The motion of electrons in molecules and condensed phase systems takes place on the attosecond timescale.
It is now possible to generate light pulses and trains of pulses with sub-femtosecond (i.e. attosecond) duration.
These technical developments have launched the field of experimental attosecond science~\cite{nobel_2023}.
Even with access to attosecond pulses, measuring electron motion with attosecond time resolution is a significant technical challenge.
Time-resolved measurements require the ability to synchronize an attosecond pulse with a second event, such as the interaction with a second laser pulse, with sub-femtosecond precision.
One method that has proven highly successful involves combining the attosecond pulse with a longer duration infrared laser pulse, and using the oscillating electric field of the infrared laser pulse to map time onto a measurable quantity such as the momentum of an emitted electron~\cite{itatani_attosecond_2002}.
The term `attosecond streaking' has been coined to describe this class of experiments, as the action of the field on the electron is analogous to the action of the time-varying voltage in a streak camera~\cite{bradley_direct_1971,itatani_attosecond_2002}.
Since the period of oscillation of an infrared laser pulse is on the few-femtosecond timescale and the action of the field on an electron depends on the phase of the oscillation, attosecond streaking experiments have emerged as a powerful probe of attosecond electron dynamics~\cite{thumm_attosecond_2015}, including measurements of photoemission delays~\cite{schultze_delay_2010}, Auger-Meitner decay~\cite{drescher_time-resolved_2002}, and characterization of attosecond pulses ~\cite{ eckle_attosecond_2008, hartmann_attosecond_2018}. 

The principal of laser streaking measurements relies on a time reference for the ultrafast process, which can be provided either by the precise timing stability~(few-attosecond) between the attosecond pulse and the infrared field, or through a single-shot self-referencing technique. 
The timing stability can be achieved when the attosecond pulse has been produced by high harmonic generation~(HHG), since the HHG emission is naturally synchronized with the infrared driving field.
Meanwhile, attosecond x-ray free-electron lasers~(XFELs) have many advantageous properties: continuous tunability of photon energy, peak powers that can approach one terawatt, and roughly Fourier-limited pulse durations~\cite{duris_tunable_2020, franz_terawatt-scale_2024}.
The inherent timing jitter between an infrared laser and the XFEL pulses is typically larger than the infrared period, in contrast to HHG-based sources.
To make use of the exceptional properties of XFELs in experiments approaching the attosecond regime, approaches that employ a single-shot time reference signal are required.
Such approaches have been explored recently, demonstrating the ability to achieve time resolution better than an optical cycle~\cite{haynes_clocking_2021, li_attosecond_2022, guo_experimental_2024, maroju_attosecond_2020, maroju_attosecond_2023, driver_attosecond_2024}.
The single-shot reference signal can be provided by photoelectrons produced from prompt ionization. 
When the duration of the XFEL pulse is much shorter than the infrared laser period, the streaking interaction can be treated in the impulsive regime, in which the streaking laser imparts the same momentum shift to all photoelectrons. 
The momentum shift of the prompt electrons therefore provides a single-shot reference for the relative x-ray/laser arrival time.
Such a time reference has been employed to achieve attosecond timing resolution for measurements of Auger-Meitner decay by characterizing the photoelectrons' momentum shift on a single-shot basis~\cite{haynes_clocking_2021, li_attosecond_2022, wang_probing_2024}.
The single-shot quantification of the momentum shift places stringent conditions on the single-shot data and risks systematic error.

Correlation-based approaches have also been employed to overcome the inherent timing jitter in laser/x-ray measurements~\cite{haynes_clocking_2020,maroju_attosecond_2020,maroju_attosecond_2023}.
Such a technique has recently been used to extract the photoemission delay in x-ray ionization~\cite{driver_attosecond_2024} and measure the delay between two attosecond pulses~\cite{guo_experimental_2024}.
Here we provide a formal treatment for this covariance analysis, focusing on measurements with a circularly polarized infrared dressing field, or ``angular streaking''. 
In addition to circumventing the requirement for single-shot analysis, this correlation-based analysis bypasses the requirement for the complex modeling of the streaking process in the retrieval of photoemission delay.

In section\,\ref{sec:model}, we introduce our mathematical model for the covariance analysis of impulsive angular streaking, in the presence of large jitter in the x-ray/infrared relative timing.
In section\,\ref{sec:phi-retrieval}, we describe and compare several methods for extracting the time delay between two impulsive processes.
These methods can be used to measure the relative photoemission delay between photoionization events produced by the same pulse, or the delay between attosecond pulses.
In section\,\ref{sec:discussions} we discuss additional considerations for the interpretation and design of attosecond angular streaking measurements.
In section\,\ref{sec:generalY} we generalize our analysis to the case of an emission process with a complex~(non-impulsive) profile in the time domain.

\section{Model}\label{sec:model}
To describe the x-ray-matter interaction in the presence of an intense laser-field with vector potential $\bm{A}(t)$, we employ the strong-field approximation to write the probability amplitude for observing a photoelectron with momentum $\bm{p}$~\cite{kitzler_quantum_2002}:
\begin{equation}
\label{eq:SFA}
b\left(\bm{p};\bm{A}\right) = \int_{t_0}^{\infty} d t e^{-i\Phi(t;\bm{p},\bm{A})}
    G(t;\bm{p}-e\bm{A}(t))~,
\end{equation}
where $\Phi(t;\bm{p},\bm{A}) \equiv \int_{t}^{+\infty}dt' (\bm{p}- e\bm{A}(t'))^2/(2m\hbar)$ is the so-called Volkov phase~\cite{wolkow_uber_1935}, $e<0$ and $m$ are the charge and mass of electron, respectively, and $G(t;\bm{p}')$ describes the electron source term for electrons with momentum $\bm{p}'$ entering the continuum at time $t$, and $t_0$ is a fixed initial time before the onset of x-ray pulse. 
Equation~(\ref{eq:SFA}) can be simplified using the stationary phase approximation to yield,  
\begin{equation}~\label{eq:SPA}
    b(\bm{p};\bm{A}) \simeq \sum_{t_\mathrm{s}\in T(\bm{p},\bm{A})} C_{\bm{p}}(t_\mathrm{s}) G(t_\mathrm{s}, \bm{p}-e\bm{A}(t_\mathrm{s}))~,
\end{equation}
where $T(\bm{p},\bm{A})$ is the collection of all $t_\mathrm{s}$ that satisfy the stationary phase condition $0=(\bm{p}-e\bm{A}(t_\mathrm{s}))^2/2 + \left.\frac{d}{dt}\right|_{t=t_s} \arg G(t, \bm{p}-e\bm{A}(t))$, and $C_{\bm{p}}(t_s)$ is a weighting factor for each stationary point. 
If the duration of source-term $G$ is much shorter than the streaking laser period $T_L$, a unique $t_\mathrm{s}$ dominates for all $\bm{p}$. 
Since the derivative of the phase $\arg[G]$ describes an energy, we can write $\left.\frac{d}{dt}\right|_{t=t_s} \arg[G]=-E_0$.
With this substitution, the stationary phase condition becomes $(\bm{p}-e\bm{A}(t_\mathrm{s}))^2/2=E_0$, which we recognize as an equation of motion for a classical electron in an external field, with kinetic energy $E_0$ at time $t_s$. 
Revisiting Eqn.\,(\ref{eq:SPA}) and the stationary phase condition, this uniqueness of $t_\mathrm{s}$ enables an approximation, demonstrating that the dressed photoelectron momentum distribution~(MD) $|b(\bm{p};\bm{A})|^2$ can be approximated by a displacement of the dressing-free~(\ie~$\bm{A}=0$) MD:
\begin{equation}\label{eq:shift}
    |b(\bm{p};\bm{A})|^2 \simeq |b(\bm{p}- \bm{k};\bm{A}=0)|^2~,
\end{equation}
where $\bm{k}\equiv e(\bm{A}(t_s)+\tau \left.\frac{d \bm{A}}{dt}\right|_{t_s})$ is the momentum shift, and $\tau=\langle\frac{\partial^2}{\partial \bm{p}'^2} \left.\arg G\right|_{t_s}\rangle$ is the momentum-averaged photoemission delay.
We call this the impulsive streaking regime and label $\bm{k}$ the streaking vector. 

\begin{figure}[htbp]
    \centering
    \includegraphics[width=\linewidth]{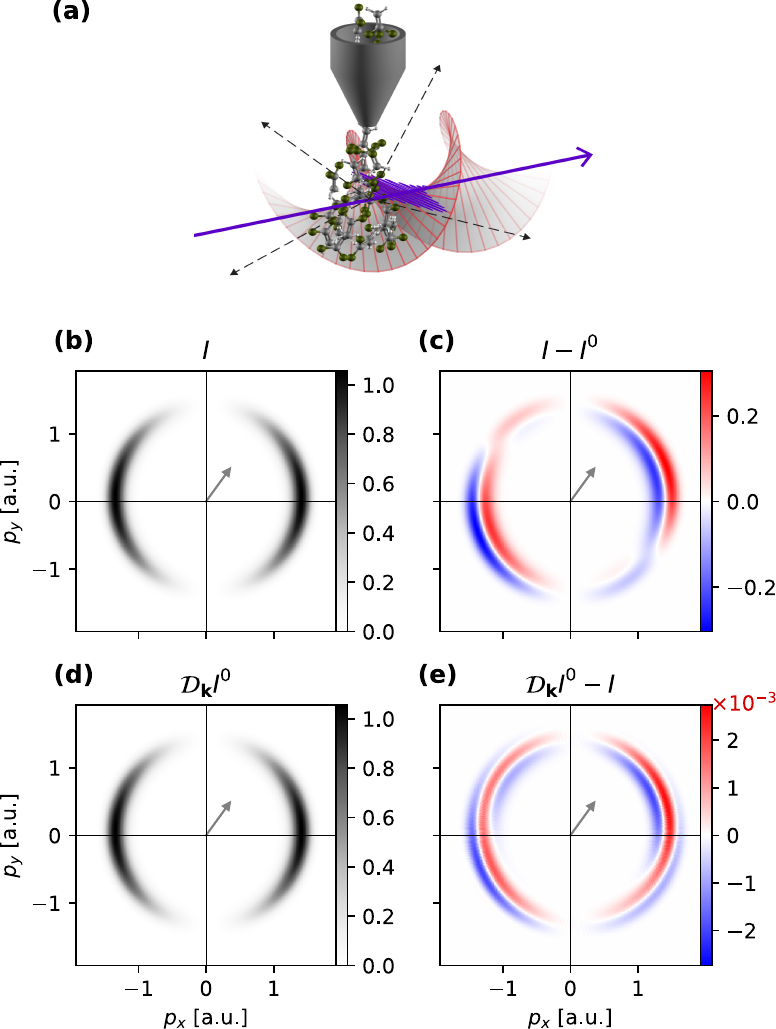}
    \caption{Illustration of impulsive regime of angular streaking.
    \textbf{(a)} In the presence of the circularly polarized streaking laser field (red), a pulse (purple) much shorter than the laser period ionizes the sample molecules, from which electrons are emitted~(black dashed). 
    \textbf{(b)} Streaked photoelectron momentum distribution~(MD) at the $p_z=0$ slice simulated with strong-field approximation~(SFA). 
    A hydrogen atom $I_P{=}13.6$\,eV is ionized by a $40.8$\,eV, $200$\,as Gaussian pulse in a $\lambda=1.85\,\mathrm{\mu m}, |\bm{A}|=0.06\,\mathrm{a.u.}$ dressing field. 
    The arrow indicates $e\bm{A}$ at the ionizing pulse's peak instant, with 10x-magnified length.
    \textbf{(c)} Difference of (b) from the dressing-free MD $I^0$, \ie~the MD when the streaking field is mistimed from the ionization pulse. 
    \textbf{(d)} The dressing-free MD $I^0$ displaced by $\bm{k}$. 
    \text{(e)} Difference between (b) and (d). 
    } 
    \label{fig:illustration}
\end{figure}

For the remainder of this work, we consider the case of a circularly polarized dressing laser propagating along the $\hat{z}$-direction, 
\begin{equation}
    \label{eqn:A(t)}
    \bm{A}(t)=A_0(t)\left[\cos\left(\omega_Lt\right)\hat{x}+\sin\left(\omega_L t\right)\hat{y}\right],
\end{equation}
where $A_0(t)$ is the slowly varying envelope of the laser field assumed to be much longer than the laser period $T_L$, and $\omega_L=2\pi/T_L$ is the angular frequency.
We consider the two-dimensional~(2D) momentum distribution $I(\bm{r})$ of the electrons, where $\bm{r}$ denotes the momentum in the $xy$-plane.  Depending on the measurement scheme, $I(\bm{r})$ can either be the projection of $|b(\bm{p})|^2$ along the $\hat{z}$ direction~\cite{li_co-axial_2018} or the slice of $|b(\bm{p})|^2$ at the $p_z=0$ plane~\cite{hartmann_attosecond_2018}. 
The model and methods introduced in this work are applicable to both measurement schemes, and a quantitative comparison is given in Sec.~\ref{sec:discussions}. 

The defining property of the impulsive regime~(Eqn.\,(\ref{eq:shift})) is the displacement of the dressing-free MD $I^0(\bm{r})$ by the streaking vector:
$I(\bm{r};\bm{A}) \simeq I^0(\bm{r} - \bm{k})\equiv \mathcal{D}_{\bm{k}} I^0(\bm{r})$, where we define the displacement operator $\mathcal{D}_{\bm{k}}$. 
Throughout this work, the superscript ``0'' indicates the MD is dressing-free~($\bm{A}=0$), \eg~$I^0(\bm{r})$.
An example MD simulated using Eqn.\,(\ref{eq:SFA}) is shown in Fig.\,\ref{fig:illustration}(b).
In this case, the photoelectrons are produced from direct ionization $G(t,\bm{p}')=-ie^{iI_Pt/\hbar}D(\bm{p}')E_X(t)$, where $I_P$ is the ionization potential, $D(\bm{p}')$ the dipole along the polarization of the ionizing pulse $E_X(t)$. 
As shown in Fig.\,\ref{fig:illustration}(c-d), the effect of the vector potential on the photoelectron MD is well approximated by the displacement operator, to the 1\% level in this case~(panel (e)).

As discussed in Sec.\,\ref{sec:introduction}, most experimental photoelectron MDs contain multiple photoemission features.
We generally denote two non-overlapping features in the MD as $X(\bm{r})$ and $Y(\bm{r})$, such as the photoelectrons produced by two ionizing pulses with different wavelengths, separated in time by less than $T_L$~\cite{guo_experimental_2024}.
Due to instabilities in the relative delay between the ionizing and streaking pulses, both $X$ and $Y$ vary randomly.
To extract the relative timing information between the features, we compute the covariance between $X$ and $Y$, 
\ie~the two-point function defined as 
\begin{equation}\label{eq:defCovXY}
    \Cov[X,Y](\bm{r}_q,\bm{r}_p) \equiv 
    \E{X(\bm{r}_q)Y(\bm{r}_p)} - \E{X(\bm{r}_q)}\E{Y(\bm{r}_p)}~,
\end{equation} 
where $\mathbb{E}[\cdot]$ refers to the expectation over all fluctuations, which is also written as $\avg{\cdot}{\equiv} \E{\cdot}$ for simplicity.
We choose the feature $X(\bm{r})$ to be the result of an impulsive process~(Eqn.\,\ref{eq:shift}), which provides a timing reference for any general ionization feature, $Y(\bm{r})$.
We denote the streaking vector of $X$ as $\bm{k}=\bm{k}_X= k(\cos\kappa\,\bm{e}_x + \sin\kappa\,\bm{e}_y)$, with amplitude $k$ and direction $\kappa$.

The key to encoding the relative timing between $X$ and $Y$ into the covariance is the randomness of the streaking direction $\kappa$.
The distribution of $\kappa$ is generally considered to be uniform $\mathcal{U}[0,2\pi]$, since the arrival time jitter spans many periods of the dressing laser~\cite{glownia_time-resolved_2010}. 
Although $\kappa$ is random, the relative timing between $X$ and $Y$ is determined by the underlying atomic or molecular physics. 
The streaking amplitude $k$ is determined by the spatial overlap and the intensity of dressing laser, which can be assumed statistically independent from $\kappa$.
The arrival time jitters may also vary $k$, but since the duration of $A_0(t)$ is much longer than $T_L$, we can still assume $k$ and $\kappa$ are independent. 
In the following, we refer to the covariance that purely results from the variations in the streaking vector $\bm{k}$ as the ``streaking covariance'':
\begin{equation}\label{eq:defK}
K[X,Y]\equiv \Cov[\E{X|\bm{k}}, \E{Y|\bm{k}}]~,
\end{equation}
where $\E{\cdot|\bm{k}}$ is the conditional expectation operator given the streaking vector, \ie~the average over all other parameters except  $\bm{k}$. 
In this way, $\E{X|\bm{k}}$ and $\E{Y|\bm{k}}$ are two random functions that only vary with $\bm{k}$, thus $K[X,Y]\neq 0$ if and only if $\bm{k}$ fluctuates.

In Sec.\,\ref{sec:impulsive-streak}-\ref{sec:discussions}, we demonstrate the covariance analysis in the situation where $Y$ is also impulsive.
The angle $\phi$ between the two streaking vectors, $\bm{k}_Y$ and $\bm{k}_X=\bm{k}$, corresponds to a time delay $\phi/\omega_L$ of the feature $Y$ referenced to $X$. 
If we assume a counter-clockwise rotation of the vector potential as specified in Eqn.\,(\ref{eqn:A(t)}),
$\phi>0$ means that $Y$ is delayed from $X$. 
Typically the time delay is shorter than $T_L$, much shorter than the evolution of the envelope function $A_0(t)$, so the two features share the same magnitude of vector potential $|\bm{A}(t_s)|$.
An interesting consequence of the stationary phase approximation in Eqn.\,(\ref{eq:shift}) is the difference in streaking amplitudes $|\bm{k}_X|{=}k$ and $|\bm{k}_Y|$ for two photoelectron features dressed by the same streaking field.
Combining Eqn.\,(\ref{eqn:A(t)}), which describes the vector potential of a circularly polarized dressing field, with the momentum shift from the stationary phase approximation $\bm{k}=e(\bm{A}(t_s) + \tau \left.\frac{d\bm{A}}{dt}\right|_{t_s})$, we find that $|\bm{k}_Y|=\lambda k$, where the factor $\lambda\equiv\sqrt{(1+(\omega_L\tau_Y)^2)/(1+(\omega_L\tau_X)^2)}$ accounts for the difference between $|\bm{k}_X|$ and $|\bm{k}_Y|$ resulting from the relative photoemission delay $\tau_Y-\tau_X$~\cite{kheifets_ionization_2022}. 
When $|\tau_Y-\tau_X|\omega_L\ll 1$\,rad, this factor is $\lambda\simeq 1$, and as $|\tau_X-\tau_Y|$ increases, the two streaking amplitudes diverge.
In this way, the streaking vector of $Y$ is given by $\bm{k}_Y = \lambda k(\cos(\kappa+\phi)\,\bm{e}_x + \sin(\kappa+\phi)\,\bm{e}_y)$.

\subsection{Impulsive Streaking Covariance}\label{sec:impulsive-streak}

We first derive the impulsive model for $K[X,Y]$ in the absence of machine fluctuations, \ie~when the only variation in the measurement is $\bm{k}$, in which case $K[X,Y]=\Cov[X,Y]$. 
As illustrated in Eqn.\,(\ref{eq:shift}), in the impulsive regime $X(\bm{r})\simeq X^0(\bm{r}-\bm{k}_X)$, and $Y(\bm{r})\simeq Y^0(\bm{r}-\bm{k}_Y)$. Substituting the Taylor expansions of $X$ and $Y$ into the definition of covariance Eqn.\,(\ref{eq:defCovXY}), we find the leading order is
\begin{multline}\label{eq:Cov-as-gradIP}
\Cov[X,Y] \simeq \Cov[\OpD_{\bm{k}_X}{X^0}, \OpD_{\bm{k}_Y}{Y^0}] \\
= \frac{\langle k^2\rangle\lambda}{2} (\nabla {X^0})^T~R(-\phi)~\nabla {Y^0}+o(k^4)~,
\end{multline}
where $\nabla$ is the 2D momentum gradient operator, $R(-\phi)$ is the 2-by-2 matrix representing the counterclockwise rotation of $-\phi$ in the $xy$-plane. 
This leading order is the inner-product between $\nabla {X^0}$ and $R(-\phi)\nabla {Y^0}$, also referred to as the gradient inner-product~(GIP),
and the delay is encoded in the rotation. 

In the impulsive regime, the streaking covariance defined in Eqn.\,(\ref{eq:defK}) is given by a bilinear operation on the unstreaked distributions, $\Cov[\OpD_{\bm{k}_X}, \OpD_{\bm{k}_Y}]{X^0}{Y^0}$.
As shown in Eqn.\,(\ref{eq:Cov-as-gradIP}), the leading order in this bilinear operator $\Cov[\OpD_{\bm{k}_X}, \OpD_{\bm{k}_Y}]$ can be written as 
\begin{equation}\label{eq:def-GIP}
M_\mathrm{GIP}\equiv \frac{\avg{k^2}\lambda}{2}(\nabla {X^0})^T\,R(-\phi)\,\nabla {Y^0} =
\frac{\avg{k^2}}{2}\hat{G}(\phi)X^0Y^0,
\end{equation}
where $\hat{G}(\phi)\equiv \lambda\sum_{i,j}R_{ij}(-\phi)\partial_i\otimes\partial_j$, with the direct product defined as $\partial_i\otimes\partial_j\equiv \frac{\partial}{\partial r_{q,i}}\frac{\partial}{\partial r_{p,j}}$ to perform partial differentiation with respect to momentum coordinates $\bm{r}_q$ and $\bm{r}_p$, respectively. 
The full expansion of $\Cov[\OpD_{\bm{k}_X}, \OpD_{\bm{k}_Y}]$ is 
\begin{widetext}
    \begin{equation}\label{eq:CovDD_fullseries}
\Cov[\OpD_{\bm{k}_X},\OpD_{\bm{k}_Y}] = \sum_{N=0}^{+\infty}2^N\gamma_N (\hat{G}(\phi) + \hat{H})^N - \sum_{n=0}^{+\infty}\sum_{m=0}^{+\infty}\gamma_n\gamma_m (\nabla^2)^n\otimes (\lambda^2\nabla^2)^m~,
\end{equation}
\end{widetext}
where  $\hat{H}\equiv\left(\nabla^2\otimes\hat{I}+ \hat{I}\otimes\lambda^2\nabla^2\right)/2$,  $\hat{I}$ is the identity operator, and $\gamma_n\equiv \avg{k^{2n}}/(2^{2n}(n!)^2)$.
According to Eqn.\,(\ref{eq:CovDD_fullseries}), the streaking covariance expands into multiple direct products between the $n_X$-th order partial derivative of $X$ and the $n_Y$-th order partial derivative of $Y$. 
We number them by the differentiation orders $(n_X+n_Y)$, \eg~the GIP term is the (1+1) order. 
The next orders are (1+3), (3+1), and (2+2), since all terms' $(n_X+n_Y)$ are even numbers, as shown by Eqn.\,(\ref{eq:CovDD_fullseries}).

 \begin{figure}[hbtp]
    \centering
    \includegraphics[width=0.7\linewidth]{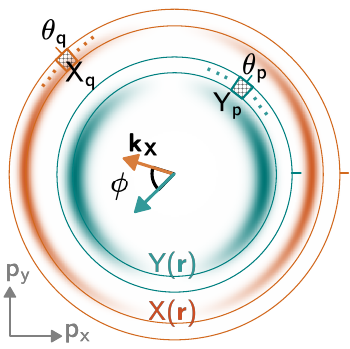}
    \caption{Notation conventions used in this work. 
    Two-dimensional MD of the continuum electron features are represented by $X(\bm{r})$~(orange shades) and $Y(\bm{r})$~(green shades).
    Reference feature $X$ is always in the impulsive regime, with streaking vector $\bm{k}=\bm{k}_X$~(orange arrow).
    When feature $Y$ is also impulsive, the angle between the two streaking vectors (arrows in respective colors) defines $\phi$.
    Angular bins are labeled by their central angular positions $\theta_q, \theta_p$ measured from $+x$ direction, with radial boundaries marked as rings in respective colors. 
    $X_q, Y_p$ denote the integral of $X(\bm{r}), Y(\bm{r})$ in the angular bins at $\theta_q, \theta_q$, respectively.
    The red arc indicates the convention for the chirality of the dressing vector potential $\bm{A}(t)$.
    }
    \label{fig:XYdef}
\end{figure}

\begin{figure*}[hbtp]
    \centering
    \includegraphics[width=0.8\linewidth]{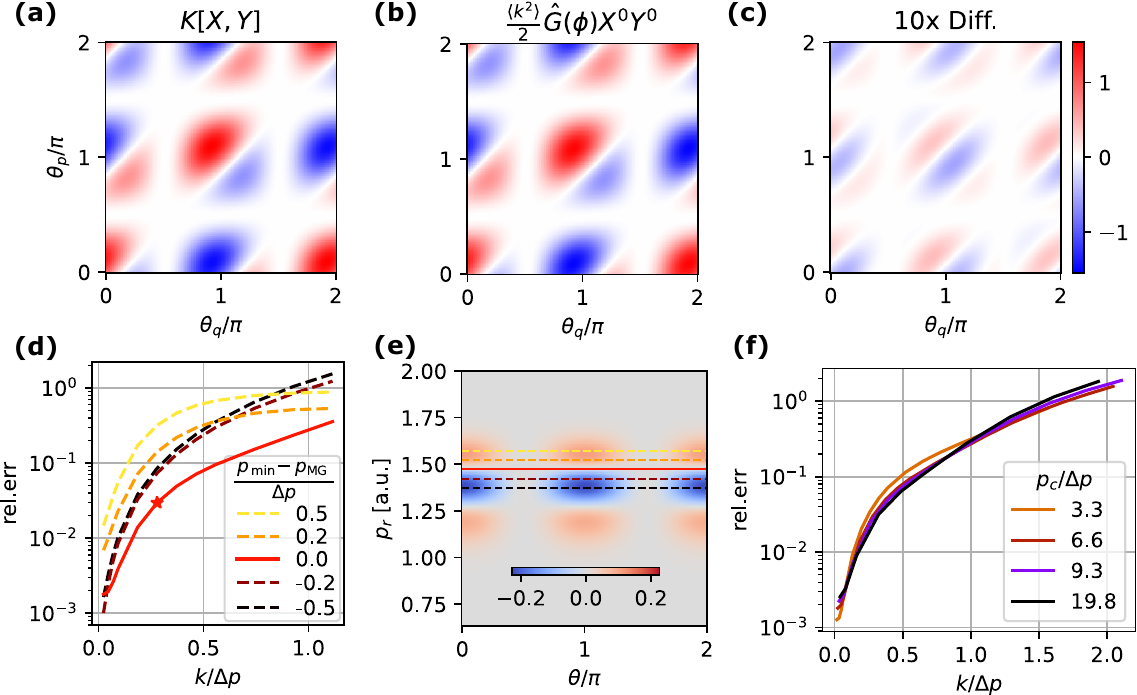}
    \caption{Streaking covariance compared to its leading order with streaking amplitude $k$. 
    Both features $X$ and $Y$ are simulated under the same condition as in Fig.\,\ref{fig:illustration}, throughout (a-e). 
    \textbf{(a)} Streaking covariance $K[X,Y]$ with $\phi=\pi/3$. 
    The lower radial boundary of the ROIs $p_{\min}$ is at the maximal gradient line $p_\mathrm{MG}$, see main text, and the upper boundary is $p_{\max}{=}2.15$\,a.u. 
    \textbf{(b)} The leading order in $K[X,Y]$, \ie~the gradient inner-product~(GIP, see Eqn.\,(\ref{eq:Cov-as-gradIP})). 
    \textbf{(c)} Difference between (a) and (b), values scaled 10x for visibility. 
    Colorbar is shared among (a-c).
    \textbf{(d)} Relative error of the GIP, depending on the normalized streaking amplitude $k/\Delta p$, with $\Delta p$ denoting the momentum width FWHM~(see main text). 
    Colors correspond to the ROI lower bound $p_{\min}$ marked in (e), among which the red corresponds to $p_{\min}{=}p_\mathrm{MG}$.
    The relative error between (a) and (b) is marked as the red star. 
    \textbf{(e)} Average differential MD from the unstreaked MD of each photoelectron feature $\E{X}-\E{X^0}$. 
    Values are normalized to $\max X^0$.
    Radial momentum $p_r{=}|\bm{r}|$.
    \textbf{(f)} Relative error of the GIP for various normalized radii $p_c/\Delta p$ of the photoelectron feature, with $p_{\min}$ chosen at the corresponding $p_\mathrm{MG}$ in each case of $p_c/\Delta p$.
    }
    \label{fig:strkCovTerms}
\end{figure*}

In a measurement, we record electron yield with finite momentum resolution. We consider the covariance between the electron yield measured in two arbitrary regions $Q$ and $P$.
Due to the bilinearity of covariance, this is identical to the regional integral of the covariance between the densities $X$ and $Y$: 
\begin{multline}
    \Cov\left[\int_{Q}d^2\bm{r}X(\bm{r}), \int_{P}d^2\bm{r}Y(\bm{r})\right] \\
    = \int_{Q}d^2\bm{r}'\int_{P}d^2\bm{r}'' \Cov[X,Y](\bm{r}', \bm{r}'')~.
\end{multline}
As illustrated in Fig.\,\ref{fig:XYdef}, we define two sets of 2D momentum regions of interest~(ROIs): $\{Q_q\}_{q=1}^{N_Q}$ for $X$ and $\{P_p\}_{p=1}^{N_P}$ for $Y$. 
We use subscripts $q, p$ on $X, Y$ as a shorthand for the regional integrals, \eg~$X_q\equiv\int_{Q_q}X(\bm{r})d^2\bm{r}$, as illustrated in Fig.\,\ref{fig:XYdef}.
Fig.\,\ref{fig:strkCovTerms}(a) shows an exemplary visualization of $K[X,Y]$, where the covariance has been calculated between $N_Q{=}N_P{=}180$ ROIs on the $p_z{=}0$ plane, each $2^\circ$-wide.
The 180 angular bins are on a ring $p_\mathrm{min} < |\bm{r}| < p_\mathrm{max}$ and labelled by their central angular positions $\theta_q$ or $\theta_p$, thus they are also referred to as angular bins. 
In Fig.\,\ref{fig:strkCovTerms}, both photoelectron features $X$ and $Y$ are simulated in the same ionization process as Fig.\,\ref{fig:illustration} but detected separately, and we set $\lambda=1$ to simulate the scenarios where the difference in photoemission delays is much shorter than the dressing field period $\omega|\tau_X-\tau_Y|\ll 1$. 
As shown in the $\phi=\pi/3$ example in Fig.\,\ref{fig:strkCovTerms}(a), one feature of $K[X,Y]$ is that the most positive part is around $\theta_p-\theta_q\oldsim \phi$. 
At small streaking amplitudes, $K[X,Y]$ is well approximated by the GIP, as shown in Fig.\,\ref{fig:strkCovTerms}(b-c).
The the relative error $\|M_\mathrm{GIP}-K[X,Y]\|_2 /\|K[X,Y]\|_2$ between panel (b) and (a) is evaluated as $0.03$, which we use to quantify the accuracy of the GIP.
The accuracy of the GIP model is insensitive to $\phi$, but it degrades as streaking amplitude $k$ increases, due to the increased contribution from the higher order terms.
As shown by the solid curve in Fig.\,\ref{fig:strkCovTerms}(c), the root-mean-squared~(rms) relative error over $\phi\in[-\pi,+\pi]$ becomes comparable to unity once $k$ exceeds the width of the photoelectron feature in momentum $\Delta p$~(quantified as the full-width at half-maximum, FWHM). 
For panels (a-b), the lower boundary $p_{\mathrm{min}}$ of the ROIs is set to the maximum point of the unstreaked MD gradient 
$p_\mathrm{MG}\equiv \arg\max_{r}\int d\theta|\nabla I^0|$
~(red solid line in Fig.\,\ref{fig:strkCovTerms}(d)), and the upper boundary is set to where the streaked MD has fallen to zero. 
This choice of limit for $p_{\min}$ optimizes the accuracy of the GIP model, because $p_\mathrm{MG}$ is close to the zero point of $\nabla^2 I^0$. 
As shown in Fig.\,\ref{fig:strkCovTerms}(c-d), when we alternate $p_{\min}$ by a fraction of $\Delta p$, the relative error is increased when $k/\Delta p <1$.

For the photoelectron features produced in direct ionization, the discrepancy between the GIP model and the streaking covariance $K[X,Y]$ generally depends on three characteristic dimensionless quantities: (1) the size of the momentum shift normalized to the momentum spread of the photoelectron feature $k/\Delta p$,  (2) the normalized radius $p_c/\Delta p$, with $p_c$ denoting the central momentum of the photoelectron feature, and (3) the ratio of the ionizing pulse duration to the streaking laser period $\Delta t_X/T_L$. 
In the impulsive regime $\Delta t_X\ll T_L$, we find that the relative error is predominantly a function of $ k /\Delta p$, with a much weaker dependence on $p_c/\Delta p$, as shown in Fig.\,\ref{fig:strkCovTerms}(f).
In the case shown in panels (a-e), both $X$ and $Y$ have a normalized radius of $p_c/\Delta p=6.6$, and when we vary $p_c/\Delta p$ by changing $\Delta t_X$ and/or the photon energy of the ionizing pulse, the relative error curve does not significantly change as long as we remain in the impulsive regime~($\Delta t_X\ll T_L$). 
The two features $X$ and $Y$ can have different $\Delta p_X, \Delta p_Y$, but we note that by defining $\Delta p$ as the geometric mean $\Delta p = \sqrt{\Delta p_X \Delta p_Y}$, the accuracy of the GIP can be well described by $k/\Delta p$.

\subsection{Contribution from Machine Fluctuations}\label{sec:machine-fluc}

We differentiate the shot-to-shot fluctuations in angular streaking experiments into two categories, the fluctuation of the streaking vector $\bm{k}$, and everything else. The fluctuations of these other parameters~(or ``machine fluctuations''), such as pulse energy and central photon energy of the ionizing pulse, also affect the single-shot MD, but the relative timing between the ionizing and dressing pulses typically does not depend on these parameters.
Thus in this work, machine fluctuations are assumed to be statistically independent from the random streaking vector.
The streaking covariance $K[X,Y]$ defined in Eqn.\,(\ref{eq:defK}) is therefore simplified by the impulsive condition as $K[X,Y]\simeq \Cov[\mathcal{D}_{\bm{k}_X}\avg{X^0},\mathcal{D}_{\bm{k}_Y}\avg{Y^0}]$. 
We note that the expected dressing-free MDs $\avg{X^0}, \avg{Y^0}$ are fixed, free from shot-to-shot variations, so the treatment of $K[X,Y]$ introduced in Sec.\,\ref{sec:impulsive-streak} are also applicable when machine fluctuations are present, by simply replacing the stable $X^0, Y^0$ with $\avg{X^0}, \avg{Y^0}$ respectively. 

Both machine fluctuations and fluctuations in $\bm{k}$ contribute to the covariance $\Cov[X,Y]$ defined in Eqn.\,(\ref{eq:defCovXY}).
According to the law of total covariance~\cite{DeGroot_probability_2010}, $\Cov[X,Y]=K[X,Y]+L[X,Y]$ consists of the streaking covariance $K[X,Y]$ defined in Eqn.\,(\ref{eq:defK}) and the contribution from machine fluctuations:
\begin{equation}\label{eq:defL}
    L[X,Y]\equiv \E{\Cov[X,Y|\bm{k}]}\simeq\E{\OpD_{\bm{k}_X}\otimes \OpD_{\bm{k}_Y}}\Cov[X^0, Y^0]~,
\end{equation}
where in the approximation, we used the impulsive streaking condition and the statistical independence between machine fluctuations and streaking vector.

\section{Methods to Retrieve Delay}\label{sec:phi-retrieval}

\begin{figure*}[hbtp]
    \centering
    \includegraphics[width=0.9\linewidth]{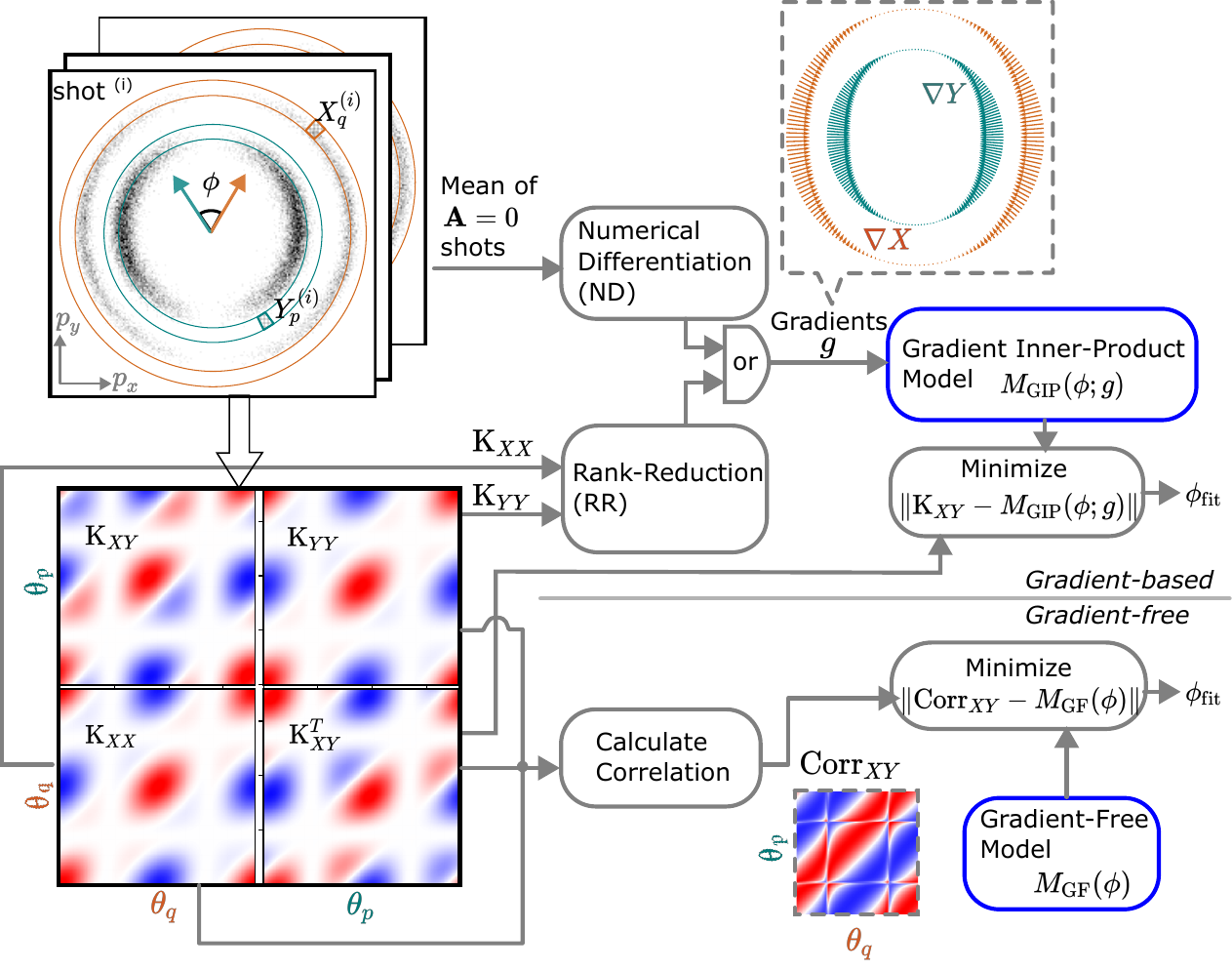}
    \caption{Overview of delay retrieval methods. 
    Two photoelectron features are imaged in momentum space over many shots~(top left). 
    Streaking vectors of the two features~(colored arrows at center) vary from shot to shot, but the angle $\phi$ between the two directions remains stable. 
    Colored rings delineate the ROI of angular bins for each feature. Based on the electron yield measured over many shots, one obtains a sample estimate of the streaking covariance matrices $\mathrm{K}_{XX}, \mathrm{K}_{XY}$ and $\mathrm{K}_{YY}$~(bottom left).
    Two approaches to gradient reconstruction~(top right) are introduced: numerical differentiation and rank-reduction, see main text.
    With either approach, the gradients~(top right inset) support the GIP model for delay retrieval from the measured $\mathrm{K}_{XY}$. 
    Alternatively, a gradient-free method~(bottom right) retrieves the delay from the sample correlation matrix $\mathrm{Corr}_{XY}$.
    }
    \label{fig:overview_delay_retrieval}
\end{figure*}

Figure\,\ref{fig:overview_delay_retrieval} illustrates three approaches for retrieving the relative angle $\phi$, which  is directly proportional to the time delay between $X$ and $Y$, from the measured momentum distribution. 
Measuring $N_s$ laser shots, we obtain a sample of electron yield MDs: $X^{(i)}, Y^{(i)}, 1{\leq}i{\leq}N_s$.
The sample covariance between the regional yields $X_q$ and $Y_p$, $(\mathrm{C}_{XY})_{qp} \equiv (\overline{X_qY_p}-\overline{X_q}~\overline{Y_p})N_s/(N_s-1)$, with $\overline{\bullet}$ the sample mean over shots,
gives an estimate of the underlying covariance $\Cov[X_q,Y_p]$. 
As introduced in Sec.\,\ref{sec:introduction}, a key feature of the covariance analysis is leveraging the angular isotropy of the $\bm{k}$ distribution to circumvent the need for single-shot knowledge of $\bm{k}$. 
While this feature avoids errors introduced control or quantification of $\bm{k}$, it prevents forming a sample estimator according to $K[X,Y]\equiv \Cov[\E{X|\bm{k}}, \E{Y|\bm{k}}]$, as an estimator of the conditional expectation $\E{\cdot |\bm{k}}$ would require binning shots by vector $\bm{k}$. 
Thus our general strategy is to remove the contribution of machine fluctuations from the sample covariance $\mathrm{C}_{XY}, \mathrm{C}_{XX}$ and $\mathrm{C}_{YY}$ to obtain a sample estimate of the streaking covariance~(or ``K-estimators'') denoted as $\mathrm{K}_{XY}, \mathrm{K}_{XX}$ and $\mathrm{K}_{YY}$.
These modelling and removal procedures are detailed in Sec.\,\ref{sec:MFdelay}.
Throughout this work, the sample estimates are denoted with subscripts~(\eg~$\mathrm{C}_{XY}$), whereas the underlying covariance and streaking covariance are with brackets~(\eg~$C[X,Y]$).

The delay retrieval methods are generally separated into two classes, gradient-based~(involving reconstruction of the MD gradient) and gradient-free. 
In the gradient-based methods, we reconstruct the density gradients of the dressing-free MD $\nabla X_q\equiv \int_{Q_q}\nabla \avg{X^0}d^2\bm{r}$ and $\nabla Y_p\equiv \int_{P_p}\nabla \avg{Y^0}d^2\bm{r}$, in order to fit the GIP model~(Eqn.\,(\ref{eq:def-GIP})) to the sample streaking covariance $\mathrm{K}_{XY}$. 
In the gradient-free method, we use the K-estimators to calculate a sample correlation matrix $\mathrm{Corr}_{XY}$ and leverage part of $\mathrm{Corr}_{XY}$ to circumvent reconstruction of the gradients. 
The procedures and benefits of these different approaches are described below.
As introduced in Sec.\,\ref{sec:introduction}, one use case of these approaches is to measure the relative photoemission delay between two features produced by the same ionizing pulse $\tau_Y-\tau_X$. 
Although $\lambda$ also depends on $\tau_Y-\tau_X$, it is much less sensitive than $\phi$ when $|\tau_Y-\tau_X|\lesssim 1/\omega_L$, so we focus on retrieving $\phi$. 
 \subsection{Gradient-based Methods}\label{sec:grad-based}
\begin{figure*}[hbtp]
    \centering
    \includegraphics[width=0.8\linewidth]{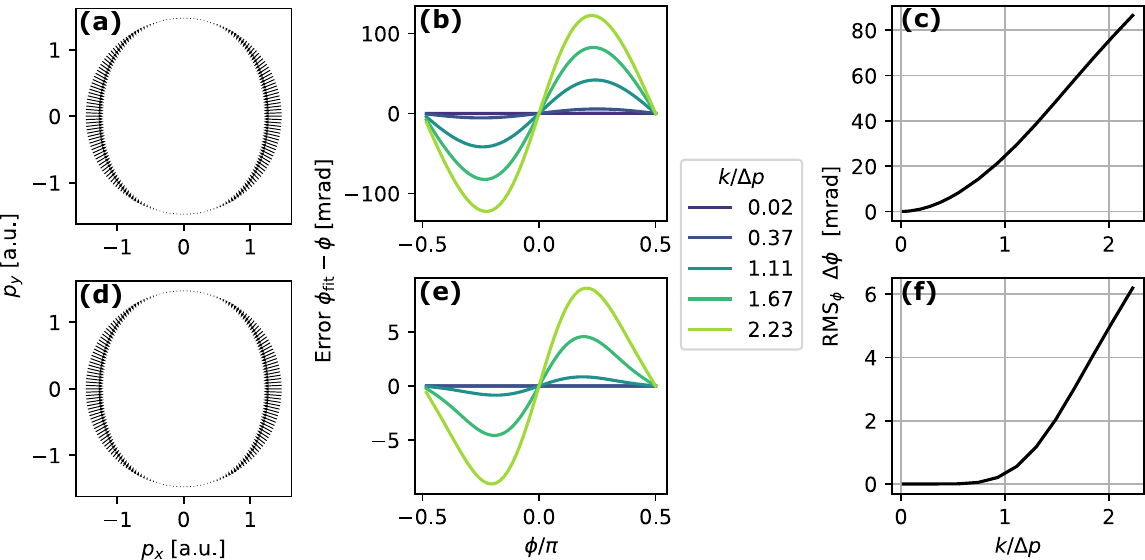}
    \caption{Accuracy of gradient-based delay retrieval methods in the noiseless limit. 
    The simulation conditions, measurement scheme, including ROI boundaries, are the same as in Fig.\,\ref{fig:strkCovTerms}(a), except with $\phi$ and $k/\Delta p$ scanned. 
    \textbf{(a-c)} With gradient reconstructed by numerical differentiation of the dressing-free MD.
    \textbf{(a)} Reconstructed gradient field. 
    \textbf{(b)} Delay retrieval error $\phi_\mathrm{fit}-\phi$ with the ND method, depending on the ground-truth $\phi$.
    Normalized streaking amplitudes $k/\Delta p$ are encoded by the color of traces.
    \textbf{(c)} The retrieval error in (b) root-mean-squared over $\phi$. 
    \textbf{(d)} Gradient reconstructed with RR, at $k/\Delta p$=0.5. 
    \textbf{(e-f)} Same as (b-c) but with RR gradients.
    }
    \label{fig:phi-extract}
\end{figure*}

To retrieve the delay, the GIP model of $K[X,Y]$ relies on the gradients $\nabla X_q, \nabla Y_p$:
\begin{small}
\begin{equation}\label{eq:phi-fit}
    \phi_\mathrm{fit} = \mathrm{argmin} \sum_{q,p}\left|
    \frac{\avg{k^2}\lambda}{2}(\nabla X_q)^T R(-\phi) \nabla Y_p  - (\mathrm{K}_{XY})_{qp}\right|^2~.
\end{equation}
\end{small}
One way to reconstruct the gradients $\nabla X_q, \nabla Y_p$ is through numerical differentiation~(ND). 
Dressing-free shots provide the sample mean of the measured distributions $\overline{X^0}$ and $\overline{Y^0}$ to estimate the expectations $\avg{X^0}$ and $\avg{Y^0}$, respectively.
With adequate momentum resolution, we can use a finite difference scheme to numerically differentiate $\overline{X^0}$, and then integrate over $Q_q$ to measure the gradient $\nabla X_q$. 
An example is shown in Fig.\,\ref{fig:phi-extract}(a) for the case in Fig.\,\ref{fig:illustration} \&\ref{fig:strkCovTerms}. 
Obtaining the gradients $\nabla X_q, \nabla Y_p$ from numerical differentiation, we fit the GIP model to the sample streaking covariance $\mathrm{K}_{XY}$, with the factor $\avg{k^2}\lambda$ treated as a free parameter in Eqn.\,(\ref{eq:phi-fit}). 
Because the amplitude ratio $\lambda$ has been absorbed into the free scaling parameter, not knowing $\lambda$ does not affect this method.
We benchmark the accuracy of the ND method in the noiseless limit: i.e. with sufficient number of shots $N_s\to\infty$ to suppress statistical noise, complete removal of machine fluctuations, and ignoring readout noise in the electron yield.
Thus we equate $\mathrm{K}_{XY}$ to the underlying $K[X,Y]$ and obtain the delay retrieval error $\Delta\phi=\phi_\mathrm{fit}-\phi$ at various ``ground-truth'' $\phi$, Fig.\,\ref{fig:phi-extract}(b). 
This systematic error is zero when $\phi$ is a multiple of $\pi/2$ and reaches a maximum at $\phi\,\sim\pm\pi/4$.
The rms error over $\phi$ increases with $k$ due to the increase in higher-order terms, as shown in panel (c), but within $k<\Delta p $ this rms error is $<0.03\,\mathrm{rad}{=}1.7^\circ$, which converts to a time delay error $<30\,\mathrm{as}$ for a $2\,\mathrm{\mu m}$-wavelength dressing field.

Another way to reconstruct the gradient is ``rank-reduction''~(RR), described in Algorithm\,\ref{algo:RR}, which employs a low-rank approximation of $K[X,X]$ and $K[Y,Y]$ to estimate the gradient.
This method can be used when each set of ROIs complete a loop on the high-momentum side of the corresponding photoelectron feature, \eg~the rings delineated in Fig.\,\ref{fig:overview_delay_retrieval}.
Within the GIP model, $K[X_{q},X_{q'}]=\avg{k^2/2}(\nabla X_{q})^T\nabla X_{q'}$, whose rank is at most 2 since $\nabla X_{q}$ is 2D. 
Thus it appears we can solve the following optimization problem given $\mathrm{K}_{XX}$:
\begin{subequations}\label{eq:rankRed}
\begin{align}
    \text{minimize }f_{\mathrm{RR}}(\xi) \equiv & \sum_{q,q'} \left|\xi_q^T\xi_{q'} - (\mathrm{K}_{XX})_{qq'} \right|^2W_{qq'}~,\label{eq:fRR}\\
    \text{subject to } l_{\mathrm{RR}}(\xi)\equiv &\sum_{q=1}^{N_Q}\left(\begin{matrix} -\sin\theta_q, & \cos\theta_q\end{matrix}\right)  \xi_q a_q= 0~,\label{eq:loop=0}
\end{align}
\end{subequations}
where $\xi=(\xi_1,\cdots,\xi_{N_Q})\in \mathbb{R}_{2{\times}N_Q}$ represents the gradient field, $a_q$ is the arc length of the angular bin at $\theta_q$ and $W$ is a $N_Q\times N_Q$ matrix weights, which can be configured to combat the effect of readout noise as described below.
The reconstructed gradient field is given by the optimal point $\xi^*$ of Eqn.\,(\ref{eq:rankRed}) up to a global factor, $\xi_q^* = \sqrt{\avg{k^2/2}}\nabla X_q$.
Applying the same procedure to $\mathrm{K}_{YY}$, we obtain the optimal point $\eta^*_p=\sqrt{\avg{k^2/2}}\lambda \nabla Y_p$.
Thus the GIP model described in Eqn. (\ref{eq:Cov-as-gradIP}) can be rewritten as  $M_\mathrm{GIP}=(\xi^*)^TR(-\phi)\eta^*$, which we then substitute into Eqn.\,(\ref{eq:phi-fit}) to retrieve $\phi_\mathrm{fit}$. 

The minimization of the objective function $f_{\mathrm{RR}}$ is closely related to principal component analysis~(PCA).
The weight matrix can be configured as uniform $W_{qq'}=1$ when the signal-to-noise ratio is high in the electron yield readout process.
When the readout signal-to-noise ratio is low, we recommend ignoring the diagonal band of $\mathrm{K}_{XX}$ by setting the diagonal band in $W$ to zero~(\eg~$W_{qq'}=1-\delta_{qq'}$).
Guidance for defining is detailed in Sec.\,\ref{sec:shotnoise}. 
With a uniform $W$, a minimal point of $f_{\mathrm{RR}}$ is given by the PCA result $\xi^{\mathrm{P}} = \left(\sqrt{s^{(1)}} v^{(1)}, \sqrt{s^{(2)}} v^{(2)}\right)^T$~\cite{eckart_approximation_1936}, where $s^{(\alpha)}$ is the $\alpha$-th largest eigenvalue of $\mathrm{K}_{XX}$ and $v^{(\alpha)}$ is the corresponding eigenvector~(in column vector form). 
When $W$ is non-uniform, \eg~$W_{qq'}=1-\delta_{qq'}$, we can minimize $f_{\mathrm{RR}}$ by optimizing $\xi^{\mathrm{P}}$ using gradient descent. 

The objective function $f_{\mathrm{RR}}$ is invariant under any global orthogonal transformation $O{:}~\xi_q\mapsto O\xi_q$, but only certain minimal points can satisfy the zero-loop constraint Eqn.\,(\ref{eq:loop=0}), which is a general property of a gradient field $\oint \nabla X^0\cdot d\bm{l} = 0$.
Thus obtaining one minimal point $\xi^{\mathrm{P}}$, we solve for an orthogonal matrix $O$ such that $l_\mathrm{RR}(O\xi^\mathrm{P})=0$
This is straightforwardly achieved by parameterizing $O$ with the rotation angle and parity. 
An alternative to the zero-loop constraint is maximizing the gradient-flux $j_{\mathrm{RR}}(O\xi^{\mathrm{P}})\equiv-\sum_{q}(\cos\theta_q,~\sin\theta_q)O\xi^{\mathrm{P}}_q a_q$ with $O$, which necessarily satisfies the zero-loop constraint and additionally breaks the remaining discrete degeneracy, as proven in Supplemental Sec.\,2.

\begin{algorithm}[H]
\caption{Rank-reduction gradient reconstruction}\label{algo:RR}
\begin{algorithmic}
\State Perform PCA on $\mathrm{K}_{XX}$, obtaining $\xi^{\mathrm{P}}\gets(\sqrt{s^{(1)}}v^{(1)}, \sqrt{s^{(2)}}v^{(2)})^T$;
\If{$W$ is not uniform}
\State Update $\xi^{\mathrm{P}}$ to minimize $f_{\mathrm{RR}}(\xi^{\mathrm{P}})$ using gradient descent;
\EndIf
\State $O^*\gets\arg\max_{O} j_{\mathrm{RR}}(O\xi^{\mathrm{P}})$ with $\xi^{\mathrm{P}}$ fixed;
\State \Return $\xi^*=O^*\xi^{\mathrm{P}}$;
\end{algorithmic}
\end{algorithm}

Similar to the results in Fig.\,\ref{fig:phi-extract}(b-c), we benchmark the RR method under various $\phi$ and $k$ in the noiseless limit as mentioned above, as shown in panels (d-f).
The gradient fields $\nabla X_q$ and $\nabla Y_p$ are reconstructed from $\mathrm{K}_{XX}$ and $\mathrm{K}_{YY}$, respectively, and we show the resultant $\nabla X_q$ in panel~(d). 
Similar to the ND method, the delay retrieval error of the RR method is minimized when $\phi$ is a multiple of $\pi/2$.
The rms error increases with $k$, but the magnitude of the error is notably smaller than with the ND method. 
As shown in Fig.\,\ref{fig:phi-extract}(f), the rms error is ${<}1$\,mrad within $k{<}\Delta p$, corresponding to ${<}1$\,as for a $2\,\mathrm{\mu m}$ dressing field.

The main reason for the higher accuracy of the RR method lies in the reconstructed ``gradient''.
The ND gradients give the first-order derivatives of the dressing-free MD $\nabla X_q$ and $\nabla Y_p$, independent from the streaking amplitude $k$.
The gradient reconstructed by the RR procedure, in contrast, deviates from the first-order derivative as $k$ increases.
To the next lowest order, the RR gradients are 
\begin{subequations}\label{eq:RRgrad_renorm}
\begin{align}
\xi_q &= \sqrt{\frac{\avg{k^2}}{2}}\int_{Q_q}d^2\bm{r}\nabla \left(1+\frac{\avg{k^4}}{8\avg{k^2}}\nabla^2\right) \avg{X^0} + o(k^4)~, \\
\eta_p &= \sqrt{\frac{\avg{k^2}}{2}}\lambda\int_{P_p} d^2\bm{r}\nabla \left(1+\frac{\avg{k^4}\lambda^2}{8\avg{k^2}}\nabla^2\right) \avg{Y^0} + o(k^4)~,
\end{align}
\end{subequations}
which include the third-order derivatives in addition to the first-order (see Supplementary Sec.\,1.2). 
As a result, when using the RR gradients, the inner product $\xi_q^T R(\phi) \eta_p$ not only captures the (1+1) order of $K[X_q,Y_p]$, but also the (1+3) and (3+1) orders.
In contrast when using the ND gradient, only the (1+1) order is modelled. 
Since the errors in the noiseless limit arise from the finite accuracy of the GIP model and are insensitive to the normalized radius $p_c/\Delta p$, the result in Fig.\,\ref{fig:phi-extract} is general across $p_c/\Delta p$.
 \subsection{Gradient-free Method}
In some experiments, gradient reconstruction may be challenging.
For example, the numerical differentiation becomes infeasible when the detector lacks angular resolution, and RR becomes infeasible when the available ROIs cannot complete a loop. 
In this case, it is often possible to identify a part of the streaking correlation matrix $R[X_q,Y_p] \equiv {K[X_q,Y_p]} / {\sqrt{K[X_q,X_q]K[Y_p,Y_p]}} $ that can be used to extract the photoemission delay without requiring knowledge of the gradient.

In the angular regions where the gradient is predominantly along the radial direction $ |r^{-1}\partial_\theta X_q| \ll |\partial_r X_q|$, $|r^{-1}\partial_\theta Y_p| \ll |\partial_r Y_p|$~(also referred to as the ``radial regions''), the streaking covariance is approximated as 
\begin{equation}\label{eq:cov_rGradOnly}
K[X_q,Y_p] \approx \frac{\avg{k^2}}{2} \cos(\phi-\theta_p+\theta_q) (\partial_rX_q) (\lambda\partial_rY_p)~.
\end{equation}
The $\cos(\phi-\theta_p+\theta_q)$ factor explains the positive ridge around $\theta_p-\theta_q\sim \phi$ in the covariance matrix. 
The radial gradients in Eqn.\,(\ref{eq:cov_rGradOnly}), including the factor $\lambda$, cancel out, \ie~$R\left[X_q, Y_p\right] \approx \cos(\phi-\theta_p+\theta_q)$. 
Therefore we define the gradient-free model as $M_\mathrm{GF} \equiv a\cos(\phi-\theta_p+\theta_q)$, with free parameters $a$ and $\phi$.
From the measured K-estimators, we calculate the sample correlation matrix $\mathrm{Corr}_{XY}=\mathrm{K}_{XY} /\sqrt{\mathrm{K}_{XX}\mathrm{K}_{YY}}$, and minimize the mean-squared error in the radial region between $\mathrm{Corr}_{XY}$ and the model $M_\mathrm{GF}$.

\begin{figure*}[hbtp]
    \centering
    \includegraphics[width=0.8\linewidth]{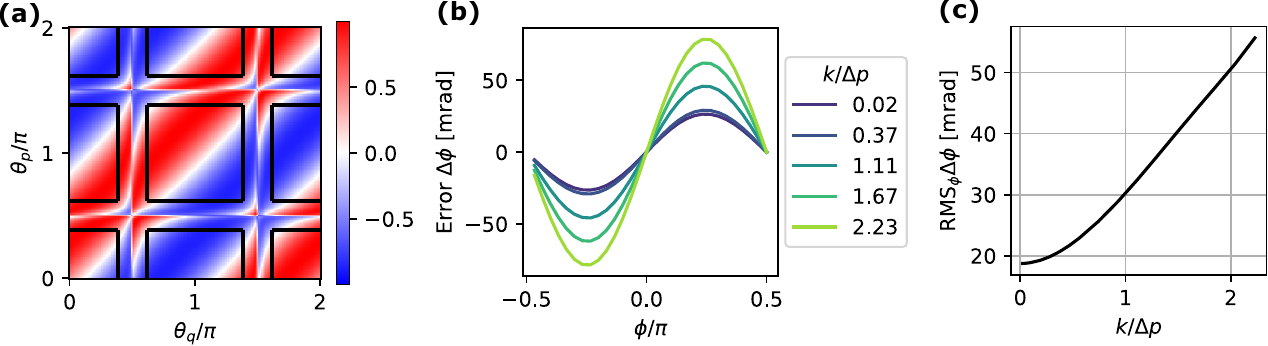}
    \caption{Accuracy of gradient-free methods in the absence of machine fluctuations. 
    The simulation conditions, measurement scheme, including ROI boundaries, are the same as in Fig.\,\ref{fig:strkCovTerms}(a). 
    \textbf{(a)} Sample correlation matrix $\mathrm{Corr}_{XY}$ at $\phi=\pi/3$. 
    The radial regions are enclosed by black rectangular boxes.
    \textbf{(b-c)} Same as Fig.\,\ref{fig:phi-extract} except using the gradient-free delay retrieval method.
    }
    \label{fig:phi-extract_gf}
\end{figure*}

The radial region is determined from the dressing-free MD. 
For the case shown in Figs.\,\ref{fig:strkCovTerms} and \ref{fig:phi-extract}, the anisotropy parameter is $\beta_2=2$, the radial region defined by $|r^{-1}\partial_\theta X_q| < 0.2 |\partial_r X_q|$ is $138^\circ$-wide around each antinode of the dipole feature. 
For lower anisotropy parameters, the radial region is larger, since an isotropic feature has no angular gradient component.
As shown in Fig.\,\ref{fig:phi-extract_gf}(a), within the radial region, the correlation matrix is well described by the gradient-free model $M_\mathrm{GF}$. 
By fitting $M_\mathrm{GF}$ to the measured $\mathrm{Corr}_{XY}$ in the radial region, we obtain $\phi_\mathrm{fit}$, and the error $\phi_\mathrm{fit}-\phi$ is shown in Fig.\,\ref{fig:phi-extract_gf}(b \& c).
Similar to the gradient-based methods, the error is zero at $\phi= 0,\pm\pi/2$, but the rms error across $\phi\in[0,2\pi]$ does not vanish when $k\to 0$.
This residual error at small streaking amplitude arises from ignoring the non-radial regions, and in the case shown in Fig.\,\ref{fig:phi-extract_gf}, it amounts to $20$\,mrad rms.  
The systematic error of the gradient-free method also increases with $k$, but the increase is slower than for the ND method. 
In this case, comparing Fig.\,\ref{fig:phi-extract_gf}(c) to Fig.\,\ref{fig:phi-extract}(c), we find the gradient-free method is more accurate than the ND method when $k\gtrsim \Delta p$, but remains less accurate than the RR method.
Another difference between the gradient-free and gradient-based methods is the dependence on $p_c/\Delta p$: the gradient-free method is generally more accurate with higher $p_c/\Delta p$, since the magnitude of the radial gradient relative to the angular gradient increases.

\section{Discussion}\label{sec:discussions}
\subsection{Machine Fluctuations}\label{sec:MFdelay}

When implementing analysis procedures that make use of fluctuations in the measured data, it is important to understand how instabilities in the measurement affect the measured correlation, in this case $\mathrm{C}_{XY}$.
When using an attosecond x-ray free electron laser there may be fluctuations in the x-ray parameters, for example the pulse energy and/or the central photon energy. 
In this section we describe how to manage the effect of these additional fluctuations in the delay extraction procedure. 

\subsubsection{Additional Contribution from Machine Fluctuations in the Impulsive Regime}
\begin{figure*}[hbtp]
    \centering
    \includegraphics[width=0.8\linewidth]{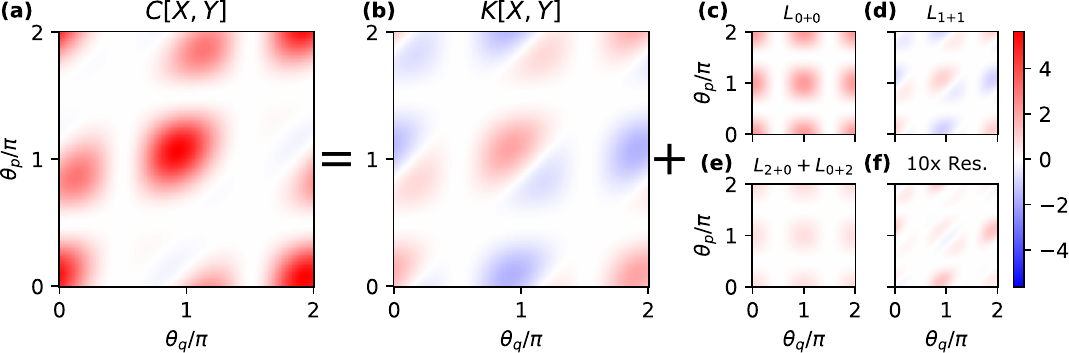}
    \caption{Decomposition of covariance into contributions from streaking effects and machine fluctuations. 
    \textbf{(a)} Total covariance $\Cov[X,Y]$ at $\phi=\pi/3$.
    In this case, the intensity and central photon energy of the ionizing pulse and streaking amplitude are randomly distributed~(see main text). 
    The average machine conditions $\avg{k}/\Delta p=0.26$, measurement scheme, including ROI boundaries, are the same as in Fig.\,\ref{fig:strkCovTerms}(a-b). 
    \textbf{(b)} Streaking covariance $K[X,Y]$.
    \textbf{(c-f)}
    The machine fluctuation $L[X,Y]$ expanded into different $(n_X+n_Y)$ orders:
    \textbf{(c)} (0+0) order $L_{0+0}=\Cov[X^0,Y^0]$, \ie~the unstreaked covariance,
    \textbf{(d)} (1+1) order $\avg{k^2/2}\hat{G}(\phi)L_{0+0}$,
    \textbf{(e)} sum of (2+0) and (0+2) orders $\avg{k^2/2}\hat{H}L_{0+0}$,
    \textbf{(f)} Higher-order terms~($n_X{+}n_Y{\geq}4$), amplified 10x for visibility.
    }
    \label{fig:K+L}
\end{figure*}

As discussed in Sec.~\ref{sec:machine-fluc}, the measured covariance is the arithmetic sum of the streaking covariance and the machine fluctuations $C[X,Y]=K[X,Y]+L[X,Y]$.
In the the impulsive regime,
$L[X,Y]\simeq\E{\OpD_{\bm{k}_X}\otimes \OpD_{\bm{k}_Y}}\Cov[X^0, Y^0]$, as written in Eqn.\,(\ref{eq:defL}).
Similar to $K[X,Y]$ in the impulsive regime, $\E{\OpD_{\bm{k}_X}\otimes \OpD_{\bm{k}_Y}}$ also expands into direct products of partial derivative operators numbered by the orders $(n_X+n_Y)$. 
In contrast to $K[X,Y]$, $L[X,Y]$ has a leading order of (0+0) given by $L_{0+0}\equiv\Cov[X^0,Y^0]$, \ie~the covariance of the unstreaked MD.
The (1+1) order of $L[X,Y]$ is given by the $\avg{k^2/2}\hat{G}(\phi)$ operator acting on the two-point function $\Cov[X^0,Y^0]$, and the sum of (0+2) and (2+0) terms amounts to $\avg{k^2/2}\hat{H}L_{0+0}$. 
The decomposition of the total covariance $\Cov[X,Y]$ in the presence of machine fluctuations is shown in Fig.\,\ref{fig:K+L}.
The parameters used in Fig.\,\ref{fig:K+L} are the same as those used in Fig.\,\ref{fig:strkCovTerms}, but machine fluctuations have been introduced in the following manner.
The central photon energy is normally distributed in 2\,eV FWHM around 40.8\,eV, the ionizing pulse energy follows a Gamma distribution $\Gamma(\alpha{=}\beta{=}2)$, and the streaking amplitude follows a Rayleigh distribution with an expected $\avg{k}=0.06\,\mathrm{a.u.}$. 
As shown in Fig.\,\ref{fig:K+L}(a), the total covariance is mostly positive, with a positive ridge at roughly $\theta_p-\theta_q\oldsim \phi$.
The streaking covariance $K[X,Y]$, shown in panel~(b), is nearly identical to Fig.\,\ref{fig:strkCovTerms}(a), since the average machine condition remains the same.
The composition of $L[X,Y]$ is shown in Fig.\,\ref{fig:K+L}(c-f). 
The unstreaked covariance $L_{0+0}$ is positive, which explains why the total covariance is overall more positive than the streaking covariance.
At the same time, $L_{0+0}$ is independent from the relative timing between the two features. 
The next order consists of the (1+1) term $\langle k^2/2\rangle \hat{G}(\phi)L_{0+0}$ and the sum of (0+2) and (2+0) terms $\langle k^2/2\rangle \hat{H}L_{0+0}$, as shown in panels (d-e).
The remaining higher-order terms in $L$ are negligible in this case, as shown in panel (f).

\subsubsection{Accounting for Machine Fluctuations in Delay Retrieval}
As mentioned in Sec.\,\ref{sec:grad-based}, the general strategy for obtaining the K-estimators~($\mathrm{K}_{XY}$,  $\mathrm{K}_{XX}$, and $\mathrm{K}_{YY}$) is to remove the contribution from machine fluctuations from the corresponding sample covariances $\mathrm{C}_{XY}, \mathrm{C}_{XX}$, and $\mathrm{C}_{YY}$. 
Here we discuss two procedures for this removal: (1) using partial covariance with respect to a fluctuating global parameter(s), and (2) subtracting the unstreaked covariance.

In the linear regime of light-matter interactions, the electron yield depends on the pulse energy of the ionizing radiation, the sample density, and the detector gain. 
The combined effect of all these parameters can be described by a global scaling factor $F$, which is uniform across momentum space and fluctuates from shot to shot.
As long as $F$ is measured on a single-shot basis, taking the partial covariance with respect to $F$ removes the correlation resulting from the linear dependence of $X,Y$ on $F$: 
\begin{equation}\label{eq:pcov-def}
\PCov[X,Y;F] \equiv \Cov[X,Y] - \Cov[X,F] \Cov[F,F]^{-1} \Cov[F,Y]~,
\end{equation}
whose corresponding sample estimate is $\mathrm{PC}_{XY;F} \equiv \mathrm{C}_{XY} - \mathrm{C}_{XF} \mathrm{C}_{FF}^{-1} \mathrm{C}_{FY}$~\cite{Johnson_Wichern_2007}.
In this case, we can decompose the two features $X, Y$ as the product of normalized distributions $U$ and $V$ with $F$, $X=FU, Y=FV$.
$U$ and $V$ depend on the streaking vector $\bm{k}$ but are statistically independent from $F$. 
Combining Eqn.\,(\ref{eq:pcov-def}) with the law of total covariance, the partial covariance is $\PCov[X,Y;F]=\avg{F^2}\Cov[U,V]=\avg{F^2}(K[U,V]+L[U,V])$.
The term $\avg{F^2}K[U,V]$ can be recognized as the streaking covariance $K[X,Y]=\avg{F}^2K[U,V]$ scaled by the factor $\avg{F^2}/\avg{F}^2\geq 1$.
The second term $\avg{F^2}L[U,V]$ is smaller than the machine fluctuation $L[X,Y]$ by a positive definite value:
\begin{equation}\label{eq:Lpcov}
L[X,Y] - \avg{F^2} L[U,V]=\Cov[F,F]~\E{\E{U|\bm{k}}\E{V|\bm{k}}}~.
\end{equation}
Therefore taking partial covariance has removed this difference that is proportional to $\Cov[F,F]$ and revealed the streaking covariance.

\begin{figure}[htbp]
    \centering
    \includegraphics[width=\linewidth]{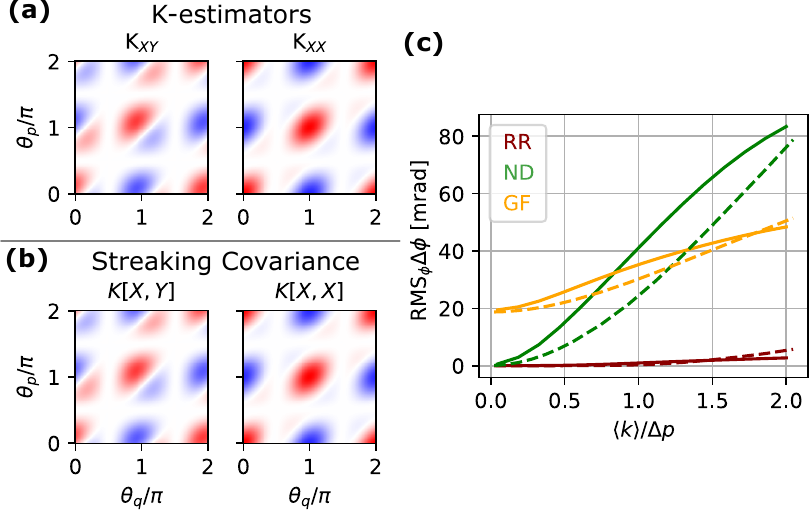}
    \caption{Impact of machine fluctuation on the accuracy of delay retrieval, in the same case as Fig.\,\ref{fig:K+L}.
    \textbf{(a)} Sample estimate of streaking covariance, see main text. 
    $\mathrm{K}_{XY}$ is again exemplified at $\phi=\pi/3$.
    \textbf{(b)} The underlying streaking covariance.
    Color scales in (a) and (b) are symmetric about zero and bounded by the maximum absolute value in each map.
    \textbf{(c)} Delay retrieval rms error as $\avg{k}$ is scanned, of the three methods: ND-gradient~(green solid), RR-gradient~(red solid), and the gradient-free method~(yellow solid). 
    Dashed curves in corresponding colors are of each method in the absence of machine fluctuations, taken from Fig.\,\ref{fig:phi-extract}(c)(f) and Fig.\,\ref{fig:phi-extract_gf}(c).
    }
    \label{fig:phi-extract_mfluc}
\end{figure}

Another method for removing effect of machine fluctuations involves calculating the difference between measurements made with the dressing field and measurements made in the absence of the dressing field.
If the sample covariance has been calculated, we use the difference $\mathrm{C}_{XY} - \mathrm{C}_{X^0Y^0}$ as $K_{XY}$.
In this case $K_{XY}$ becomes the sum of $K\left[X,Y\right]$ and all terms of the machine fluctuation covariance $L\left[X,Y\right]$ higher than $L_{0+0} = \mathrm{C}_{X^0Y^0}$, as shown in Fig.\,\ref{fig:K+L}.
An improved estimate of $K\left[X,Y\right]$ can be calculated using the partial covariance, $\mathrm{K}_{XY} = \mathrm{PC}_{XY;F} - \mathrm{PC}_{X^0Y^0;F}$.
In Fig.\,\ref{fig:phi-extract_mfluc}, we demonstrate the performance of $\mathrm{K}_{XY}$ as an estimate of 
the underlying streaking covariance $K[X,Y]$ in this case.
Panels (a) and (b) compare the sample estimates $\mathrm{K}_{XY}, \mathrm{K}_{XX}$ to the respective underlying streaking covariance $K[X,Y], K[X,X]$.
As shown in panel (c), the accuracy of all three delay retrieval methods~(solid curves) described in Sec.\,\ref{sec:phi-retrieval} is generally worse than without machine fluctuations~(dashed curves), but the change in the error is less than 20\,mrad within $\avg{k}/\Delta p <2$. 
This change in systematic error arises from the residual machine fluctuation contribution in the K-estimators.
We note that to apply this subtraction procedure, it is important to acquire comparable amount of unstreaked shots as the streaked shots, to ensure the statistical noise from the unstreaked covariance does not dominate the noise in the subtraction result.

\subsubsection{Gradient Validation in Rank Reduction}

\begin{figure*}[hbtp]
    \centering
    \includegraphics[width=\linewidth]{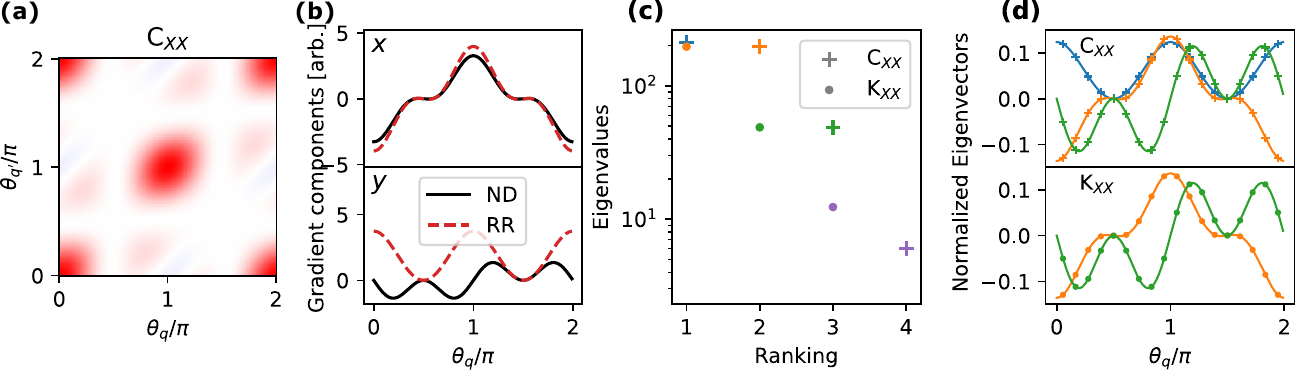}
    \caption{Validation of RR gradient in the presence of machine fluctuations, in the same case as Fig.\,\ref{fig:K+L}. 
    \textbf{(a)}~Sample covariance $\mathrm{C}_{XX}$.
    \textbf{(b)}~Gradient reconstructed from $\mathrm{C}_{XX}$ with the rank-2 RR procedure~(red dashed) is qualitatively different from the ND gradient~(black solid).
    \textbf{(c)}~Leading eigenvalues of $\mathrm{C}_{XX}$~(crosses) and $\mathrm{K}_{XX}$~(dots). Colors represent the characters of the corresponding eigenvectors in (d).
    \textbf{(d)}~Eigenvectors displayed in matching markers and colors with (c).
    }
    \label{fig:RR_mfluc}
\end{figure*}

The residual machine fluctuation contribution in $\mathrm{K}_{XX}$ is a potential systematic issue for the RR method, because the optimization problem Eqn.\,(\ref{eq:rankRed}) finds a rank-2 approximation of $\mathrm{K}_{XX}$.
This is motivated by the fact that the gradient inner-product in $K[X,X]$ has a rank of 2, but the residual machine fluctuations mix with the streaking covariance. 
Therefore in the presence of significant machine fluctuations, the gradient reconstructed with RR requires a validation.
If the RR gradient differs significantly from the ND gradient, it should be considered invalid, and delay retrieval should not proceed using such a gradient. 
We can quantify the similarity between RR gradient $\xi^*$ and ND gradient $g=(\nabla X_1,\cdots,\nabla X_{N_Q})$ by the average cosine similarity $S_C(a,b)\equiv |ab^T|/(\|a\|\|b\|)$ of the components $\mathcal{S}\equiv  (S_C(\xi^{*x},g^x) + S_C(\xi^{*y},g^y))/2$. For instance, applying the RR procedure to the $\mathrm{C}_{XX}$ shown in Fig.\,\ref{fig:RR_mfluc}(a) without any removal of the machine fluctuations, the resultant RR gradient is significantly different from the ND gradient, as shown in panel~(b). 
The average cosine similarity is $\mathcal{S}=0.5$ resulting from similarity in $x$ component but orthogonality in $y$.
Given that there are two components~($x$ and $y$) in the gradient field, we suggest a suitable threshold at $\mathcal{S}>(1/2+1)/2=3/4$ for the validity of RR gradient. 
A higher threshold is less desirable, because a valid RR gradient deviates from the first-order partial derivative and outperforms the ND method in delay retrieval accuracy, as pointed out by Eqn.\,(\ref{eq:RRgrad_renorm}).

When an acceptable similarity between the RR gradient and ND gradient cannot be obtained, it is possible to generalize to a rank-3 approximation, \ie~to minimize $f_\mathrm{RR}(\zeta)$ with $\zeta\in \mathbb{R}_{3\times N_Q}$.
We obtain a minimal point $\zeta^{\mathrm{P}}$, subject to the constraint that the rows of $\zeta^{\mathrm{P}}$ are orthogonal.
We refer to the row vectors in $\zeta^{\mathrm{P}}$ as principal components.
For each pair of principal components, we stack them as the $\xi^\mathrm{P}\in\mathbb{R}_{2\times N_Q}$, maximize the flux $j_\mathrm{RR}(O\xi^\mathrm{P})$ as in Algorithm\,\ref{algo:RR}. 
One should validate the reconstructed gradient $\xi^*$ with the ND gradient. 
When no pair of principal components can result in a valid $\xi^*$, the RR method is not recommended.
If there are multiple pairs that results in a valid $\xi^*$, we choose the valid n $\xi^*$ with the largest spectral norm.

When the measured dressing-free MD has inversion symmetry, $X^0(\bm{r})=X^0(-\bm{r})$, each principal component is either even $f(-\bm{r})=f(\bm{r})$ or odd $f(-\bm{r})=-f(\bm{r})$.
This parity serves as a useful guide for selecting the pair of principal components:  both $x$ and $y$ components of the gradient field must be odd.
As shown in Fig.\,\ref{fig:RR_mfluc}(c-d), the largest principal component of $\mathrm{C}_{XX}$~(blue) has even parity, whereas the second~(orange) and third~(green) are odd. 
After removing the machine fluctuation contribution from $\mathrm{C}_{XX}$, the top two components of the resultant $\mathrm{K}_{XX}$ both have odd parity, and strongly resemble the second and third components of $\mathrm{C}_{XX}$.
A rank-3 approximation of $\mathrm{C}_{XX}$ with the aforementioned validation procedure results in the same RR gradient field as the reconstruction with $\mathrm{K}_{XX}$.

 \subsection{Shot Noise}\label{sec:shotnoise}
Shot noise is ubiquitous in electron spectroscopy, arising from the particle nature of electrons. 
The measured electron yield in a momentum region $Q$ follows a Poisson distribution, with expectation $\int_Q I(\bm{r})d\bm{r}$, where the MD, $I(\bm{r})$ varies from shot to shot.
To understand the effect of shot noise on our measurement scheme, we denote $E_\mathrm{fx}$ as the expected electron yield at each set of ROIs.
We also refer to this quantity as the ``flux'', since it represents an average number electron counts per shot.
Here we investigate the impact of shot noise on the delay retrieval methods, assuming both sets of ROIs $\{Q_q\}$ and $\{P_p\}$ receive the same electron flux. 
We simulate $N_s$ measured shots by Poisson-sampling the MD, and apply the delay retrieval methods to this simulated data set. 
We repeat the Poisson sampling 10 times for each $E_\mathrm{fx}$, and the mean $\overline{\Delta\phi}{=}\overline{\phi_\mathrm{fit}}-\phi$ and standard deviation $\sigma[\Delta\phi]{=}\sigma[\phi_\mathrm{fit}]$ over the repetitions quantify the systematic error and the statistical error, respectively.

\begin{figure*}[hbtp]
    \centering
    \includegraphics[width=\linewidth]{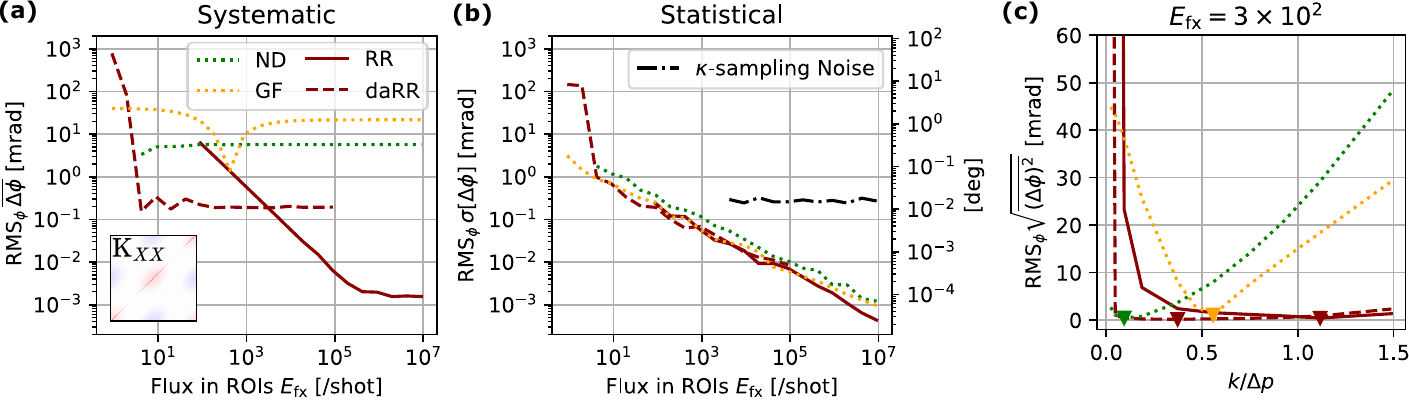}
    \caption{Effects of shot noise on the delay retrieval, in the absence of machine fluctuations. 
    Shot noise is introduced to the case shown in Fig.\,\ref{fig:phi-extract}. 
    For each level of electron flux, $N_s=7.2\times 10^5$ shots are simulated with shot noise and uniformly distributed $\kappa$, to test the delay retrieval methods at various $\phi$: ND-gradient~(green dotted), RR-gradient~(red solid), diagonal-agnostic RR-gradient~(red dashed), and the gradient-free method~(yellow dotted).
    Over 10 repetitions, the mean $\overline{\Delta \phi}$ and standard deviation $\sigma[\Delta \phi]$ quantify the systematic error and statistical error, respectively.
    \textbf{(a)} Systematic error at $k/\Delta p{=}0.5$, of the methods
    The inset in (a) shows the K-estimate $\mathrm{K}_{XX}$ at $E_\mathrm{fx}=18$. 
    \textbf{(b)} Statistical error at $k/\Delta p{=}0.5$.
    When $\kappa$ is sampled from $\mathcal{U}[0,2\pi]$ instead of strictly uniform, the statistical error of the gradient-based methods asymptotes to a noise floor due to finite $N_s$~(black dot-dashed).
    \textbf{(c)} Total error of the methods $(\overline{(\Delta \phi)^2})^{1/2}$ over $k/\Delta p$, at a fixed $E_\mathrm{fx}=3\times 10^2$. 
    For each method, the triangle marks the minimum.
    }
    \label{fig:counting_noise}
\end{figure*}

Since shot noise is typically independent across non-intersecting regions,
it does not generally induce systematic error in $\mathrm{K}_{XY}$.
There is an exception in the vicinity of the diagonal entries in $\mathrm{K}_{XX}$ and $\mathrm{K}_{YY}$, since the corresponding pairs of ROIs have correlated shot noise. 
These entries in $\mathrm{K}_{XX}$ and $\mathrm{K}_{YY}$ are positively biased because the shot noise is unaccounted for in our model. 
For example, when regions $Q_{1}$ and $Q_{2}$ both include a contribution from the same detector pixel, $(\mathrm{K}_{XX})_{12}$ is affected by such positive bias, \ie~$\avg{(\mathrm{K}_{XX})_{12}}-K[X_1,X_2]>0$.
The critical distance between the pair of ROIs is the noise correlation length of the detector.
When the distance between two regions is shorter than this noise correlation length, a bias can be introduced.

The ND method relies on $\overline{X^0}, \overline{Y^0}$, and $\mathrm{K}_{XY}$, as illustrated in Fig.\,\ref{fig:overview_delay_retrieval}, all of which are free from bias induced by shot noise. 
Thus the systematic error of the ND method is independent of shot noise but remains at the noiseless limit, as shown by the green trace in Fig.\,\ref{fig:counting_noise}(a). 
On the other hand, the RR method is affected by the positive bias in part of the K-estimators that arises from shot noise, as visualized in the inset of Fig.\,\ref{fig:counting_noise}(a). 
The impact of the bias can be mitigated by assigning zero weights to these positively biased entries of $\mathrm{K}_{XX}$ in the RR procedure.
In simulation, we can set $W_{qq'}=1-\delta_{qq'}$, since it is straightforward to ensure the independence of the Poisson-sampling process between non-intersecting regions.
We refer to this variant of the RR method, which is also free from the systematic errors arising from shot noise, as the diagonal-agnostic rank-reduction~(daRR) method.
This variation improves the accuracy of the delay extraction in low-count scenarios. 
In high-count scenarios, the daRR method is less accurate than the RR method, due to the small induced bias in the reconstructed gradient. 
As shown in  Fig.\,\ref{fig:counting_noise}(a), the point at which the RR method becomes more accurate than the daRR methods is $E_\mathrm{fx}/N_Q\approx 10$ for the parameters considered.
As $E_\mathrm{fx}\to\infty$, the systematic error of the RR method diminishes as $\propto 1/E_\mathrm{fx}$, approaching the noiseless limit.

The bias in $\mathrm{K}_{XX}$ and $\mathrm{K}_{YY}$ also affects the gradient-free delay retrieval method.
Here, shot noise affects the sample correlation matrix $\mathrm{Corr}_{XY}$ \textit{via} the positive bias in  $(\mathrm{K}_{XX})_{qq}$ and $(\mathrm{K}_{YY})_{pp}$.
In the case of $k/\Delta p=0.5, p_c/\Delta p=6.6$,  the systematic error of the gradient-free method exhibits two plateaus at both the low-count and the high-count ends, with a dip in between, as shown in Fig.\,\ref{fig:counting_noise}(a).
At high counts we approach the noiseless limit, where the error $\Delta\phi=\phi_\mathrm{fit}-\phi$ has the same sign as $\phi$, as shown in Fig.\,\ref{fig:phi-extract_gf}. In the low-count region, in contrast, the error $\Delta\phi$ exhibits opposite sign to $\phi$. Between the two plateaus, when the sign of $\Delta\phi$ flips, a minimum systematic error is found.
However, the exact $E_\mathrm{fx}$ for which the minimal systematic error occurs strongly depends on $k/\Delta p$ and $p_c/\Delta p$.
Therefore, the flux providing the minimal error can only be accurately identified when a thorough characterization of the system parameters is possible.

The accuracy of the delay retrieval, quantified by the systematic error $\overline{\Delta\phi}$, is insensitive to the number of shots $N_s$.
This is because the accuracy of $\overline{X^0}, \overline{Y^0}$, and K-estimators,  depends on the electron flux and not $N_s$, when only shot noise is considered.
In contrast, precision of the delay retrieval, quantified by the statistical error $\sigma[\Delta\phi]$, depends on $N_s$.
This is because the variance of $\overline{X^0}, \overline{Y^0}$, and K-estimators scales as ${\propto}1/N_s$ when the measurement is repeated over a large number of shots $N_s$.
Shot noise contributes to the variance in a way that is inversely proportional to the total electron counts detected $N_sE_\mathrm{fx}$.
As shown by the colored traces in Fig.\,\ref{fig:counting_noise}(b), with shot noise alone, the statistical error of the retrieved delay scales as $\propto 1/\sqrt{N_sE_\mathrm{fx}}$. 
In addition to shot noise, the variance of $\overline{X^0}, \overline{Y^0}$, and K-estimators, also include the sampling noise of the streaking direction $\kappa$, \ie~the non-uniformity of the distribution of a finite number of measured $\kappa$, which is $\propto 1/N_s$ for large $N_s$. 
Therefore the statistical error shrinks as $\sqrt{\mathcal{C}_\kappa/N_s+ \mathcal{C}_E/(N_sE_\mathrm{fx})}$ when both $N_s$ and $E_\mathrm{fx}$ are large, where $\mathcal{C}_\kappa$ and $\mathcal{C}_E$ are constants independent of $N_s$ and $E_\mathrm{fx}$.
In the case shown in Fig.\,\ref{fig:counting_noise}(a\& b), $k/\Delta p=0.5$, and these constants are $\mathcal{C}_E=2.0\,\mathrm{rad}^2$ and $\mathcal{C}_\kappa=0.045\,\mathrm{rad}^2$.
The sampling noise of $\kappa$ does not introduce systematic error in the delay retrieval, as it preserves the accuracy of $\overline{X^0}, \overline{Y^0}$, and the K-estimators.

The total error of the retrieved delay, defined as $\sqrt{\overline{(\Delta\phi)^2}}=\sqrt{(\overline{\Delta\phi})^2 + \sigma[\Delta\phi]^2}$ which combines the systematic and statistical errors, also depends on the streaking amplitude $k$.
Figure~\ref{fig:counting_noise}(c) shows this total error varies with $k/\Delta p$ at a fixed electron flux $E_\mathrm{fx}=3\times 10^2$. 
At large values of $k/\Delta p$, the total error grows with $k$ due to the increasing systematic errors as discussed in section~\ref{sec:phi-retrieval}.
However, as $k/\Delta p$ approaches zero, the total error increases due to the presence of noise.
This is because the streaking covariance, which scales as $k^2$ in the small amplitude limit~(Eqn.~\ref{eq:def-GIP}), is overwhelmed by the noise. 
The dependence on $k/\Delta p$ shown in Fig.\,\ref{fig:counting_noise}(c) represents the typical behavior of the total error in delay retrieval, although the exact optimal streaking amplitude also depends on other parameters such as $E_\mathrm{fx}$ and $N_s$.

 \subsection{Delay Fluctuation}
When employing angular streaking to measure the time-delay between two ionizing pulses, the instabilities in the delay impacts the measured covariance between the two photoelectron features produced by the respective pulses. 
The delay fluctuation results in variation of $\phi$. 
In our framework, it is straightforward to incorporate fluctuations in $\phi$, so long as the flucuations are independent from the rest of machine fluctuations. 
Since $\phi$ is the relative angle between the streaking directions of the two features, each individual MD is independent from $\phi$, thus $\E{I|\phi}=\E{I}$ and $\Cov[\E{X|\phi},\E{Y|\phi}]=0$.
Applying the law of total covariance:
\begin{align}
\Cov[X,Y] &= \E{\Cov[X,Y|\phi]} + \Cov[\E{X|\phi},\E{Y|\phi}] \nonumber\\
&= \E{\Cov[X,Y|\phi]}~,
\end{align}
we only retain the conditional covariance $\Cov[X,Y|\phi]$ averaged over $\phi$.
Then the measured covariance is simply the ensemble average of the covariance for each $\phi$.

To further illustrate the effect of a small delay jitter in the covariance analysis, we look into the scenario where $\phi$ follows a normal distribution with mean value $\phi_0$ and standard deviation $\delta\phi$. 
We found that this normal distribution of delay results in a damping factor on the GIP model $\E{M_\mathrm{GIP}(\phi)} = e^{-\delta\phi^2/2} M_\mathrm{GIP}(\phi_0)$. 
This damping factor $e^{-\delta\phi^2/2}$ has an intuitively interpretation: Increasing the delay jitter reduces the correlation between the streaking directions of the two features $X$ and $Y$, which reduces the magnitude of the streaking covariance $K[X,Y]$. 
On the other hand, $K[X,X]$ and $K[Y,Y]$ are not affected by the delay jitter $\delta\phi$, since they originate from a single ionizing pulse and thus have no dependence on $\phi$. 
Incorporating a global scaling factor as a free parameter of the gradient-based and/or gradient-free model, we can retrieve the delay.
We can also estimate the delay jitter by comparing the magnitude of $\mathrm{K}_{XY}$ to $\mathrm{K}_{XX}$ and $\mathrm{K}_{YY}$~\cite{guo_experimental_2024}.
 \subsection{Instrument Specific Issues}
\subsubsection{Momentum Projection}

\begin{figure*}[htbp]
    \centering
    \includegraphics[width=0.8\linewidth]{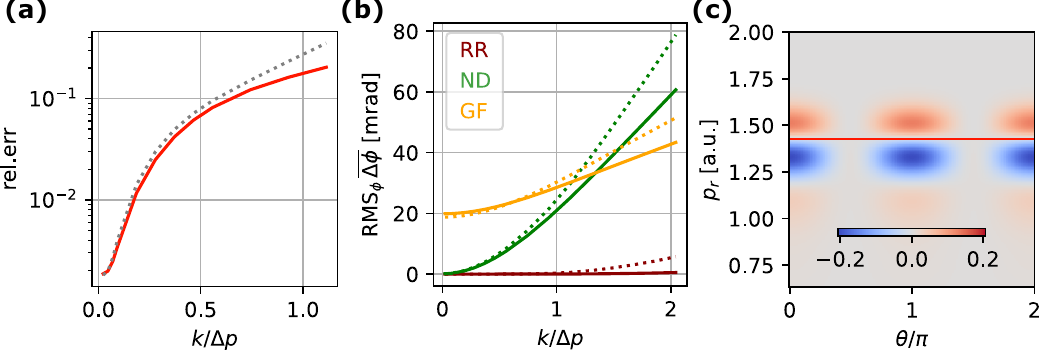}
    \caption{Effects of momentum projection. 
    \textbf{(a)} Relative error of the GIP term from the streaking covariance with the projection~(red solid) vs. the slicing~(grey dotted) measurement scheme, the latter of which is same as the $p_{\min} =p_\mathrm{MG}$ trace in Fig.\,\ref{fig:strkCovTerms}(c).
    \textbf{(b)} Delay retrieval systematic error for the three methods: ND-gradient~(green), RR-gradient~(brown), and gradient-free~(yellow), comparing the projection~(solid) vs.~the slicing~(dotted, same as the dashed in Fig.\,\ref{fig:phi-extract_mfluc}(c)). 
    \textbf{(c)} Difference of the average streaked MD from the unstreaked MD of each photoelectron feature $\E{X}-\E{X^0}$. 
    Values are normalized to $\max X^0$.
    ROI lower boundary for panels (a-b) is the maximal gradient line $p_{\mathrm{MG}}$ of the projected unstreaked MD~(red solid in panel (c)).
    }
    \label{fig:pzproj}
\end{figure*}

The framework presented above for interpreting photoelectron momentum distributions in the impulsive streaking regime can be applied to both projected and sliced MDs.
A projected MD can be measured with a VMI-type spectrometer~\cite{li_co-axial_2018}, and a sliced MD can be measured with an array of Time-of-Flight spectrometers~(ToFs)~\cite{walter_multi_2021}.
The two schemes perform nearly identically, in terms of the accuracy of the model and the performance of the delay retrieval methods.
The GIP model is slightly more accurate for the projected MD, as shown in Fig.\,\ref{fig:pzproj}(a).
As a result, the systematic error of delay retrieval methods is slightly lower, as shown in panel~(b).
The primary reason for the improved accuracy is that the higher order~($(n_X+n_Y){=}4$) terms, beyond GIP term, in streaking covariance is reduced by the project along $p_z$.
The reduction of the higher-order terms is a result of the reduced curvature of the dressing-free MD (\eg~$\nabla^2\avg{X^0}$), when the MD is projected.  
At the same time, difference in accuracy and precision between the two detection modalities are minimal.  
 
\subsubsection{Angular Sparsity}
\begin{figure*}[hbtp]
    \centering
    \includegraphics[width=0.9\linewidth]{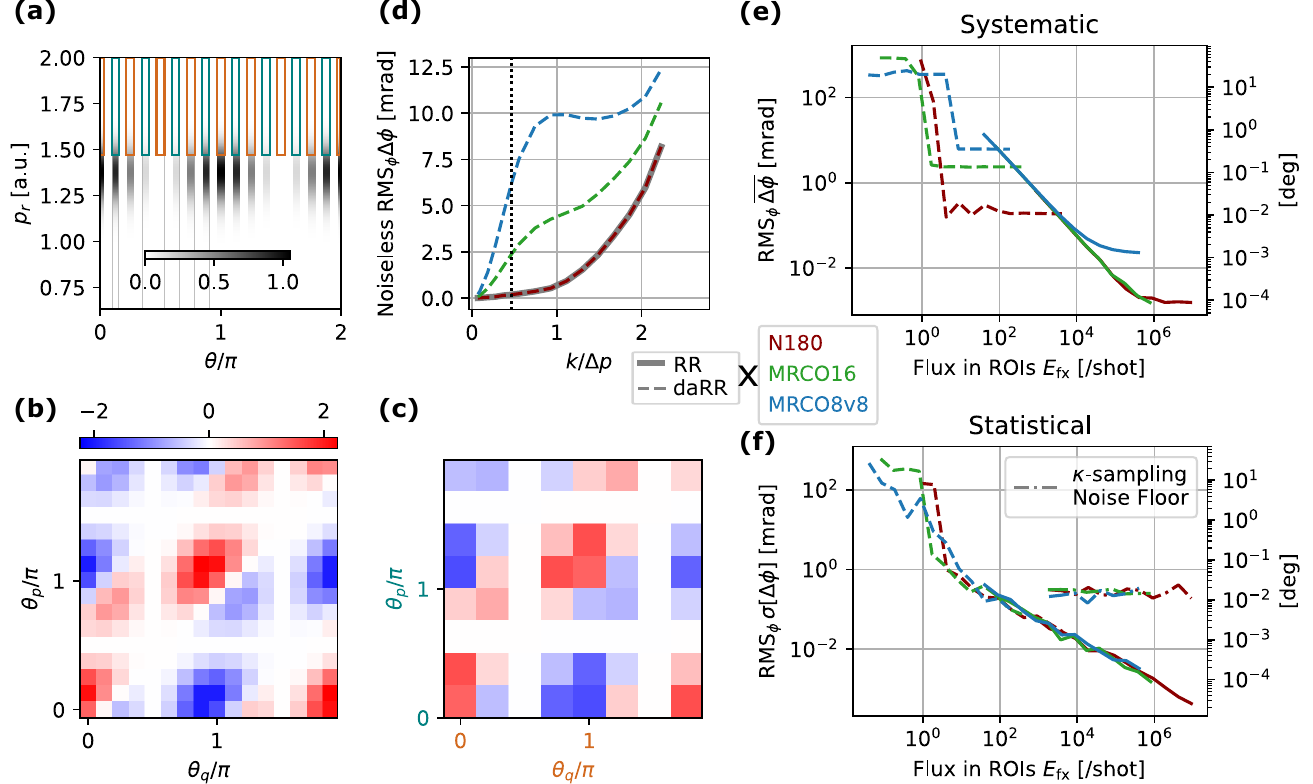}
    \caption{Effects of sparse angular sampling on the RR delay retrieval methods. 
    \textbf{(a)} The dressing-free MD collected by an array of 16 spectrometers, from the same ionization process as Fig.\,\ref{fig:illustration}. 
    Angular bins are grouped into the even~(orange rectangles) and odd~(cyan rectangles) ones. 
    \textbf{(b)} Streaking covariance $K[X,Y]$ at $\phi=\pi/3$, $k/\Delta p=0.5$, under the MRCO16 measurement scheme, see main text. 
    Radial boundaries are the same as for Fig.\,\ref{fig:strkCovTerms}(a).
    \textbf{(c)} Same as (b) but in the MRCO8v8 scheme where $X$ and $Y$ are detected by the even and odd spectrometers, respectively. 
    \textbf{(d-f)} Comparison of the three schemes~(N180 in red, MRCO16 in green, MRCO8v8 in blue) on the  RR~(solid) and daRR~(dashed) delay retrieval methods. 
    \textbf{(d)} Systematic error in the noiseless limit: of the RR method~(grey solid), almost the same for the three schemes, and of the daRR method~(dashed, colors corresponding to the schemes). 
    \textbf{(e)} In the presence of shot noise controlled by the electron flux $E_\mathrm{fx}$, systematic error at $k/\Delta p=0.5$~(black dotted line in (d)). 
    \textbf{(f)} The solid and dashed traces are the same as (e) except for statistical errors of the RR and daRR methods, respectively.
    The dot-dashed traces are asymptotic statistical errors after introducing $\kappa$ sampling noise in addition to shot noise, with colors corresponding to the schemes. 
    }
    \label{fig:MRCO}
\end{figure*}

Measurement of the sliced~(\ie~$p_z=0$) MD is often done using an array of time-of-flight spectrometers~(ToFs), as in the aforementioned device, MRCO, which consists of $16$~ToFs, each collecting $\sim 0.2$\% of the full $4\pi$ solid angle~\cite{walter_multi_2021}. 
We consider the impact of angular sparsity in the sampling of the MD on the delay retrieval methods.
The $\sim 0.2$\% angular sampling is notably more sparse than the measurement schemes employed above.To simulate this angular sampling scheme, we integrate an angular window representing each ToF, these windows are shown for the dressing-free MD in Fig.\,\ref{fig:MRCO}(a).  
As in the previous tests, we use the optimal lower bound for the ROI of the momentum, $p_{\min}=p_{\mathrm{MG}}$, to optimize the accuracy of the GIP model.
We can then calculate the K-estimators: $\mathrm{K}_{XY}, \mathrm{K}_{XX}, \mathrm{K}_{YY}$ with the limited angular resolution. 
We compare two measurement schemes, one where both $X$ and $Y$ are detected by all 16 ToFs $N_Q{=}N_P{=}16$, denoted MRCO16, and a second scheme where the ToFs are split into two interleaving subsets $N_Q{=}N_P{=}8$ for $X$ and $Y$ respectively, denoted MRCO8v8.
As a reference we use a densely sampled measurement scheme, denoted as ``N180'' in Fig.\,\ref{fig:strkCovTerms}-\ref{fig:counting_noise}.
Examples of the computed $\mathrm{K}_{XY}$ are shown in Fig.\,\ref{fig:MRCO}(b) and (c), for the MRCO16 and MRCO8v8 schemes, respectively. 
In both cases, the positive ridge around $\theta_p{-}\theta_q\sim \phi$ persists regardless of sampling scheme. 

We characterize the impact of the angular sparsity in the measurement on the delay retrieval methods described in Sec.~\ref{sec:phi-retrieval}.
We focus specifically on the RR and daRR methods, initially in the noiseless limit.
We find that the systematic error of the RR method is independent of the sampling scheme~($<0.1$\,mrad difference), as shown by the solid curves in Fig.\,\ref{fig:MRCO}(d). 
In contrast, the systematic error of the daRR method increases with increasing angular sparsity, as shown by the dashed curves in Fig.\,\ref{fig:MRCO}(d).
The number of elements in the $\mathrm{K}_{XX}$ matrix scales quadratically with $N_Q$, where as the number of diagonal elements scales linearly. 
As a result, when the daRR method ignores the diagonal elements in $\mathrm{K}_{XX}$, a smaller value of $N_Q$ ignores relatively more information in $\mathrm{K}_{XX}$. 
For the N180 scheme, the daRR method performs nearly the same as the RR method in the noiseless limit, within $0.2$\,mrad difference. 
Another observation from panel~(d) is that the systematic error of daRR is roughly proportional to $N_Q^{-1}k/\Delta p$, within $k/\Delta p<0.6$, where the error if the daRR method is approximately linear with $k$.

When we introduce shot noise to the simulations, the three measurement schemes~(N180, MRCO16, MRCO8v8) share several other similarities in terms of the delay retrieval performance. 
Decreasing the electron flux in each set of ROIs, $E_\mathrm{fx}$, from the noiseless limit, we see a region where the systematic error of RR method scale as $\propto 1/E_\mathrm{fx}$ and then exceeds the daRR method, as shown by the solid curves in Fig.\,\ref{fig:MRCO}(e). 
This behavior is exhibited under all three schemes, although the values of $E_\mathrm{fx}$ below which the daRR method outperforms the RR method are different.
As $E_\mathrm{fx}$ decreases further, the systematic error of the daRR method remains insensitive to $E_\mathrm{fx}$ until the electron counts become excessively scarce $E_\mathrm{fx} \lesssim 5$. 
On the other hand, the statistical error of both the RR and the daRR methods follows the $1/\sqrt{N_sE_\mathrm{fx}}$ scaling, as shown by the colored traces in panel~(f).
Remarkably, in this asymptotic scaling region the statistical error curves for the different measurement schemes overlap each other, rather than having an offset between them. 
Meanwhile, in the systematic error curves of the RR method, the $\propto 1/E_\mathrm{fx}$ scaling regions also overlap with each other when compared at the same $E_\mathrm{fx}$.
Note that $E_\mathrm{fx}$ represents the total electron counts detected in a set of angular bins, thus achieving the same $E_\mathrm{fx}$ requires a higher total number of electrons generated per shot when the measurement scheme is more sparse.
Another common feature for measurement schemes is the value of $E_\mathrm{fx}/N_Q$ where the RR and daRR methods are equally accurate.
For the $k/\Delta p=0.5$ case shown in panel~(e), this cross-over is at $E_\mathrm{fx}/N_Q\approx 10$.
This agreement is universal so long as $k/\Delta p<0.6$ and the RR method has not reached the noiseless limit.

Next, we consider sampling noise of $\kappa$ in addition to shot noise.
Due to this sampling noise of $\kappa$, when the number of shots $N_s$ is fixed and $E_\mathrm{fx}$ is high, the statistical error of retrieved delay approaches the lower bound $\sigma[\Delta\phi]\approx\sqrt{\mathcal{C}_\kappa/N_s}$ for both the RR and daRR methods, similar as in Fig.\,\ref{fig:counting_noise}(b).
As shown by the dot-dashed lines in Fig.\,\ref{fig:MRCO}(f), this lower bound at $\sqrt{\mathcal{C}_\kappa/N_s}$ is roughly the same under different levels of angular sparsity. 
Meanwhile, the overlap of the colored curves in Fig.\,\ref{fig:MRCO}(f) indicates that $\mathcal{C}_E$ is also almost independent of the angular sparsity.
Thus in the presence of both shot noise and sampling noise of $\kappa$, the asymptote of the statistical error $\sigma[\Delta\phi]\approx \sqrt{\mathcal{C}_E/(N_sE_\mathrm{fx}) + \mathcal{C}_\kappa/N_s}$ is not significantly affected by the angular sparsity.

Recall that systematic errors are insensitive to $N_s$, the overlap in part of the systematic error curves indicates that in order to retrieve the delay accurately, it is crucial to control the electron counts per shot. 
Acquiring more shots would not improve the accuracy of the delay retrieval but only improve the precision.

\section{Encoding Arbitrary Signals}\label{sec:generalY}
In this section, we extend our discussion to the situation where the emission process approaches or exceeds the period of the dressing laser field, and thus the emission is no longer impulsive.
We will remain in the limit that the emission process is shorter than the duration of the dressing field envelope.
For this method to work, we still require an impulsive emission process to provide a reference feature.
Moreover, these two emission processes should be correlated for the instantaneous process to serve as a reference.
Similar to the conventions used above, we will refer to the impulsive feature as feature $X$, and the longer timescale emission process will produce feature $Y$.

Owing to the periodic nature of the interaction of the laser field, 
the time-dependence of the emission is encoded in the Fourier coefficients of the measured electron momentum distribution~(MD):
\begin{equation}
\mathcal{Y}_m(\bm{r};k)\equiv \int_0^{2\pi} \E{Y(\bm{r};\bm{k})| \bm{k}}e^{-im\kappa} \frac{d\kappa}{2\pi}
\end{equation}
Here $Y(\bm{r};\bm{k})$ is the yield of electrons belonging to feature $Y$ at detector coordinate $\bm{r}$ for a streaking vector $\bm{k}$, which is similar to the conventions used above. 
In writing this expression, we have taken into account the machine fluctuations by using $\E{Y|\bm{k}}$. 
Applying the law of total covariance, we can write $K[X,Y]$ as 
\begin{widetext}
\begin{align}
    K[X,Y] &\simeq \sum_{m=1}^{+\infty}\left(\avg{\mu_{m}(k) \mathcal{Y}_m(k)} + \mathrm{c.c.}\right)  + \Cov\left[\mu_0(k), \mathcal{Y}_{0}({k})\right] \label{eq:genericYs-strkCov} \\
    \mu_m(k) &\equiv \frac{1}{m!}\left(\frac{-k\partial_+}{\sqrt 2}\right)^m {}_0F_1\left(m+1,\frac{k^2\nabla^2}{4}\right) \avg{X^0},~ m\geq 0\label{eq:genericYs-sensitivity}
\end{align}
\end{widetext}
where $\avg{X^0}$ is the dressing-free MD averaged over machine fluctuations, $\partial_+=(\frac{\partial}{\partial x}+i\frac{\partial}{\partial y})/\sqrt{2}$ is an operator that acts on a MD, and ${}_0F_1(z,x)\equiv \sum_{n=0}^{\infty} ((z-1)! x^n)/((z+n-1)! n!)$ is the confluence hypergeometric limit function~\cite{petkovsek_A_1996}. It is clear from Eqn.\,(\ref{eq:genericYs-sensitivity}) that all Fourier components, $\mathcal{Y}_m$, are encoded in $K[X,Y]$, but not always with the same sensitivity.
This indicates a signal which is periodic with $\kappa$ is encoded in the streaking covariance with the impulsive reference feature $X$ in the same way as the encoding of $\E{Y(\bm{r};\bm{k})|\bm{k}}$ described in Eqn.\,(\ref{eq:genericYs-strkCov}).
This encoding in all Fourier components is a general property of a periodic signal.

In applying the law of total covariance to partition $K[X,Y]$, the conditional variable was the streaking amplitude $k$.
In writing Eqn.\,(\ref{eq:genericYs-strkCov}) we have partitioned the terms into two terms, one that results from the shot-to-shot variations of $\kappa$~(the sum over $m{\neq}0$) and the other that results from variations in $k$.
The streaking covariance $K[X,Y]$ depends linearly on the $|m|{\geq}1$ components, with sensitivity $\mu_m(k)$ given by Eqn.\,(\ref{eq:genericYs-sensitivity}).
In contrast, the $m{=}0$ component of $Y$ only affects the second term $\Cov\left[\mu_0(k), \mathcal{Y}_{0}({k})\right]$, which indicates that $K[X,Y]$ is sensitive to the covariance of $\mathcal{Y}_0$ and $\mu_0$ arising from the variation of the streaking amplitude $k$.
Note that in the small streaking regime~($k<\Delta p_X$ of the reference feature), the leading order of the sensitivity is $\mu_m\approx\frac{k^m}{m!2^{m/2}}\partial_{+}^m\avg{X^0}$, where the minimal order of differentiation on $\avg{X^0}$ is $m$. 
Thus only the components $\mathcal{Y}_{\pm 1}$ are coupled to the gradient of the reference feature $\nabla \avg{X^0}$.
This indicates that the delay retrieval methods discussed in the preceding sections are only sensitive to the $|m|=1$ components.
In the cases where feature $Y$ is in the impulsive regime, as discussed in the preceding sections, Eqn.\,(\ref{eq:genericYs-strkCov}) reduces to Eqn.\,(\ref{eq:CovDD_fullseries}) after rearranging the terms, as shown in Supplemental Sec.\,4.1.

To illustrate Eqn.\,(\ref{eq:genericYs-strkCov}), we simulate the MD for Auger-Meitner~(AM) decay of a molecule in a superposition of two core-ionized states, prepared by $K$-shell ionization with a $0.2$\,fs FWHM pulse~\cite{wang_probing_2024}.
The two intermediate cationic states have an energetic separation of $\Delta I_\mathrm{P}=7\hbar\omega_L$, and the AM decay process couples these two states to a common dication state.
In our simulation, the dressing-free AM electrons have a central momentum of $4.34$\,a.u. and $4.38$\,a.u. The emitted electrons are dressed by 
a $|\bm{A}|=0.1$\,a.u., $cT_L=1.85\,\mathrm{\mu m}$ dressing field~(equivalent to Fig.\,\ref{fig:counting_noise}) and the MD considered is the $p_z=0$ slice.
Other details of the simulation of the AM distribution are provided in Supplemental Sec.\,VI.
The decay of this superposition of core-ionized states results in a 
characteristic modulation of $Y$ over $\kappa$ at a period of $2\pi/7$, which results from interference between the two pathways \textit{via} different intermediate states~\cite{wang_probing_2024}. 
This modulation appears across a wide range of AM momenta.
Here we have assumed that $Y^0$ is angularly isotropic.
This means that the MD of AM electrons at different angular positions $\theta_p$ are equivalent up to an offset in $\kappa$, \ie~$Y(\theta_p, r;\kappa,k)=Y(0, r;\kappa-\theta_p,k)$, so the yield is also modulated over angular position $\theta_p$.

\begin{figure*}[hbtp]
    \centering
    \includegraphics[width=0.9\linewidth]{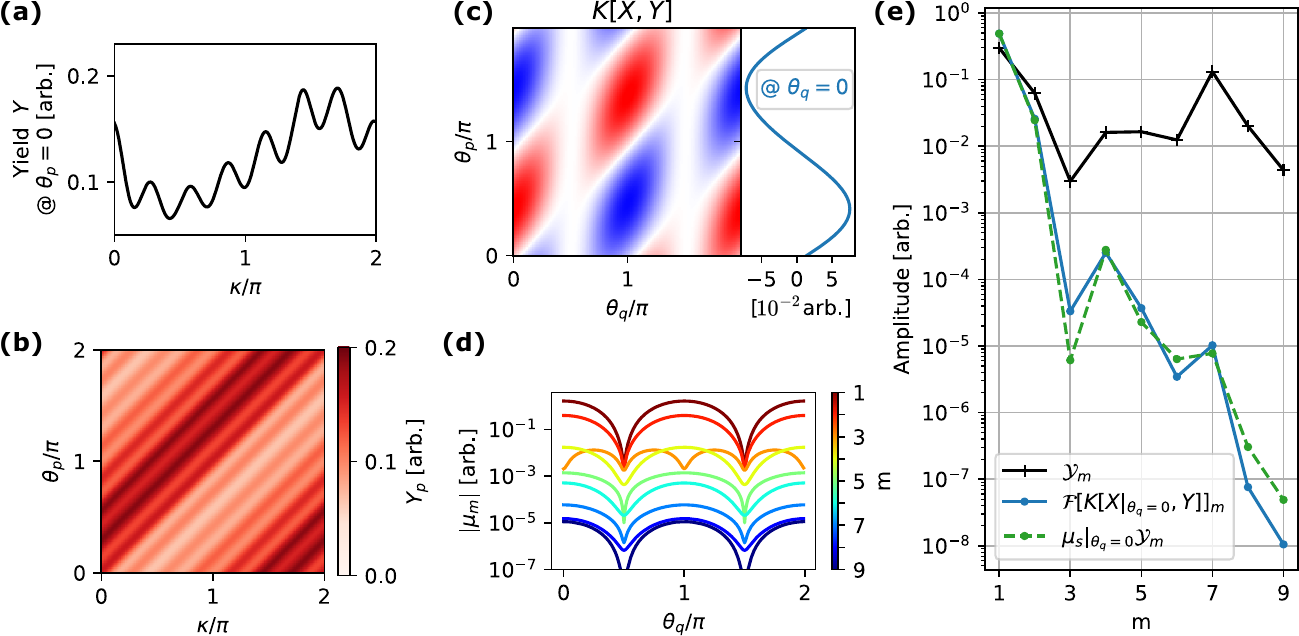}
    \caption{Encoding a $\kappa$-dependent signal in the covariance with an impulsive reference feature.
    \textbf{(a)} 
    Auger-Meitner~(AM) electron yield in the $2^\circ$ angular bin at $\theta_p=0$, over streaking direction $\kappa$ of the reference feature, simulated at $|\bm{A}|{=}0.1$\,a.u.
    \textbf{(b)} AM yield $Y_p$ across 180 angular bins.
    \textbf{(c)} Streaking covariance of the AM yield $Y_p$ and the photoelectron feature employed in Fig.\,\ref{fig:counting_noise}.
    Right inset shows the $\theta_q=0$ column of this map,  \ie~$K[X|_{\theta_q{=}0}, Y]$.
    \textbf{(d)} Amplitude of sensitivity $|\mu_m|$ at $k{=}0.1$\,a.u., in the same angular bins as in Fig.\,\ref{fig:counting_noise}. 
    \textbf{(e)} The Fourier coefficient amplitudes of the streaking covariance at $\theta_q=0$ that has been shown in (c)~(blue solid), compared to amplitude of the product $|\mu_m \mathcal{Y}_m|$~(green dashed) between sensitivity  $\mu_m$ and the AM yield Fourier coefficients $\mathcal{Y}_m$.
    For reference, $|\mathcal{Y}_m|$ is shown in black solid.
    }
    \label{fig:genericY}
\end{figure*}

In Fig.\ref{fig:genericY} we analyze the MD by partitioning the feature $Y$ into $N_P=180$ angular bins, between radial boundaries $p_\mathrm{min}{=}4.40$\,a.u.~and $p_\mathrm{max}{=}4.50$\,a.u..
Figure \ref{fig:genericY}(a) shows the integral of $Y(\bm{r};\bm{k})$, in the angular bin at $\theta_p=0$, as a function of the streaking directions of the reference feature $\kappa$.
As shown in Fig.\,\ref{fig:genericY}(b), the yield in these angular bins $Y_p$ is clearly modulated with both $\kappa$ and $\theta_p$. 
We compute the streaking covariance $K[X,Y]$ between $Y$ and the impulsive reference feature $X$. This reference feature $X$ is simulated in the same dressing field~($k=0.1\,\mathrm{a.u.}\approx 0.5\Delta p_X$), and the ROIs $\{Q_q\}$ are remain between $p_\mathrm{MG}$ and $p_{\max}{=}2.15$\,a.u. as in Fig.\,\ref{fig:counting_noise}.
The $2\pi/7$-period modulation is barely visible in $K[X,Y]$, as shown in Fig.\,\ref{fig:genericY}(c).
This attenuation of the $2\pi/7$-period modulation results from the low sensitivity of $K[X,Y]$ to $\mathcal{Y}_7$. We can study the sensitivity of $K[X,Y]$ to the components $\mathcal{Y}_m$ by computing $\mu_m$ according to Eqn.\,(\ref{eq:genericYs-sensitivity}) and integrating over the angular bins $Q_q$.
We show the magnitude of the resultant sensitivities in Fig.\,\ref{fig:genericY}(d), which demonstrate a decrease with increasing $m$.
Such dependence on $m$ is determined by $k, \Delta p_X$, and the ROIs on feature $X$, as described below.
The products of multiplying these calculated $\mu_m$ to the corresponding Fourier components $\mathcal{Y}_m$ yield good agreements with the Fourier coefficients of $K[X,Y]$ with respect to $\theta_p$, which is visualized for the $\theta_q=0$ slice in panel~(e), corroborating Eqn.\,(\ref{eq:genericYs-sensitivity}).

We notice that in the small streaking regime, according to Equations~(\ref{eq:genericYs-strkCov})(\ref{eq:genericYs-sensitivity}), the sensitivity is given by $\int_{Q_q} \mu_m d^2\bm{r}\approx \frac{k^m}{m!2^{m/2}}\int_{Q_q} \partial_+^m\avg{X^0} d^2\bm{r}$, which is proportional to the regional integral of a $m$-th order derivative of the reference feature. 
Therefore, the scaling of the sensitivity over $m$ depends on the streaking amplitude relative to the characteristic momentum scale $\delta p$ far below which the variations in $\avg{X^0}$ become negligible, \eg~the width $\Delta p_X$ in the example shown in Fig.\,\ref{fig:genericY}. In the limit of small streaking, the magnitude of sensitivity $|\mu_m|$ scales as $o((k/\delta p)^m/m!)$, which rapidly decreases with increasing $m$.
Even in cases where $k$ is comparable to $\delta p$, as it is the case in Fig.\,\ref{fig:genericY}, the streaking covariance tends to understate the high-order Fourier components due to the $1/m!$ factor in the sensitivity.
This also indicates that to improve the sensitivity to high-order components, it is advisable to intensify the dressing field and to reduce the smoothness of the reference feature to increase $k/\delta p$.

The bilinearity of covariance also leads to the independence of sensitivity $\mu_m$ on feature $Y$.
Rather, $\mu_m$ is determined by the reference feature $X$ and the streaking amplitude $k$, as shown in Eqn.~(\ref{eq:genericYs-sensitivity}).
Therefore, when detecting $Y$ as a function of $\kappa$ by analyzing the covariance with an impulsive reference feature $X$, the sensitivity to certain Fourier components $\mathcal{Y}_m$ can be designed in simulation, by choosing the streaking amplitude $k$, the shape of $X^0$ and the position of ROIs $\{Q_q\}$, without detailed prior knowledge in $Y$.

\section{Concluding Remarks}\label{sec:conclusion}
In this work, we present a comprehensive analysis of modelling covariance in the impulsive regime of angular streaking experiments. 
The displacement of the electron momentum distribution~(MD) provides a tight connection between the dressing-free MD and the dressed MD. 
Such connection establishes universal structures in the composition of streaking covariance that are common across different MDs, regardless of their exact shape. 
Building on this robust framework, we have developed methods for retrieving temporal information from angular streaking measurements. 
Specifically for the situations where both features $X$ and $Y$ in the MD are impulsive, we have proposed and evaluated three methods for retrieving relative time delays between $X$ and $Y$: numerical differentiation (ND), rank-reduction (RR), and a gradient-free approach. 
Each method offers certain advantages while also depending on experimental conditions and data quality.
Several key insights are obtained in our study:
(1) Both the streaking effect and the machine fluctuations contribute to the covariance that can be estimated from measurements. 
Proper removal of the machine fluctuation contribution is crucial for isolating the streaking covariance. 
(2) Shot noise of the electrons impacts the delay retrieval precision, but not significantly on the accuracy of the ND method or the diagonal-agnostic variant of the RR method. 
(3) Our investigation of angular sparsity in electron detection has shown that in the high-count scenarios, achieving accurate delay retrieval is possible even with limited angular sampling.
(4) Ae of an arbitrary periodic  $Y$signabe encoded to thecan he streaking covarwith a concurrent impulsive streaking signal $X$, where the sensitivity to different Fourier components of $Y$ is only determined by the dressing-free MD $X^0$ and the dressing fieldiance .

By providing a detailed understanding of the covariance structure in angular streaking experiments, our work enables more accurate and robust temporal measurements in a wide range of experimental scenarios.
Future work could focus on 
(1) 
Developing delay retrieval methods that incorporate higher orders into the model of streaking covariance, in addition to the GIP model;
(2) Higher order corrections for the removal of machine fluctuation contribution, \eg~the $n_X+n_Y=2$ terms shown in Fig.\,\ref{fig:K+L}(c-d), which can be measured by applying finite difference schemes to $\mathrm{C}_{X^0Y^0}$ or $\mathrm{PC}_{X^0Y^0;F}$, provided with adequate momentum resolution in all four dimensions of $(\bm{r}_q,\bm{r}_p)$;
(3) Developing adaptive algorithms that can optimize sensitivity to specific Fourier components of the non-impulsive signals.
In conclusion, this work provides a solid framework for leveraging covariance analysis in angular streaking experiments, paving the way for advances in ultrafast science and attosecond metrology.
 \section*{Acknowledgements}
This work is primarily supported by the US DOE, Office of Science, Office of Basic Energy Sciences (BES), Chemical Sciences, Geosciences, and Biosciences Division (CSGB). 
Z.G. and A.M. acknowledge support from the Accelerator and Detector Research Program of the Department of Energy, Basic Energy Sciences division. 
Z.G. also acknowledge support from Robert Siemann Fellowship of Stanford University. 
 \bibliography{referencesJPC, aux}

\begin{thebibliography}{28}%
\makeatletter
\providecommand \@ifxundefined [1]{%
 \@ifx{#1\undefined}
}%
\providecommand \@ifnum [1]{%
 \ifnum #1\expandafter \@firstoftwo
 \else \expandafter \@secondoftwo
 \fi
}%
\providecommand \@ifx [1]{%
 \ifx #1\expandafter \@firstoftwo
 \else \expandafter \@secondoftwo
 \fi
}%
\providecommand \natexlab [1]{#1}%
\providecommand \enquote  [1]{``#1''}%
\providecommand \bibnamefont  [1]{#1}%
\providecommand \bibfnamefont [1]{#1}%
\providecommand \citenamefont [1]{#1}%
\providecommand \href@noop [0]{\@secondoftwo}%
\providecommand \href [0]{\begingroup \@sanitize@url \@href}%
\providecommand \@href[1]{\@@startlink{#1}\@@href}%
\providecommand \@@href[1]{\endgroup#1\@@endlink}%
\providecommand \@sanitize@url [0]{\catcode `\\12\catcode `\$12\catcode
  `\&12\catcode `\#12\catcode `\^12\catcode `\_12\catcode `\%12\relax}%
\providecommand \@@startlink[1]{}%
\providecommand \@@endlink[0]{}%
\providecommand \url  [0]{\begingroup\@sanitize@url \@url }%
\providecommand \@url [1]{\endgroup\@href {#1}{\urlprefix }}%
\providecommand \urlprefix  [0]{URL }%
\providecommand \Eprint [0]{\href }%
\providecommand \doibase [0]{https://doi.org/}%
\providecommand \selectlanguage [0]{\@gobble}%
\providecommand \bibinfo  [0]{\@secondoftwo}%
\providecommand \bibfield  [0]{\@secondoftwo}%
\providecommand \translation [1]{[#1]}%
\providecommand \BibitemOpen [0]{}%
\providecommand \bibitemStop [0]{}%
\providecommand \bibitemNoStop [0]{.\EOS\space}%
\providecommand \EOS [0]{\spacefactor3000\relax}%
\providecommand \BibitemShut  [1]{\csname bibitem#1\endcsname}%
\let\auto@bib@innerbib\@empty
\bibitem [{\citenamefont {{The Nobel Committee for
  Physics}}(2023)}]{nobel_2023}%
  \BibitemOpen
  \bibfield  {author} {\bibinfo {author} {\bibnamefont {{The Nobel Committee
  for Physics}}},\ }\href
  {https://www.nobelprize.org/prizes/physics/2023/advanced-information/}
  {\bibinfo {title} {Scientific background to the nobel prize in physics 2023}}
  (\bibinfo {year} {2023})\BibitemShut {NoStop}%
\bibitem [{\citenamefont {Itatani}\ \emph {et~al.}(2002)\citenamefont
  {Itatani}, \citenamefont {Quéré}, \citenamefont {Yudin}, \citenamefont
  {Ivanov}, \citenamefont {Krausz},\ and\ \citenamefont
  {Corkum}}]{itatani_attosecond_2002}%
  \BibitemOpen
  \bibfield  {author} {\bibinfo {author} {\bibfnamefont {J.}~\bibnamefont
  {Itatani}}, \bibinfo {author} {\bibfnamefont {F.}~\bibnamefont {Quéré}},
  \bibinfo {author} {\bibfnamefont {G.~L.}\ \bibnamefont {Yudin}}, \bibinfo
  {author} {\bibfnamefont {M.~Y.}\ \bibnamefont {Ivanov}}, \bibinfo {author}
  {\bibfnamefont {F.}~\bibnamefont {Krausz}},\ and\ \bibinfo {author}
  {\bibfnamefont {P.~B.}\ \bibnamefont {Corkum}},\ }\bibfield  {title}
  {\bibinfo {title} {Attosecond {Streak} {Camera}},\ }\href
  {https://doi.org/10.1103/PhysRevLett.88.173903} {\bibfield  {journal}
  {\bibinfo  {journal} {Physical Review Letters}\ }\textbf {\bibinfo {volume}
  {88}},\ \bibinfo {pages} {173903} (\bibinfo {year} {2002})}\BibitemShut
  {NoStop}%
\bibitem [{\citenamefont {Bradley}\ \emph {et~al.}(1971)\citenamefont
  {Bradley}, \citenamefont {Liddy},\ and\ \citenamefont
  {Sleat}}]{bradley_direct_1971}%
  \BibitemOpen
  \bibfield  {author} {\bibinfo {author} {\bibfnamefont {D.}~\bibnamefont
  {Bradley}}, \bibinfo {author} {\bibfnamefont {B.}~\bibnamefont {Liddy}},\
  and\ \bibinfo {author} {\bibfnamefont {W.}~\bibnamefont {Sleat}},\ }\bibfield
   {title} {\bibinfo {title} {Direct linear measurement of ultrashort light
  pulses with a picosecond streak camera},\ }\href
  {https://doi.org/https://doi.org/10.1016/0030-4018(71)90252-5} {\bibfield
  {journal} {\bibinfo  {journal} {Optics Communications}\ }\textbf {\bibinfo
  {volume} {2}},\ \bibinfo {pages} {391} (\bibinfo {year} {1971})}\BibitemShut
  {NoStop}%
\bibitem [{\citenamefont {Thumm}\ \emph {et~al.}(2015)\citenamefont {Thumm},
  \citenamefont {Liao}, \citenamefont {Bothschafter}, \citenamefont
  {S{\"u}{\ss}mann}, \citenamefont {Kling},\ and\ \citenamefont
  {Kienberger}}]{thumm_attosecond_2015}%
  \BibitemOpen
  \bibfield  {author} {\bibinfo {author} {\bibfnamefont {U.}~\bibnamefont
  {Thumm}}, \bibinfo {author} {\bibfnamefont {Q.}~\bibnamefont {Liao}},
  \bibinfo {author} {\bibfnamefont {E.~M.}\ \bibnamefont {Bothschafter}},
  \bibinfo {author} {\bibfnamefont {F.}~\bibnamefont {S{\"u}{\ss}mann}},
  \bibinfo {author} {\bibfnamefont {M.~F.}\ \bibnamefont {Kling}},\ and\
  \bibinfo {author} {\bibfnamefont {R.}~\bibnamefont {Kienberger}},\ }\bibinfo
  {title} {Attosecond physics: Attosecond streaking spectroscopy of atoms and
  solids},\ in\ \href
  {https://doi.org/https://doi.org/10.1002/9781119009719.ch13} {\emph {\bibinfo
  {booktitle} {Photonics}}}\ (\bibinfo  {publisher} {John Wiley \& Sons, Ltd},\
  \bibinfo {year} {2015})\ Chap.~\bibinfo {chapter} {13}, pp.\ \bibinfo {pages}
  {387--441},\ \Eprint
  {https://arxiv.org/abs/https://onlinelibrary.wiley.com/doi/pdf/10.1002/9781119009719.ch13}
  {https://onlinelibrary.wiley.com/doi/pdf/10.1002/9781119009719.ch13}
  \BibitemShut {NoStop}%
\bibitem [{\citenamefont {Schultze}\ \emph {et~al.}(2010)\citenamefont
  {Schultze}, \citenamefont {Fieß}, \citenamefont {Karpowicz}, \citenamefont
  {Gagnon}, \citenamefont {Korbman}, \citenamefont {Hofstetter}, \citenamefont
  {Neppl}, \citenamefont {Cavalieri}, \citenamefont {Komninos}, \citenamefont
  {Mercouris}, \citenamefont {Nicolaides}, \citenamefont {Pazourek},
  \citenamefont {Nagele}, \citenamefont {Feist}, \citenamefont {Burgdörfer},
  \citenamefont {Azzeer}, \citenamefont {Ernstorfer}, \citenamefont
  {Kienberger}, \citenamefont {Kleineberg}, \citenamefont {Goulielmakis},
  \citenamefont {Krausz},\ and\ \citenamefont
  {Yakovlev}}]{schultze_delay_2010}%
  \BibitemOpen
  \bibfield  {author} {\bibinfo {author} {\bibfnamefont {M.}~\bibnamefont
  {Schultze}}, \bibinfo {author} {\bibfnamefont {M.}~\bibnamefont {Fieß}},
  \bibinfo {author} {\bibfnamefont {N.}~\bibnamefont {Karpowicz}}, \bibinfo
  {author} {\bibfnamefont {J.}~\bibnamefont {Gagnon}}, \bibinfo {author}
  {\bibfnamefont {M.}~\bibnamefont {Korbman}}, \bibinfo {author} {\bibfnamefont
  {M.}~\bibnamefont {Hofstetter}}, \bibinfo {author} {\bibfnamefont
  {S.}~\bibnamefont {Neppl}}, \bibinfo {author} {\bibfnamefont {A.~L.}\
  \bibnamefont {Cavalieri}}, \bibinfo {author} {\bibfnamefont {Y.}~\bibnamefont
  {Komninos}}, \bibinfo {author} {\bibfnamefont {T.}~\bibnamefont {Mercouris}},
  \bibinfo {author} {\bibfnamefont {C.~A.}\ \bibnamefont {Nicolaides}},
  \bibinfo {author} {\bibfnamefont {R.}~\bibnamefont {Pazourek}}, \bibinfo
  {author} {\bibfnamefont {S.}~\bibnamefont {Nagele}}, \bibinfo {author}
  {\bibfnamefont {J.}~\bibnamefont {Feist}}, \bibinfo {author} {\bibfnamefont
  {J.}~\bibnamefont {Burgdörfer}}, \bibinfo {author} {\bibfnamefont {A.~M.}\
  \bibnamefont {Azzeer}}, \bibinfo {author} {\bibfnamefont {R.}~\bibnamefont
  {Ernstorfer}}, \bibinfo {author} {\bibfnamefont {R.}~\bibnamefont
  {Kienberger}}, \bibinfo {author} {\bibfnamefont {U.}~\bibnamefont
  {Kleineberg}}, \bibinfo {author} {\bibfnamefont {E.}~\bibnamefont
  {Goulielmakis}}, \bibinfo {author} {\bibfnamefont {F.}~\bibnamefont
  {Krausz}},\ and\ \bibinfo {author} {\bibfnamefont {V.~S.}\ \bibnamefont
  {Yakovlev}},\ }\bibfield  {title} {{\selectlanguage {en}\bibinfo {title}
  {Delay in {Photoemission}}},\ }\href
  {https://doi.org/10.1126/science.1189401} {\bibfield  {journal} {\bibinfo
  {journal} {Science}\ }\textbf {\bibinfo {volume} {328}},\ \bibinfo {pages}
  {1658} (\bibinfo {year} {2010})}\BibitemShut {NoStop}%
\bibitem [{\citenamefont {Drescher}\ \emph {et~al.}(2002)\citenamefont
  {Drescher}, \citenamefont {Hentschel}, \citenamefont {Kienberger},
  \citenamefont {Uiberacker}, \citenamefont {Yakovlev}, \citenamefont
  {Scrinzi}, \citenamefont {Westerwalbesloh}, \citenamefont {Kleineberg},
  \citenamefont {Heinzmann},\ and\ \citenamefont
  {Krausz}}]{drescher_time-resolved_2002}%
  \BibitemOpen
  \bibfield  {author} {\bibinfo {author} {\bibfnamefont {M.}~\bibnamefont
  {Drescher}}, \bibinfo {author} {\bibfnamefont {M.}~\bibnamefont {Hentschel}},
  \bibinfo {author} {\bibfnamefont {R.}~\bibnamefont {Kienberger}}, \bibinfo
  {author} {\bibfnamefont {M.}~\bibnamefont {Uiberacker}}, \bibinfo {author}
  {\bibfnamefont {V.}~\bibnamefont {Yakovlev}}, \bibinfo {author}
  {\bibfnamefont {A.}~\bibnamefont {Scrinzi}}, \bibinfo {author} {\bibfnamefont
  {T.}~\bibnamefont {Westerwalbesloh}}, \bibinfo {author} {\bibfnamefont
  {U.}~\bibnamefont {Kleineberg}}, \bibinfo {author} {\bibfnamefont
  {U.}~\bibnamefont {Heinzmann}},\ and\ \bibinfo {author} {\bibfnamefont
  {F.}~\bibnamefont {Krausz}},\ }\bibfield  {title} {{\selectlanguage
  {en}\bibinfo {title} {Time-resolved atomic inner-shell spectroscopy}},\
  }\href {https://doi.org/10.1038/nature01143} {\bibfield  {journal} {\bibinfo
  {journal} {Nature}\ }\textbf {\bibinfo {volume} {419}},\ \bibinfo {pages}
  {803} (\bibinfo {year} {2002})}\BibitemShut {NoStop}%
\bibitem [{\citenamefont {Eckle}\ \emph {et~al.}(2008)\citenamefont {Eckle},
  \citenamefont {Pfeiffer}, \citenamefont {Cirelli}, \citenamefont {Staudte},
  \citenamefont {Dörner}, \citenamefont {Muller}, \citenamefont {Büttiker},\
  and\ \citenamefont {Keller}}]{eckle_attosecond_2008}%
  \BibitemOpen
  \bibfield  {author} {\bibinfo {author} {\bibfnamefont {P.}~\bibnamefont
  {Eckle}}, \bibinfo {author} {\bibfnamefont {A.~N.}\ \bibnamefont {Pfeiffer}},
  \bibinfo {author} {\bibfnamefont {C.}~\bibnamefont {Cirelli}}, \bibinfo
  {author} {\bibfnamefont {A.}~\bibnamefont {Staudte}}, \bibinfo {author}
  {\bibfnamefont {R.}~\bibnamefont {Dörner}}, \bibinfo {author} {\bibfnamefont
  {H.~G.}\ \bibnamefont {Muller}}, \bibinfo {author} {\bibfnamefont
  {M.}~\bibnamefont {Büttiker}},\ and\ \bibinfo {author} {\bibfnamefont
  {U.}~\bibnamefont {Keller}},\ }\bibfield  {title} {{\selectlanguage
  {en}\bibinfo {title} {Attosecond {Ionization} and {Tunneling} {Delay} {Time}
  {Measurements} in {Helium}}},\ }\href
  {https://doi.org/10.1126/science.1163439} {\bibfield  {journal} {\bibinfo
  {journal} {Science}\ }\textbf {\bibinfo {volume} {322}},\ \bibinfo {pages}
  {1525} (\bibinfo {year} {2008})}\BibitemShut {NoStop}%
\bibitem [{\citenamefont {Hartmann}\ \emph {et~al.}(2018)\citenamefont
  {Hartmann}, \citenamefont {Hartmann}, \citenamefont {Heider}, \citenamefont
  {Wagner}, \citenamefont {Ilchen}, \citenamefont {Buck}, \citenamefont
  {Lindahl}, \citenamefont {Benko}, \citenamefont {Grünert}, \citenamefont
  {Krzywinski}, \citenamefont {Liu}, \citenamefont {Lutman}, \citenamefont
  {Marinelli}, \citenamefont {Maxwell}, \citenamefont {Miahnahri},
  \citenamefont {Moeller}, \citenamefont {Planas}, \citenamefont {Robinson},
  \citenamefont {Kazansky}, \citenamefont {Kabachnik}, \citenamefont
  {Viefhaus}, \citenamefont {Feurer}, \citenamefont {Kienberger}, \citenamefont
  {Coffee},\ and\ \citenamefont {Helml}}]{hartmann_attosecond_2018}%
  \BibitemOpen
  \bibfield  {author} {\bibinfo {author} {\bibfnamefont {N.}~\bibnamefont
  {Hartmann}}, \bibinfo {author} {\bibfnamefont {G.}~\bibnamefont {Hartmann}},
  \bibinfo {author} {\bibfnamefont {R.}~\bibnamefont {Heider}}, \bibinfo
  {author} {\bibfnamefont {M.~S.}\ \bibnamefont {Wagner}}, \bibinfo {author}
  {\bibfnamefont {M.}~\bibnamefont {Ilchen}}, \bibinfo {author} {\bibfnamefont
  {J.}~\bibnamefont {Buck}}, \bibinfo {author} {\bibfnamefont {A.~O.}\
  \bibnamefont {Lindahl}}, \bibinfo {author} {\bibfnamefont {C.}~\bibnamefont
  {Benko}}, \bibinfo {author} {\bibfnamefont {J.}~\bibnamefont {Grünert}},
  \bibinfo {author} {\bibfnamefont {J.}~\bibnamefont {Krzywinski}}, \bibinfo
  {author} {\bibfnamefont {J.}~\bibnamefont {Liu}}, \bibinfo {author}
  {\bibfnamefont {A.~A.}\ \bibnamefont {Lutman}}, \bibinfo {author}
  {\bibfnamefont {A.}~\bibnamefont {Marinelli}}, \bibinfo {author}
  {\bibfnamefont {T.}~\bibnamefont {Maxwell}}, \bibinfo {author} {\bibfnamefont
  {A.~A.}\ \bibnamefont {Miahnahri}}, \bibinfo {author} {\bibfnamefont {S.~P.}\
  \bibnamefont {Moeller}}, \bibinfo {author} {\bibfnamefont {M.}~\bibnamefont
  {Planas}}, \bibinfo {author} {\bibfnamefont {J.}~\bibnamefont {Robinson}},
  \bibinfo {author} {\bibfnamefont {A.~K.}\ \bibnamefont {Kazansky}}, \bibinfo
  {author} {\bibfnamefont {N.~M.}\ \bibnamefont {Kabachnik}}, \bibinfo {author}
  {\bibfnamefont {J.}~\bibnamefont {Viefhaus}}, \bibinfo {author}
  {\bibfnamefont {T.}~\bibnamefont {Feurer}}, \bibinfo {author} {\bibfnamefont
  {R.}~\bibnamefont {Kienberger}}, \bibinfo {author} {\bibfnamefont {R.~N.}\
  \bibnamefont {Coffee}},\ and\ \bibinfo {author} {\bibfnamefont
  {W.}~\bibnamefont {Helml}},\ }\bibfield  {title} {{\selectlanguage
  {en}\bibinfo {title} {Attosecond time–energy structure of {X}-ray
  free-electron laser pulses}},\ }\href
  {https://doi.org/10.1038/s41566-018-0107-6} {\bibfield  {journal} {\bibinfo
  {journal} {Nature Photonics}\ }\textbf {\bibinfo {volume} {12}},\ \bibinfo
  {pages} {215} (\bibinfo {year} {2018})}\BibitemShut {NoStop}%
\bibitem [{\citenamefont {Duris}\ \emph {et~al.}(2020)\citenamefont {Duris},
  \citenamefont {Li}, \citenamefont {Driver}, \citenamefont {Champenois},
  \citenamefont {MacArthur}, \citenamefont {Lutman}, \citenamefont {Zhang},
  \citenamefont {Rosenberger}, \citenamefont {Aldrich}, \citenamefont {Coffee},
  \citenamefont {Coslovich}, \citenamefont {Decker}, \citenamefont {Glownia},
  \citenamefont {Hartmann}, \citenamefont {Helml}, \citenamefont {Kamalov},
  \citenamefont {Knurr}, \citenamefont {Krzywinski}, \citenamefont {Lin},
  \citenamefont {Marangos}, \citenamefont {Nantel}, \citenamefont {Natan},
  \citenamefont {O’Neal}, \citenamefont {Shivaram}, \citenamefont {Walter},
  \citenamefont {Wang}, \citenamefont {Welch}, \citenamefont {Wolf},
  \citenamefont {Xu}, \citenamefont {Kling}, \citenamefont {Bucksbaum},
  \citenamefont {Zholents}, \citenamefont {Huang}, \citenamefont {Cryan},\ and\
  \citenamefont {Marinelli}}]{duris_tunable_2020}%
  \BibitemOpen
  \bibfield  {author} {\bibinfo {author} {\bibfnamefont {J.}~\bibnamefont
  {Duris}}, \bibinfo {author} {\bibfnamefont {S.}~\bibnamefont {Li}}, \bibinfo
  {author} {\bibfnamefont {T.}~\bibnamefont {Driver}}, \bibinfo {author}
  {\bibfnamefont {E.~G.}\ \bibnamefont {Champenois}}, \bibinfo {author}
  {\bibfnamefont {J.~P.}\ \bibnamefont {MacArthur}}, \bibinfo {author}
  {\bibfnamefont {A.~A.}\ \bibnamefont {Lutman}}, \bibinfo {author}
  {\bibfnamefont {Z.}~\bibnamefont {Zhang}}, \bibinfo {author} {\bibfnamefont
  {P.}~\bibnamefont {Rosenberger}}, \bibinfo {author} {\bibfnamefont {J.~W.}\
  \bibnamefont {Aldrich}}, \bibinfo {author} {\bibfnamefont {R.}~\bibnamefont
  {Coffee}}, \bibinfo {author} {\bibfnamefont {G.}~\bibnamefont {Coslovich}},
  \bibinfo {author} {\bibfnamefont {F.-J.}\ \bibnamefont {Decker}}, \bibinfo
  {author} {\bibfnamefont {J.~M.}\ \bibnamefont {Glownia}}, \bibinfo {author}
  {\bibfnamefont {G.}~\bibnamefont {Hartmann}}, \bibinfo {author}
  {\bibfnamefont {W.}~\bibnamefont {Helml}}, \bibinfo {author} {\bibfnamefont
  {A.}~\bibnamefont {Kamalov}}, \bibinfo {author} {\bibfnamefont
  {J.}~\bibnamefont {Knurr}}, \bibinfo {author} {\bibfnamefont
  {J.}~\bibnamefont {Krzywinski}}, \bibinfo {author} {\bibfnamefont {M.-F.}\
  \bibnamefont {Lin}}, \bibinfo {author} {\bibfnamefont {J.~P.}\ \bibnamefont
  {Marangos}}, \bibinfo {author} {\bibfnamefont {M.}~\bibnamefont {Nantel}},
  \bibinfo {author} {\bibfnamefont {A.}~\bibnamefont {Natan}}, \bibinfo
  {author} {\bibfnamefont {J.~T.}\ \bibnamefont {O’Neal}}, \bibinfo {author}
  {\bibfnamefont {N.}~\bibnamefont {Shivaram}}, \bibinfo {author}
  {\bibfnamefont {P.}~\bibnamefont {Walter}}, \bibinfo {author} {\bibfnamefont
  {A.~L.}\ \bibnamefont {Wang}}, \bibinfo {author} {\bibfnamefont {J.~J.}\
  \bibnamefont {Welch}}, \bibinfo {author} {\bibfnamefont {T.~J.~A.}\
  \bibnamefont {Wolf}}, \bibinfo {author} {\bibfnamefont {J.~Z.}\ \bibnamefont
  {Xu}}, \bibinfo {author} {\bibfnamefont {M.~F.}\ \bibnamefont {Kling}},
  \bibinfo {author} {\bibfnamefont {P.~H.}\ \bibnamefont {Bucksbaum}}, \bibinfo
  {author} {\bibfnamefont {A.}~\bibnamefont {Zholents}}, \bibinfo {author}
  {\bibfnamefont {Z.}~\bibnamefont {Huang}}, \bibinfo {author} {\bibfnamefont
  {J.~P.}\ \bibnamefont {Cryan}},\ and\ \bibinfo {author} {\bibfnamefont
  {A.}~\bibnamefont {Marinelli}},\ }\bibfield  {title} {{\selectlanguage
  {en}\bibinfo {title} {Tunable isolated attosecond {X}-ray pulses with
  gigawatt peak power from a free-electron laser}},\ }\href
  {https://doi.org/10.1038/s41566-019-0549-5} {\bibfield  {journal} {\bibinfo
  {journal} {Nature Photonics}\ }\textbf {\bibinfo {volume} {14}},\ \bibinfo
  {pages} {30} (\bibinfo {year} {2020})},\ \bibinfo {note} {number: 1
  Publisher: Nature Publishing Group}\BibitemShut {NoStop}%
\bibitem [{\citenamefont {Franz}\ \emph {et~al.}(2024)\citenamefont {Franz},
  \citenamefont {Li}, \citenamefont {Driver}, \citenamefont {Robles},
  \citenamefont {Cesar}, \citenamefont {Isele}, \citenamefont {Guo},
  \citenamefont {Wang}, \citenamefont {Duris}, \citenamefont {Larsen},
  \citenamefont {Glownia}, \citenamefont {Cheng}, \citenamefont {Hoffmann},
  \citenamefont {Li}, \citenamefont {Lin}, \citenamefont {Kamalov},
  \citenamefont {Obaid}, \citenamefont {Summers}, \citenamefont {Sudar},
  \citenamefont {Thierstein}, \citenamefont {Zhang}, \citenamefont {Kling},
  \citenamefont {Huang}, \citenamefont {Cryan},\ and\ \citenamefont
  {Marinelli}}]{franz_terawatt-scale_2024}%
  \BibitemOpen
  \bibfield  {author} {\bibinfo {author} {\bibfnamefont {P.}~\bibnamefont
  {Franz}}, \bibinfo {author} {\bibfnamefont {S.}~\bibnamefont {Li}}, \bibinfo
  {author} {\bibfnamefont {T.}~\bibnamefont {Driver}}, \bibinfo {author}
  {\bibfnamefont {R.~R.}\ \bibnamefont {Robles}}, \bibinfo {author}
  {\bibfnamefont {D.}~\bibnamefont {Cesar}}, \bibinfo {author} {\bibfnamefont
  {E.}~\bibnamefont {Isele}}, \bibinfo {author} {\bibfnamefont
  {Z.}~\bibnamefont {Guo}}, \bibinfo {author} {\bibfnamefont {J.}~\bibnamefont
  {Wang}}, \bibinfo {author} {\bibfnamefont {J.~P.}\ \bibnamefont {Duris}},
  \bibinfo {author} {\bibfnamefont {K.}~\bibnamefont {Larsen}}, \bibinfo
  {author} {\bibfnamefont {J.~M.}\ \bibnamefont {Glownia}}, \bibinfo {author}
  {\bibfnamefont {X.}~\bibnamefont {Cheng}}, \bibinfo {author} {\bibfnamefont
  {M.~C.}\ \bibnamefont {Hoffmann}}, \bibinfo {author} {\bibfnamefont
  {X.}~\bibnamefont {Li}}, \bibinfo {author} {\bibfnamefont {M.-F.}\
  \bibnamefont {Lin}}, \bibinfo {author} {\bibfnamefont {A.}~\bibnamefont
  {Kamalov}}, \bibinfo {author} {\bibfnamefont {R.}~\bibnamefont {Obaid}},
  \bibinfo {author} {\bibfnamefont {A.}~\bibnamefont {Summers}}, \bibinfo
  {author} {\bibfnamefont {N.}~\bibnamefont {Sudar}}, \bibinfo {author}
  {\bibfnamefont {E.}~\bibnamefont {Thierstein}}, \bibinfo {author}
  {\bibfnamefont {Z.}~\bibnamefont {Zhang}}, \bibinfo {author} {\bibfnamefont
  {M.~F.}\ \bibnamefont {Kling}}, \bibinfo {author} {\bibfnamefont
  {Z.}~\bibnamefont {Huang}}, \bibinfo {author} {\bibfnamefont {J.~P.}\
  \bibnamefont {Cryan}},\ and\ \bibinfo {author} {\bibfnamefont
  {A.}~\bibnamefont {Marinelli}},\ }\bibfield  {title} {{\selectlanguage
  {en}\bibinfo {title} {Terawatt-scale attosecond {X}-ray pulses from a
  cascaded superradiant free-electron laser}},\ }\href
  {https://doi.org/10.1038/s41566-024-01427-w} {\bibfield  {journal} {\bibinfo
  {journal} {Nature Photonics}\ }\textbf {\bibinfo {volume} {18}},\ \bibinfo
  {pages} {698} (\bibinfo {year} {2024})},\ \bibinfo {note} {publisher: Nature
  Publishing Group}\BibitemShut {NoStop}%
\bibitem [{\citenamefont {Haynes}\ \emph {et~al.}(2021)\citenamefont {Haynes},
  \citenamefont {Wurzer}, \citenamefont {Schletter}, \citenamefont {Al-Haddad},
  \citenamefont {Blaga}, \citenamefont {Bostedt}, \citenamefont {Bozek},
  \citenamefont {Bromberger}, \citenamefont {Bucher}, \citenamefont {Camper},
  \citenamefont {Carron}, \citenamefont {Coffee}, \citenamefont {Costello},
  \citenamefont {DiMauro}, \citenamefont {Ding}, \citenamefont {Ferguson},
  \citenamefont {Grguraš}, \citenamefont {Helml}, \citenamefont {Hoffmann},
  \citenamefont {Ilchen}, \citenamefont {Jalas}, \citenamefont {Kabachnik},
  \citenamefont {Kazansky}, \citenamefont {Kienberger}, \citenamefont {Maier},
  \citenamefont {Maxwell}, \citenamefont {Mazza}, \citenamefont {Meyer},
  \citenamefont {Park}, \citenamefont {Robinson}, \citenamefont {Roedig},
  \citenamefont {Schlarb}, \citenamefont {Singla}, \citenamefont {Tellkamp},
  \citenamefont {Walker}, \citenamefont {Zhang}, \citenamefont {Doumy},
  \citenamefont {Behrens},\ and\ \citenamefont
  {Cavalieri}}]{haynes_clocking_2021}%
  \BibitemOpen
  \bibfield  {author} {\bibinfo {author} {\bibfnamefont {D.~C.}\ \bibnamefont
  {Haynes}}, \bibinfo {author} {\bibfnamefont {M.}~\bibnamefont {Wurzer}},
  \bibinfo {author} {\bibfnamefont {A.}~\bibnamefont {Schletter}}, \bibinfo
  {author} {\bibfnamefont {A.}~\bibnamefont {Al-Haddad}}, \bibinfo {author}
  {\bibfnamefont {C.}~\bibnamefont {Blaga}}, \bibinfo {author} {\bibfnamefont
  {C.}~\bibnamefont {Bostedt}}, \bibinfo {author} {\bibfnamefont
  {J.}~\bibnamefont {Bozek}}, \bibinfo {author} {\bibfnamefont
  {H.}~\bibnamefont {Bromberger}}, \bibinfo {author} {\bibfnamefont
  {M.}~\bibnamefont {Bucher}}, \bibinfo {author} {\bibfnamefont
  {A.}~\bibnamefont {Camper}}, \bibinfo {author} {\bibfnamefont
  {S.}~\bibnamefont {Carron}}, \bibinfo {author} {\bibfnamefont
  {R.}~\bibnamefont {Coffee}}, \bibinfo {author} {\bibfnamefont {J.~T.}\
  \bibnamefont {Costello}}, \bibinfo {author} {\bibfnamefont {L.~F.}\
  \bibnamefont {DiMauro}}, \bibinfo {author} {\bibfnamefont {Y.}~\bibnamefont
  {Ding}}, \bibinfo {author} {\bibfnamefont {K.}~\bibnamefont {Ferguson}},
  \bibinfo {author} {\bibfnamefont {I.}~\bibnamefont {Grguraš}}, \bibinfo
  {author} {\bibfnamefont {W.}~\bibnamefont {Helml}}, \bibinfo {author}
  {\bibfnamefont {M.~C.}\ \bibnamefont {Hoffmann}}, \bibinfo {author}
  {\bibfnamefont {M.}~\bibnamefont {Ilchen}}, \bibinfo {author} {\bibfnamefont
  {S.}~\bibnamefont {Jalas}}, \bibinfo {author} {\bibfnamefont {N.~M.}\
  \bibnamefont {Kabachnik}}, \bibinfo {author} {\bibfnamefont {A.~K.}\
  \bibnamefont {Kazansky}}, \bibinfo {author} {\bibfnamefont {R.}~\bibnamefont
  {Kienberger}}, \bibinfo {author} {\bibfnamefont {A.~R.}\ \bibnamefont
  {Maier}}, \bibinfo {author} {\bibfnamefont {T.}~\bibnamefont {Maxwell}},
  \bibinfo {author} {\bibfnamefont {T.}~\bibnamefont {Mazza}}, \bibinfo
  {author} {\bibfnamefont {M.}~\bibnamefont {Meyer}}, \bibinfo {author}
  {\bibfnamefont {H.}~\bibnamefont {Park}}, \bibinfo {author} {\bibfnamefont
  {J.}~\bibnamefont {Robinson}}, \bibinfo {author} {\bibfnamefont
  {C.}~\bibnamefont {Roedig}}, \bibinfo {author} {\bibfnamefont
  {H.}~\bibnamefont {Schlarb}}, \bibinfo {author} {\bibfnamefont
  {R.}~\bibnamefont {Singla}}, \bibinfo {author} {\bibfnamefont
  {F.}~\bibnamefont {Tellkamp}}, \bibinfo {author} {\bibfnamefont {P.~A.}\
  \bibnamefont {Walker}}, \bibinfo {author} {\bibfnamefont {K.}~\bibnamefont
  {Zhang}}, \bibinfo {author} {\bibfnamefont {G.}~\bibnamefont {Doumy}},
  \bibinfo {author} {\bibfnamefont {C.}~\bibnamefont {Behrens}},\ and\ \bibinfo
  {author} {\bibfnamefont {A.~L.}\ \bibnamefont {Cavalieri}},\ }\bibfield
  {title} {{\selectlanguage {en}\bibinfo {title} {Clocking {Auger}
  electrons}},\ }\href {https://doi.org/10.1038/s41567-020-01111-0} {\bibfield
  {journal} {\bibinfo  {journal} {Nature Physics}\ ,\ \bibinfo {pages} {1}}
  (\bibinfo {year} {2021})},\ \bibinfo {note} {publisher: Nature Publishing
  Group}\BibitemShut {NoStop}%
\bibitem [{\citenamefont {Li}\ \emph {et~al.}(2022)\citenamefont {Li},
  \citenamefont {Driver}, \citenamefont {Rosenberger}, \citenamefont
  {Champenois}, \citenamefont {Duris}, \citenamefont {Al-Haddad}, \citenamefont
  {Averbukh}, \citenamefont {Barnard}, \citenamefont {Berrah}, \citenamefont
  {Bostedt}, \citenamefont {Bucksbaum}, \citenamefont {Coffee}, \citenamefont
  {DiMauro}, \citenamefont {Fang}, \citenamefont {Garratt}, \citenamefont
  {Gatton}, \citenamefont {Guo}, \citenamefont {Hartmann}, \citenamefont
  {Haxton}, \citenamefont {Helml}, \citenamefont {Huang}, \citenamefont
  {LaForge}, \citenamefont {Kamalov}, \citenamefont {Knurr}, \citenamefont
  {Lin}, \citenamefont {Lutman}, \citenamefont {MacArthur}, \citenamefont
  {Marangos}, \citenamefont {Nantel}, \citenamefont {Natan}, \citenamefont
  {Obaid}, \citenamefont {O’Neal}, \citenamefont {Shivaram}, \citenamefont
  {Schori}, \citenamefont {Walter}, \citenamefont {Wang}, \citenamefont {Wolf},
  \citenamefont {Zhang}, \citenamefont {Kling}, \citenamefont {Marinelli},\
  and\ \citenamefont {Cryan}}]{li_attosecond_2022}%
  \BibitemOpen
  \bibfield  {author} {\bibinfo {author} {\bibfnamefont {S.}~\bibnamefont
  {Li}}, \bibinfo {author} {\bibfnamefont {T.}~\bibnamefont {Driver}}, \bibinfo
  {author} {\bibfnamefont {P.}~\bibnamefont {Rosenberger}}, \bibinfo {author}
  {\bibfnamefont {E.~G.}\ \bibnamefont {Champenois}}, \bibinfo {author}
  {\bibfnamefont {J.}~\bibnamefont {Duris}}, \bibinfo {author} {\bibfnamefont
  {A.}~\bibnamefont {Al-Haddad}}, \bibinfo {author} {\bibfnamefont
  {V.}~\bibnamefont {Averbukh}}, \bibinfo {author} {\bibfnamefont {J.~C.~T.}\
  \bibnamefont {Barnard}}, \bibinfo {author} {\bibfnamefont {N.}~\bibnamefont
  {Berrah}}, \bibinfo {author} {\bibfnamefont {C.}~\bibnamefont {Bostedt}},
  \bibinfo {author} {\bibfnamefont {P.~H.}\ \bibnamefont {Bucksbaum}}, \bibinfo
  {author} {\bibfnamefont {R.~N.}\ \bibnamefont {Coffee}}, \bibinfo {author}
  {\bibfnamefont {L.~F.}\ \bibnamefont {DiMauro}}, \bibinfo {author}
  {\bibfnamefont {L.}~\bibnamefont {Fang}}, \bibinfo {author} {\bibfnamefont
  {D.}~\bibnamefont {Garratt}}, \bibinfo {author} {\bibfnamefont
  {A.}~\bibnamefont {Gatton}}, \bibinfo {author} {\bibfnamefont
  {Z.}~\bibnamefont {Guo}}, \bibinfo {author} {\bibfnamefont {G.}~\bibnamefont
  {Hartmann}}, \bibinfo {author} {\bibfnamefont {D.}~\bibnamefont {Haxton}},
  \bibinfo {author} {\bibfnamefont {W.}~\bibnamefont {Helml}}, \bibinfo
  {author} {\bibfnamefont {Z.}~\bibnamefont {Huang}}, \bibinfo {author}
  {\bibfnamefont {A.~C.}\ \bibnamefont {LaForge}}, \bibinfo {author}
  {\bibfnamefont {A.}~\bibnamefont {Kamalov}}, \bibinfo {author} {\bibfnamefont
  {J.}~\bibnamefont {Knurr}}, \bibinfo {author} {\bibfnamefont {M.-F.}\
  \bibnamefont {Lin}}, \bibinfo {author} {\bibfnamefont {A.~A.}\ \bibnamefont
  {Lutman}}, \bibinfo {author} {\bibfnamefont {J.~P.}\ \bibnamefont
  {MacArthur}}, \bibinfo {author} {\bibfnamefont {J.~P.}\ \bibnamefont
  {Marangos}}, \bibinfo {author} {\bibfnamefont {M.}~\bibnamefont {Nantel}},
  \bibinfo {author} {\bibfnamefont {A.}~\bibnamefont {Natan}}, \bibinfo
  {author} {\bibfnamefont {R.}~\bibnamefont {Obaid}}, \bibinfo {author}
  {\bibfnamefont {J.~T.}\ \bibnamefont {O’Neal}}, \bibinfo {author}
  {\bibfnamefont {N.~H.}\ \bibnamefont {Shivaram}}, \bibinfo {author}
  {\bibfnamefont {A.}~\bibnamefont {Schori}}, \bibinfo {author} {\bibfnamefont
  {P.}~\bibnamefont {Walter}}, \bibinfo {author} {\bibfnamefont {A.~L.}\
  \bibnamefont {Wang}}, \bibinfo {author} {\bibfnamefont {T.~J.~A.}\
  \bibnamefont {Wolf}}, \bibinfo {author} {\bibfnamefont {Z.}~\bibnamefont
  {Zhang}}, \bibinfo {author} {\bibfnamefont {M.~F.}\ \bibnamefont {Kling}},
  \bibinfo {author} {\bibfnamefont {A.}~\bibnamefont {Marinelli}},\ and\
  \bibinfo {author} {\bibfnamefont {J.~P.}\ \bibnamefont {Cryan}},\ }\bibfield
  {title} {\bibinfo {title} {Attosecond coherent electron motion in
  {Auger}-{Meitner} decay},\ }\href {https://doi.org/10.1126/science.abj2096}
  {\bibfield  {journal} {\bibinfo  {journal} {Science}\ }\textbf {\bibinfo
  {volume} {375}},\ \bibinfo {pages} {285} (\bibinfo {year} {2022})},\ \bibinfo
  {note} {publisher: American Association for the Advancement of
  Science}\BibitemShut {NoStop}%
\bibitem [{\citenamefont {Guo}\ \emph {et~al.}(2024)\citenamefont {Guo},
  \citenamefont {Driver}, \citenamefont {Beauvarlet}, \citenamefont {Cesar},
  \citenamefont {Duris}, \citenamefont {Franz}, \citenamefont {Alexander},
  \citenamefont {Bohler}, \citenamefont {Bostedt}, \citenamefont {Averbukh},
  \citenamefont {Cheng}, \citenamefont {DiMauro}, \citenamefont {Doumy},
  \citenamefont {Forbes}, \citenamefont {Gessner}, \citenamefont {Glownia},
  \citenamefont {Isele}, \citenamefont {Kamalov}, \citenamefont {Larsen},
  \citenamefont {Li}, \citenamefont {Li}, \citenamefont {Lin}, \citenamefont
  {McCracken}, \citenamefont {Obaid}, \citenamefont {O’Neal}, \citenamefont
  {Robles}, \citenamefont {Rolles}, \citenamefont {Ruberti}, \citenamefont
  {Rudenko}, \citenamefont {Slaughter}, \citenamefont {Sudar}, \citenamefont
  {Thierstein}, \citenamefont {Tuthill}, \citenamefont {Ueda}, \citenamefont
  {Wang}, \citenamefont {Wang}, \citenamefont {Wang}, \citenamefont {Weber},
  \citenamefont {Wolf}, \citenamefont {Young}, \citenamefont {Zhang},
  \citenamefont {Bucksbaum}, \citenamefont {Marangos}, \citenamefont {Kling},
  \citenamefont {Huang}, \citenamefont {Walter}, \citenamefont {Inhester},
  \citenamefont {Berrah}, \citenamefont {Cryan},\ and\ \citenamefont
  {Marinelli}}]{guo_experimental_2024}%
  \BibitemOpen
  \bibfield  {author} {\bibinfo {author} {\bibfnamefont {Z.}~\bibnamefont
  {Guo}}, \bibinfo {author} {\bibfnamefont {T.}~\bibnamefont {Driver}},
  \bibinfo {author} {\bibfnamefont {S.}~\bibnamefont {Beauvarlet}}, \bibinfo
  {author} {\bibfnamefont {D.}~\bibnamefont {Cesar}}, \bibinfo {author}
  {\bibfnamefont {J.}~\bibnamefont {Duris}}, \bibinfo {author} {\bibfnamefont
  {P.~L.}\ \bibnamefont {Franz}}, \bibinfo {author} {\bibfnamefont
  {O.}~\bibnamefont {Alexander}}, \bibinfo {author} {\bibfnamefont
  {D.}~\bibnamefont {Bohler}}, \bibinfo {author} {\bibfnamefont
  {C.}~\bibnamefont {Bostedt}}, \bibinfo {author} {\bibfnamefont
  {V.}~\bibnamefont {Averbukh}}, \bibinfo {author} {\bibfnamefont
  {X.}~\bibnamefont {Cheng}}, \bibinfo {author} {\bibfnamefont {L.~F.}\
  \bibnamefont {DiMauro}}, \bibinfo {author} {\bibfnamefont {G.}~\bibnamefont
  {Doumy}}, \bibinfo {author} {\bibfnamefont {R.}~\bibnamefont {Forbes}},
  \bibinfo {author} {\bibfnamefont {O.}~\bibnamefont {Gessner}}, \bibinfo
  {author} {\bibfnamefont {J.~M.}\ \bibnamefont {Glownia}}, \bibinfo {author}
  {\bibfnamefont {E.}~\bibnamefont {Isele}}, \bibinfo {author} {\bibfnamefont
  {A.}~\bibnamefont {Kamalov}}, \bibinfo {author} {\bibfnamefont {K.~A.}\
  \bibnamefont {Larsen}}, \bibinfo {author} {\bibfnamefont {S.}~\bibnamefont
  {Li}}, \bibinfo {author} {\bibfnamefont {X.}~\bibnamefont {Li}}, \bibinfo
  {author} {\bibfnamefont {M.-F.}\ \bibnamefont {Lin}}, \bibinfo {author}
  {\bibfnamefont {G.~A.}\ \bibnamefont {McCracken}}, \bibinfo {author}
  {\bibfnamefont {R.}~\bibnamefont {Obaid}}, \bibinfo {author} {\bibfnamefont
  {J.~T.}\ \bibnamefont {O’Neal}}, \bibinfo {author} {\bibfnamefont {R.~R.}\
  \bibnamefont {Robles}}, \bibinfo {author} {\bibfnamefont {D.}~\bibnamefont
  {Rolles}}, \bibinfo {author} {\bibfnamefont {M.}~\bibnamefont {Ruberti}},
  \bibinfo {author} {\bibfnamefont {A.}~\bibnamefont {Rudenko}}, \bibinfo
  {author} {\bibfnamefont {D.~S.}\ \bibnamefont {Slaughter}}, \bibinfo {author}
  {\bibfnamefont {N.~S.}\ \bibnamefont {Sudar}}, \bibinfo {author}
  {\bibfnamefont {E.}~\bibnamefont {Thierstein}}, \bibinfo {author}
  {\bibfnamefont {D.}~\bibnamefont {Tuthill}}, \bibinfo {author} {\bibfnamefont
  {K.}~\bibnamefont {Ueda}}, \bibinfo {author} {\bibfnamefont {E.}~\bibnamefont
  {Wang}}, \bibinfo {author} {\bibfnamefont {A.~L.}\ \bibnamefont {Wang}},
  \bibinfo {author} {\bibfnamefont {J.}~\bibnamefont {Wang}}, \bibinfo {author}
  {\bibfnamefont {T.}~\bibnamefont {Weber}}, \bibinfo {author} {\bibfnamefont
  {T.~J.~A.}\ \bibnamefont {Wolf}}, \bibinfo {author} {\bibfnamefont
  {L.}~\bibnamefont {Young}}, \bibinfo {author} {\bibfnamefont
  {Z.}~\bibnamefont {Zhang}}, \bibinfo {author} {\bibfnamefont {P.~H.}\
  \bibnamefont {Bucksbaum}}, \bibinfo {author} {\bibfnamefont {J.~P.}\
  \bibnamefont {Marangos}}, \bibinfo {author} {\bibfnamefont {M.~F.}\
  \bibnamefont {Kling}}, \bibinfo {author} {\bibfnamefont {Z.}~\bibnamefont
  {Huang}}, \bibinfo {author} {\bibfnamefont {P.}~\bibnamefont {Walter}},
  \bibinfo {author} {\bibfnamefont {L.}~\bibnamefont {Inhester}}, \bibinfo
  {author} {\bibfnamefont {N.}~\bibnamefont {Berrah}}, \bibinfo {author}
  {\bibfnamefont {J.~P.}\ \bibnamefont {Cryan}},\ and\ \bibinfo {author}
  {\bibfnamefont {A.}~\bibnamefont {Marinelli}},\ }\bibfield  {title}
  {{\selectlanguage {en}\bibinfo {title} {Experimental demonstration of
  attosecond pump–probe spectroscopy with an {X}-ray free-electron laser}},\
  }\href {https://doi.org/10.1038/s41566-024-01419-w} {\bibfield  {journal}
  {\bibinfo  {journal} {Nature Photonics}\ }\textbf {\bibinfo {volume} {18}},\
  \bibinfo {pages} {691} (\bibinfo {year} {2024})},\ \bibinfo {note}
  {publisher: Nature Publishing Group}\BibitemShut {NoStop}%
\bibitem [{\citenamefont {Maroju}\ \emph {et~al.}(2020)\citenamefont {Maroju},
  \citenamefont {Grazioli}, \citenamefont {Di~Fraia}, \citenamefont {Moioli},
  \citenamefont {Ertel}, \citenamefont {Ahmadi}, \citenamefont {Plekan},
  \citenamefont {Finetti}, \citenamefont {Allaria}, \citenamefont {Giannessi},
  \citenamefont {De~Ninno}, \citenamefont {Spezzani}, \citenamefont {Penco},
  \citenamefont {Spampinati}, \citenamefont {Demidovich}, \citenamefont
  {Danailov}, \citenamefont {Borghes}, \citenamefont {Kourousias},
  \citenamefont {Sanches Dos~Reis}, \citenamefont {Billé}, \citenamefont
  {Lutman}, \citenamefont {Squibb}, \citenamefont {Feifel}, \citenamefont
  {Carpeggiani}, \citenamefont {Reduzzi}, \citenamefont {Mazza}, \citenamefont
  {Meyer}, \citenamefont {Bengtsson}, \citenamefont {Ibrakovic}, \citenamefont
  {Simpson}, \citenamefont {Mauritsson}, \citenamefont {Csizmadia},
  \citenamefont {Dumergue}, \citenamefont {Kühn}, \citenamefont
  {Nandiga~Gopalakrishna}, \citenamefont {You}, \citenamefont {Ueda},
  \citenamefont {Labeye}, \citenamefont {Bækhøj}, \citenamefont {Schafer},
  \citenamefont {Gryzlova}, \citenamefont {Grum-Grzhimailo}, \citenamefont
  {Prince}, \citenamefont {Callegari},\ and\ \citenamefont
  {Sansone}}]{maroju_attosecond_2020}%
  \BibitemOpen
  \bibfield  {author} {\bibinfo {author} {\bibfnamefont {P.~K.}\ \bibnamefont
  {Maroju}}, \bibinfo {author} {\bibfnamefont {C.}~\bibnamefont {Grazioli}},
  \bibinfo {author} {\bibfnamefont {M.}~\bibnamefont {Di~Fraia}}, \bibinfo
  {author} {\bibfnamefont {M.}~\bibnamefont {Moioli}}, \bibinfo {author}
  {\bibfnamefont {D.}~\bibnamefont {Ertel}}, \bibinfo {author} {\bibfnamefont
  {H.}~\bibnamefont {Ahmadi}}, \bibinfo {author} {\bibfnamefont
  {O.}~\bibnamefont {Plekan}}, \bibinfo {author} {\bibfnamefont
  {P.}~\bibnamefont {Finetti}}, \bibinfo {author} {\bibfnamefont
  {E.}~\bibnamefont {Allaria}}, \bibinfo {author} {\bibfnamefont
  {L.}~\bibnamefont {Giannessi}}, \bibinfo {author} {\bibfnamefont
  {G.}~\bibnamefont {De~Ninno}}, \bibinfo {author} {\bibfnamefont
  {C.}~\bibnamefont {Spezzani}}, \bibinfo {author} {\bibfnamefont
  {G.}~\bibnamefont {Penco}}, \bibinfo {author} {\bibfnamefont
  {S.}~\bibnamefont {Spampinati}}, \bibinfo {author} {\bibfnamefont
  {A.}~\bibnamefont {Demidovich}}, \bibinfo {author} {\bibfnamefont {M.~B.}\
  \bibnamefont {Danailov}}, \bibinfo {author} {\bibfnamefont {R.}~\bibnamefont
  {Borghes}}, \bibinfo {author} {\bibfnamefont {G.}~\bibnamefont {Kourousias}},
  \bibinfo {author} {\bibfnamefont {C.~E.}\ \bibnamefont {Sanches Dos~Reis}},
  \bibinfo {author} {\bibfnamefont {F.}~\bibnamefont {Billé}}, \bibinfo
  {author} {\bibfnamefont {A.~A.}\ \bibnamefont {Lutman}}, \bibinfo {author}
  {\bibfnamefont {R.~J.}\ \bibnamefont {Squibb}}, \bibinfo {author}
  {\bibfnamefont {R.}~\bibnamefont {Feifel}}, \bibinfo {author} {\bibfnamefont
  {P.}~\bibnamefont {Carpeggiani}}, \bibinfo {author} {\bibfnamefont
  {M.}~\bibnamefont {Reduzzi}}, \bibinfo {author} {\bibfnamefont
  {T.}~\bibnamefont {Mazza}}, \bibinfo {author} {\bibfnamefont
  {M.}~\bibnamefont {Meyer}}, \bibinfo {author} {\bibfnamefont
  {S.}~\bibnamefont {Bengtsson}}, \bibinfo {author} {\bibfnamefont
  {N.}~\bibnamefont {Ibrakovic}}, \bibinfo {author} {\bibfnamefont {E.~R.}\
  \bibnamefont {Simpson}}, \bibinfo {author} {\bibfnamefont {J.}~\bibnamefont
  {Mauritsson}}, \bibinfo {author} {\bibfnamefont {T.}~\bibnamefont
  {Csizmadia}}, \bibinfo {author} {\bibfnamefont {M.}~\bibnamefont {Dumergue}},
  \bibinfo {author} {\bibfnamefont {S.}~\bibnamefont {Kühn}}, \bibinfo
  {author} {\bibfnamefont {H.}~\bibnamefont {Nandiga~Gopalakrishna}}, \bibinfo
  {author} {\bibfnamefont {D.}~\bibnamefont {You}}, \bibinfo {author}
  {\bibfnamefont {K.}~\bibnamefont {Ueda}}, \bibinfo {author} {\bibfnamefont
  {M.}~\bibnamefont {Labeye}}, \bibinfo {author} {\bibfnamefont {J.~E.}\
  \bibnamefont {Bækhøj}}, \bibinfo {author} {\bibfnamefont {K.~J.}\
  \bibnamefont {Schafer}}, \bibinfo {author} {\bibfnamefont {E.~V.}\
  \bibnamefont {Gryzlova}}, \bibinfo {author} {\bibfnamefont {A.~N.}\
  \bibnamefont {Grum-Grzhimailo}}, \bibinfo {author} {\bibfnamefont {K.~C.}\
  \bibnamefont {Prince}}, \bibinfo {author} {\bibfnamefont {C.}~\bibnamefont
  {Callegari}},\ and\ \bibinfo {author} {\bibfnamefont {G.}~\bibnamefont
  {Sansone}},\ }\bibfield  {title} {{\selectlanguage {en}\bibinfo {title}
  {Attosecond pulse shaping using a seeded free-electron laser}},\ }\href
  {https://doi.org/10.1038/s41586-020-2005-6} {\bibfield  {journal} {\bibinfo
  {journal} {Nature}\ }\textbf {\bibinfo {volume} {578}},\ \bibinfo {pages}
  {386} (\bibinfo {year} {2020})}\BibitemShut {NoStop}%
\bibitem [{\citenamefont {Maroju}\ \emph {et~al.}(2023)\citenamefont {Maroju},
  \citenamefont {Di~Fraia}, \citenamefont {Plekan}, \citenamefont {Bonanomi},
  \citenamefont {Merzuk}, \citenamefont {Busto}, \citenamefont {Makos},
  \citenamefont {Schmoll}, \citenamefont {Shah}, \citenamefont {Ribič},
  \citenamefont {Giannessi}, \citenamefont {De~Ninno}, \citenamefont
  {Spezzani}, \citenamefont {Penco}, \citenamefont {Demidovich}, \citenamefont
  {Danailov}, \citenamefont {Coreno}, \citenamefont {Zangrando}, \citenamefont
  {Simoncig}, \citenamefont {Manfredda}, \citenamefont {Squibb}, \citenamefont
  {Feifel}, \citenamefont {Bengtsson}, \citenamefont {Simpson}, \citenamefont
  {Csizmadia}, \citenamefont {Dumergue}, \citenamefont {Kühn}, \citenamefont
  {Ueda}, \citenamefont {Li}, \citenamefont {Schafer}, \citenamefont
  {Frassetto}, \citenamefont {Poletto}, \citenamefont {Prince}, \citenamefont
  {Mauritsson}, \citenamefont {Callegari},\ and\ \citenamefont
  {Sansone}}]{maroju_attosecond_2023}%
  \BibitemOpen
  \bibfield  {author} {\bibinfo {author} {\bibfnamefont {P.~K.}\ \bibnamefont
  {Maroju}}, \bibinfo {author} {\bibfnamefont {M.}~\bibnamefont {Di~Fraia}},
  \bibinfo {author} {\bibfnamefont {O.}~\bibnamefont {Plekan}}, \bibinfo
  {author} {\bibfnamefont {M.}~\bibnamefont {Bonanomi}}, \bibinfo {author}
  {\bibfnamefont {B.}~\bibnamefont {Merzuk}}, \bibinfo {author} {\bibfnamefont
  {D.}~\bibnamefont {Busto}}, \bibinfo {author} {\bibfnamefont
  {I.}~\bibnamefont {Makos}}, \bibinfo {author} {\bibfnamefont
  {M.}~\bibnamefont {Schmoll}}, \bibinfo {author} {\bibfnamefont
  {R.}~\bibnamefont {Shah}}, \bibinfo {author} {\bibfnamefont {P.~R.}\
  \bibnamefont {Ribič}}, \bibinfo {author} {\bibfnamefont {L.}~\bibnamefont
  {Giannessi}}, \bibinfo {author} {\bibfnamefont {G.}~\bibnamefont {De~Ninno}},
  \bibinfo {author} {\bibfnamefont {C.}~\bibnamefont {Spezzani}}, \bibinfo
  {author} {\bibfnamefont {G.}~\bibnamefont {Penco}}, \bibinfo {author}
  {\bibfnamefont {A.}~\bibnamefont {Demidovich}}, \bibinfo {author}
  {\bibfnamefont {M.}~\bibnamefont {Danailov}}, \bibinfo {author}
  {\bibfnamefont {M.}~\bibnamefont {Coreno}}, \bibinfo {author} {\bibfnamefont
  {M.}~\bibnamefont {Zangrando}}, \bibinfo {author} {\bibfnamefont
  {A.}~\bibnamefont {Simoncig}}, \bibinfo {author} {\bibfnamefont
  {M.}~\bibnamefont {Manfredda}}, \bibinfo {author} {\bibfnamefont {R.~J.}\
  \bibnamefont {Squibb}}, \bibinfo {author} {\bibfnamefont {R.}~\bibnamefont
  {Feifel}}, \bibinfo {author} {\bibfnamefont {S.}~\bibnamefont {Bengtsson}},
  \bibinfo {author} {\bibfnamefont {E.~R.}\ \bibnamefont {Simpson}}, \bibinfo
  {author} {\bibfnamefont {T.}~\bibnamefont {Csizmadia}}, \bibinfo {author}
  {\bibfnamefont {M.}~\bibnamefont {Dumergue}}, \bibinfo {author}
  {\bibfnamefont {S.}~\bibnamefont {Kühn}}, \bibinfo {author} {\bibfnamefont
  {K.}~\bibnamefont {Ueda}}, \bibinfo {author} {\bibfnamefont {J.}~\bibnamefont
  {Li}}, \bibinfo {author} {\bibfnamefont {K.~J.}\ \bibnamefont {Schafer}},
  \bibinfo {author} {\bibfnamefont {F.}~\bibnamefont {Frassetto}}, \bibinfo
  {author} {\bibfnamefont {L.}~\bibnamefont {Poletto}}, \bibinfo {author}
  {\bibfnamefont {K.~C.}\ \bibnamefont {Prince}}, \bibinfo {author}
  {\bibfnamefont {J.}~\bibnamefont {Mauritsson}}, \bibinfo {author}
  {\bibfnamefont {C.}~\bibnamefont {Callegari}},\ and\ \bibinfo {author}
  {\bibfnamefont {G.}~\bibnamefont {Sansone}},\ }\bibfield  {title}
  {{\selectlanguage {en}\bibinfo {title} {Attosecond coherent control of
  electronic wave packets in two-colour photoionization using a novel timing
  tool for seeded free-electron laser}},\ }\href
  {https://doi.org/10.1038/s41566-022-01127-3} {\bibfield  {journal} {\bibinfo
  {journal} {Nature Photonics}\ }\textbf {\bibinfo {volume} {17}},\ \bibinfo
  {pages} {200} (\bibinfo {year} {2023})},\ \bibinfo {note} {number: 2
  Publisher: Nature Publishing Group}\BibitemShut {NoStop}%
\bibitem [{\citenamefont {Driver}\ \emph {et~al.}(2024)\citenamefont {Driver},
  \citenamefont {Mountney}, \citenamefont {Wang}, \citenamefont {Ortmann},
  \citenamefont {Al-Haddad}, \citenamefont {Berrah}, \citenamefont {Bostedt},
  \citenamefont {Champenois}, \citenamefont {DiMauro}, \citenamefont {Duris},
  \citenamefont {Garratt}, \citenamefont {Glownia}, \citenamefont {Guo},
  \citenamefont {Haxton}, \citenamefont {Isele}, \citenamefont {Ivanov},
  \citenamefont {Ji}, \citenamefont {Kamalov}, \citenamefont {Li},
  \citenamefont {Lin}, \citenamefont {Marangos}, \citenamefont {Obaid},
  \citenamefont {O’Neal}, \citenamefont {Rosenberger}, \citenamefont
  {Shivaram}, \citenamefont {Wang}, \citenamefont {Walter}, \citenamefont
  {Wolf}, \citenamefont {Wörner}, \citenamefont {Zhang}, \citenamefont
  {Bucksbaum}, \citenamefont {Kling}, \citenamefont {Landsman}, \citenamefont
  {Lucchese}, \citenamefont {Emmanouilidou}, \citenamefont {Marinelli},\ and\
  \citenamefont {Cryan}}]{driver_attosecond_2024}%
  \BibitemOpen
  \bibfield  {author} {\bibinfo {author} {\bibfnamefont {T.}~\bibnamefont
  {Driver}}, \bibinfo {author} {\bibfnamefont {M.}~\bibnamefont {Mountney}},
  \bibinfo {author} {\bibfnamefont {J.}~\bibnamefont {Wang}}, \bibinfo {author}
  {\bibfnamefont {L.}~\bibnamefont {Ortmann}}, \bibinfo {author} {\bibfnamefont
  {A.}~\bibnamefont {Al-Haddad}}, \bibinfo {author} {\bibfnamefont
  {N.}~\bibnamefont {Berrah}}, \bibinfo {author} {\bibfnamefont
  {C.}~\bibnamefont {Bostedt}}, \bibinfo {author} {\bibfnamefont {E.~G.}\
  \bibnamefont {Champenois}}, \bibinfo {author} {\bibfnamefont {L.~F.}\
  \bibnamefont {DiMauro}}, \bibinfo {author} {\bibfnamefont {J.}~\bibnamefont
  {Duris}}, \bibinfo {author} {\bibfnamefont {D.}~\bibnamefont {Garratt}},
  \bibinfo {author} {\bibfnamefont {J.~M.}\ \bibnamefont {Glownia}}, \bibinfo
  {author} {\bibfnamefont {Z.}~\bibnamefont {Guo}}, \bibinfo {author}
  {\bibfnamefont {D.}~\bibnamefont {Haxton}}, \bibinfo {author} {\bibfnamefont
  {E.}~\bibnamefont {Isele}}, \bibinfo {author} {\bibfnamefont
  {I.}~\bibnamefont {Ivanov}}, \bibinfo {author} {\bibfnamefont
  {J.}~\bibnamefont {Ji}}, \bibinfo {author} {\bibfnamefont {A.}~\bibnamefont
  {Kamalov}}, \bibinfo {author} {\bibfnamefont {S.}~\bibnamefont {Li}},
  \bibinfo {author} {\bibfnamefont {M.-F.}\ \bibnamefont {Lin}}, \bibinfo
  {author} {\bibfnamefont {J.~P.}\ \bibnamefont {Marangos}}, \bibinfo {author}
  {\bibfnamefont {R.}~\bibnamefont {Obaid}}, \bibinfo {author} {\bibfnamefont
  {J.~T.}\ \bibnamefont {O’Neal}}, \bibinfo {author} {\bibfnamefont
  {P.}~\bibnamefont {Rosenberger}}, \bibinfo {author} {\bibfnamefont {N.~H.}\
  \bibnamefont {Shivaram}}, \bibinfo {author} {\bibfnamefont {A.~L.}\
  \bibnamefont {Wang}}, \bibinfo {author} {\bibfnamefont {P.}~\bibnamefont
  {Walter}}, \bibinfo {author} {\bibfnamefont {T.~J.~A.}\ \bibnamefont {Wolf}},
  \bibinfo {author} {\bibfnamefont {H.~J.}\ \bibnamefont {Wörner}}, \bibinfo
  {author} {\bibfnamefont {Z.}~\bibnamefont {Zhang}}, \bibinfo {author}
  {\bibfnamefont {P.~H.}\ \bibnamefont {Bucksbaum}}, \bibinfo {author}
  {\bibfnamefont {M.~F.}\ \bibnamefont {Kling}}, \bibinfo {author}
  {\bibfnamefont {A.~S.}\ \bibnamefont {Landsman}}, \bibinfo {author}
  {\bibfnamefont {R.~R.}\ \bibnamefont {Lucchese}}, \bibinfo {author}
  {\bibfnamefont {A.}~\bibnamefont {Emmanouilidou}}, \bibinfo {author}
  {\bibfnamefont {A.}~\bibnamefont {Marinelli}},\ and\ \bibinfo {author}
  {\bibfnamefont {J.~P.}\ \bibnamefont {Cryan}},\ }\bibfield  {title}
  {{\selectlanguage {en}\bibinfo {title} {Attosecond delays in {X}-ray
  molecular ionization}},\ }\href {https://doi.org/10.1038/s41586-024-07771-9}
  {\bibfield  {journal} {\bibinfo  {journal} {Nature}\ }\textbf {\bibinfo
  {volume} {632}},\ \bibinfo {pages} {762} (\bibinfo {year} {2024})},\ \bibinfo
  {note} {publisher: Nature Publishing Group}\BibitemShut {NoStop}%
\bibitem [{\citenamefont {Wang}\ \emph {et~al.}(2024)\citenamefont {Wang},
  \citenamefont {Driver}, \citenamefont {Franz}, \citenamefont {Koloren\v{c}},
  \citenamefont {Thierstein}, \citenamefont {Robles}, \citenamefont {Isele},
  \citenamefont {Guo}, \citenamefont {Cesar}, \citenamefont {Alexander},
  \citenamefont {Beauvarlet}, \citenamefont {Borne}, \citenamefont {Cheng},
  \citenamefont {DiMauro}, \citenamefont {Duris}, \citenamefont {Glownia},
  \citenamefont {Gra{\ss}l}, \citenamefont {Hockett}, \citenamefont {Hoffman},
  \citenamefont {Kamalov}, \citenamefont {Larsen}, \citenamefont {Li},
  \citenamefont {Li}, \citenamefont {Lin}, \citenamefont {Obaid}, \citenamefont
  {Rosenberger}, \citenamefont {Walter}, \citenamefont {Wolf}, \citenamefont
  {Marangos}, \citenamefont {Kling}, \citenamefont {Bucksbaum}, \citenamefont
  {Marinelli},\ and\ \citenamefont {Cryan}}]{wang_probing_2024}%
  \BibitemOpen
  \bibfield  {author} {\bibinfo {author} {\bibfnamefont {J.}~\bibnamefont
  {Wang}}, \bibinfo {author} {\bibfnamefont {T.}~\bibnamefont {Driver}},
  \bibinfo {author} {\bibfnamefont {P.~L.}\ \bibnamefont {Franz}}, \bibinfo
  {author} {\bibfnamefont {P.}~\bibnamefont {Koloren\v{c}}}, \bibinfo {author}
  {\bibfnamefont {E.}~\bibnamefont {Thierstein}}, \bibinfo {author}
  {\bibfnamefont {R.~R.}\ \bibnamefont {Robles}}, \bibinfo {author}
  {\bibfnamefont {E.}~\bibnamefont {Isele}}, \bibinfo {author} {\bibfnamefont
  {Z.}~\bibnamefont {Guo}}, \bibinfo {author} {\bibfnamefont {D.}~\bibnamefont
  {Cesar}}, \bibinfo {author} {\bibfnamefont {O.}~\bibnamefont {Alexander}},
  \bibinfo {author} {\bibfnamefont {S.}~\bibnamefont {Beauvarlet}}, \bibinfo
  {author} {\bibfnamefont {K.}~\bibnamefont {Borne}}, \bibinfo {author}
  {\bibfnamefont {X.}~\bibnamefont {Cheng}}, \bibinfo {author} {\bibfnamefont
  {L.~F.}\ \bibnamefont {DiMauro}}, \bibinfo {author} {\bibfnamefont
  {J.}~\bibnamefont {Duris}}, \bibinfo {author} {\bibfnamefont {J.~M.}\
  \bibnamefont {Glownia}}, \bibinfo {author} {\bibfnamefont {M.}~\bibnamefont
  {Gra{\ss}l}}, \bibinfo {author} {\bibfnamefont {P.}~\bibnamefont {Hockett}},
  \bibinfo {author} {\bibfnamefont {M.}~\bibnamefont {Hoffman}}, \bibinfo
  {author} {\bibfnamefont {A.}~\bibnamefont {Kamalov}}, \bibinfo {author}
  {\bibfnamefont {K.~A.}\ \bibnamefont {Larsen}}, \bibinfo {author}
  {\bibfnamefont {S.}~\bibnamefont {Li}}, \bibinfo {author} {\bibfnamefont
  {X.}~\bibnamefont {Li}}, \bibinfo {author} {\bibfnamefont {M.-F.}\
  \bibnamefont {Lin}}, \bibinfo {author} {\bibfnamefont {R.}~\bibnamefont
  {Obaid}}, \bibinfo {author} {\bibfnamefont {P.}~\bibnamefont {Rosenberger}},
  \bibinfo {author} {\bibfnamefont {P.}~\bibnamefont {Walter}}, \bibinfo
  {author} {\bibfnamefont {T.~J.}\ \bibnamefont {Wolf}}, \bibinfo {author}
  {\bibfnamefont {J.~P.}\ \bibnamefont {Marangos}}, \bibinfo {author}
  {\bibfnamefont {M.~F.}\ \bibnamefont {Kling}}, \bibinfo {author}
  {\bibfnamefont {P.~H.}\ \bibnamefont {Bucksbaum}}, \bibinfo {author}
  {\bibfnamefont {A.}~\bibnamefont {Marinelli}},\ and\ \bibinfo {author}
  {\bibfnamefont {J.~P.}\ \bibnamefont {Cryan}},\ }\bibfield  {title} {\bibinfo
  {title} {Probing electronic coherence between core-level vacancies at
  different atomic sites},\ }\href@noop {} {\bibfield  {journal} {\bibinfo
  {journal} {Physical Review X (accepted)}\ } (\bibinfo {year}
  {2024})}\BibitemShut {NoStop}%
\bibitem [{\citenamefont {Haynes}\ \emph {et~al.}(2020)\citenamefont {Haynes},
  \citenamefont {Wurzer}, \citenamefont {Schletter}, \citenamefont {Al-Haddad},
  \citenamefont {Blaga}, \citenamefont {Bostedt}, \citenamefont {Bozek},
  \citenamefont {Bucher}, \citenamefont {Camper}, \citenamefont {Carron},
  \citenamefont {Coffee}, \citenamefont {Costello}, \citenamefont {DiMauro},
  \citenamefont {Ding}, \citenamefont {Ferguson}, \citenamefont {Grguraš},
  \citenamefont {Helml}, \citenamefont {Hoffmann}, \citenamefont {Ilchen},
  \citenamefont {Jalas}, \citenamefont {Kabachnik}, \citenamefont {Kazansky},
  \citenamefont {Kienberger}, \citenamefont {Maier}, \citenamefont {Maxwell},
  \citenamefont {Mazza}, \citenamefont {Meyer}, \citenamefont {Park},
  \citenamefont {Robinson}, \citenamefont {Roedig}, \citenamefont {Schlarb},
  \citenamefont {Singla}, \citenamefont {Tellkamp}, \citenamefont {Zhang},
  \citenamefont {Doumy}, \citenamefont {Behrens},\ and\ \citenamefont
  {Cavalieri}}]{haynes_clocking_2020}%
  \BibitemOpen
  \bibfield  {author} {\bibinfo {author} {\bibfnamefont {D.~C.}\ \bibnamefont
  {Haynes}}, \bibinfo {author} {\bibfnamefont {M.}~\bibnamefont {Wurzer}},
  \bibinfo {author} {\bibfnamefont {A.}~\bibnamefont {Schletter}}, \bibinfo
  {author} {\bibfnamefont {A.}~\bibnamefont {Al-Haddad}}, \bibinfo {author}
  {\bibfnamefont {C.}~\bibnamefont {Blaga}}, \bibinfo {author} {\bibfnamefont
  {C.}~\bibnamefont {Bostedt}}, \bibinfo {author} {\bibfnamefont
  {J.}~\bibnamefont {Bozek}}, \bibinfo {author} {\bibfnamefont
  {M.}~\bibnamefont {Bucher}}, \bibinfo {author} {\bibfnamefont
  {A.}~\bibnamefont {Camper}}, \bibinfo {author} {\bibfnamefont
  {S.}~\bibnamefont {Carron}}, \bibinfo {author} {\bibfnamefont
  {R.}~\bibnamefont {Coffee}}, \bibinfo {author} {\bibfnamefont {J.~T.}\
  \bibnamefont {Costello}}, \bibinfo {author} {\bibfnamefont {L.~F.}\
  \bibnamefont {DiMauro}}, \bibinfo {author} {\bibfnamefont {Y.}~\bibnamefont
  {Ding}}, \bibinfo {author} {\bibfnamefont {K.}~\bibnamefont {Ferguson}},
  \bibinfo {author} {\bibfnamefont {I.}~\bibnamefont {Grguraš}}, \bibinfo
  {author} {\bibfnamefont {W.}~\bibnamefont {Helml}}, \bibinfo {author}
  {\bibfnamefont {M.~C.}\ \bibnamefont {Hoffmann}}, \bibinfo {author}
  {\bibfnamefont {M.}~\bibnamefont {Ilchen}}, \bibinfo {author} {\bibfnamefont
  {S.}~\bibnamefont {Jalas}}, \bibinfo {author} {\bibfnamefont {N.~M.}\
  \bibnamefont {Kabachnik}}, \bibinfo {author} {\bibfnamefont {A.~K.}\
  \bibnamefont {Kazansky}}, \bibinfo {author} {\bibfnamefont {R.}~\bibnamefont
  {Kienberger}}, \bibinfo {author} {\bibfnamefont {A.~R.}\ \bibnamefont
  {Maier}}, \bibinfo {author} {\bibfnamefont {T.}~\bibnamefont {Maxwell}},
  \bibinfo {author} {\bibfnamefont {T.}~\bibnamefont {Mazza}}, \bibinfo
  {author} {\bibfnamefont {M.}~\bibnamefont {Meyer}}, \bibinfo {author}
  {\bibfnamefont {H.}~\bibnamefont {Park}}, \bibinfo {author} {\bibfnamefont
  {J.~S.}\ \bibnamefont {Robinson}}, \bibinfo {author} {\bibfnamefont
  {C.}~\bibnamefont {Roedig}}, \bibinfo {author} {\bibfnamefont
  {H.}~\bibnamefont {Schlarb}}, \bibinfo {author} {\bibfnamefont
  {R.}~\bibnamefont {Singla}}, \bibinfo {author} {\bibfnamefont
  {F.}~\bibnamefont {Tellkamp}}, \bibinfo {author} {\bibfnamefont
  {K.}~\bibnamefont {Zhang}}, \bibinfo {author} {\bibfnamefont
  {G.}~\bibnamefont {Doumy}}, \bibinfo {author} {\bibfnamefont
  {C.}~\bibnamefont {Behrens}},\ and\ \bibinfo {author} {\bibfnamefont {A.~L.}\
  \bibnamefont {Cavalieri}},\ }\bibfield  {title} {\bibinfo {title} {Clocking
  {Auger} {Electrons}},\ }\href {http://arxiv.org/abs/2003.10398} {\bibfield
  {journal} {\bibinfo  {journal} {arXiv:2003.10398 [physics]}\ } (\bibinfo
  {year} {2020})},\ \bibinfo {note} {arXiv: 2003.10398}\BibitemShut {NoStop}%
\bibitem [{\citenamefont {Kitzler}\ \emph {et~al.}(2002)\citenamefont
  {Kitzler}, \citenamefont {Milosevic}, \citenamefont {Scrinzi}, \citenamefont
  {Krausz},\ and\ \citenamefont {Brabec}}]{kitzler_quantum_2002}%
  \BibitemOpen
  \bibfield  {author} {\bibinfo {author} {\bibfnamefont {M.}~\bibnamefont
  {Kitzler}}, \bibinfo {author} {\bibfnamefont {N.}~\bibnamefont {Milosevic}},
  \bibinfo {author} {\bibfnamefont {A.}~\bibnamefont {Scrinzi}}, \bibinfo
  {author} {\bibfnamefont {F.}~\bibnamefont {Krausz}},\ and\ \bibinfo {author}
  {\bibfnamefont {T.}~\bibnamefont {Brabec}},\ }\bibfield  {title} {\bibinfo
  {title} {Quantum {Theory} of {Attosecond} {XUV} {Pulse} {Measurement} by
  {Laser} {Dressed} {Photoionization}},\ }\href
  {https://doi.org/10.1103/PhysRevLett.88.173904} {\bibfield  {journal}
  {\bibinfo  {journal} {Physical Review Letters}\ }\textbf {\bibinfo {volume}
  {88}},\ \bibinfo {pages} {173904} (\bibinfo {year} {2002})}\BibitemShut
  {NoStop}%
\bibitem [{\citenamefont {Wolkow}(1935)}]{wolkow_uber_1935}%
  \BibitemOpen
  \bibfield  {author} {\bibinfo {author} {\bibfnamefont {D.~M.}\ \bibnamefont
  {Wolkow}},\ }\bibfield  {title} {{\selectlanguage {de}\bibinfo {title} {Über
  eine {Klasse} von {Lösungen} der {Diracschen} {Gleichung}}},\ }\href
  {https://doi.org/10.1007/BF01331022} {\bibfield  {journal} {\bibinfo
  {journal} {Zeitschrift für Physik}\ }\textbf {\bibinfo {volume} {94}},\
  \bibinfo {pages} {250} (\bibinfo {year} {1935})}\BibitemShut {NoStop}%
\bibitem [{\citenamefont {Li}\ \emph {et~al.}(2018)\citenamefont {Li},
  \citenamefont {Champenois}, \citenamefont {Coffee}, \citenamefont {Guo},
  \citenamefont {Hegazy}, \citenamefont {Kamalov}, \citenamefont {Natan},
  \citenamefont {O’Neal}, \citenamefont {Osipov}, \citenamefont {Owens},
  \citenamefont {Ray}, \citenamefont {Rich}, \citenamefont {Walter},
  \citenamefont {Marinelli},\ and\ \citenamefont {Cryan}}]{li_co-axial_2018}%
  \BibitemOpen
  \bibfield  {author} {\bibinfo {author} {\bibfnamefont {S.}~\bibnamefont
  {Li}}, \bibinfo {author} {\bibfnamefont {E.~G.}\ \bibnamefont {Champenois}},
  \bibinfo {author} {\bibfnamefont {R.}~\bibnamefont {Coffee}}, \bibinfo
  {author} {\bibfnamefont {Z.}~\bibnamefont {Guo}}, \bibinfo {author}
  {\bibfnamefont {K.}~\bibnamefont {Hegazy}}, \bibinfo {author} {\bibfnamefont
  {A.}~\bibnamefont {Kamalov}}, \bibinfo {author} {\bibfnamefont
  {A.}~\bibnamefont {Natan}}, \bibinfo {author} {\bibfnamefont
  {J.}~\bibnamefont {O’Neal}}, \bibinfo {author} {\bibfnamefont
  {T.}~\bibnamefont {Osipov}}, \bibinfo {author} {\bibfnamefont
  {M.}~\bibnamefont {Owens}}, \bibinfo {author} {\bibfnamefont
  {D.}~\bibnamefont {Ray}}, \bibinfo {author} {\bibfnamefont {D.}~\bibnamefont
  {Rich}}, \bibinfo {author} {\bibfnamefont {P.}~\bibnamefont {Walter}},
  \bibinfo {author} {\bibfnamefont {A.}~\bibnamefont {Marinelli}},\ and\
  \bibinfo {author} {\bibfnamefont {J.~P.}\ \bibnamefont {Cryan}},\ }\bibfield
  {title} {\bibinfo {title} {A co-axial velocity map imaging spectrometer for
  electrons},\ }\href {https://doi.org/10.1063/1.5046192} {\bibfield  {journal}
  {\bibinfo  {journal} {AIP Advances}\ }\textbf {\bibinfo {volume} {8}},\
  \bibinfo {pages} {115308} (\bibinfo {year} {2018})},\ \bibinfo {note}
  {publisher: American Institute of Physics}\BibitemShut {NoStop}%
\bibitem [{\citenamefont {Glownia}\ \emph {et~al.}(2010)\citenamefont
  {Glownia}, \citenamefont {Cryan}, \citenamefont {Andreasson}, \citenamefont
  {Belkacem}, \citenamefont {Berrah}, \citenamefont {Blaga}, \citenamefont
  {Bostedt}, \citenamefont {Bozek}, \citenamefont {DiMauro}, \citenamefont
  {Fang}, \citenamefont {Frisch}, \citenamefont {Gessner}, \citenamefont
  {Gühr}, \citenamefont {Hajdu}, \citenamefont {Hertlein}, \citenamefont
  {Hoener}, \citenamefont {Huang}, \citenamefont {Kornilov}, \citenamefont
  {Marangos}, \citenamefont {March}, \citenamefont {McFarland}, \citenamefont
  {Merdji}, \citenamefont {Petrovic}, \citenamefont {Raman}, \citenamefont
  {Ray}, \citenamefont {Reis}, \citenamefont {Trigo}, \citenamefont {White},
  \citenamefont {White}, \citenamefont {Wilcox}, \citenamefont {Young},
  \citenamefont {Coffee},\ and\ \citenamefont
  {Bucksbaum}}]{glownia_time-resolved_2010}%
  \BibitemOpen
  \bibfield  {author} {\bibinfo {author} {\bibfnamefont {J.~M.}\ \bibnamefont
  {Glownia}}, \bibinfo {author} {\bibfnamefont {J.}~\bibnamefont {Cryan}},
  \bibinfo {author} {\bibfnamefont {J.}~\bibnamefont {Andreasson}}, \bibinfo
  {author} {\bibfnamefont {A.}~\bibnamefont {Belkacem}}, \bibinfo {author}
  {\bibfnamefont {N.}~\bibnamefont {Berrah}}, \bibinfo {author} {\bibfnamefont
  {C.~I.}\ \bibnamefont {Blaga}}, \bibinfo {author} {\bibfnamefont
  {C.}~\bibnamefont {Bostedt}}, \bibinfo {author} {\bibfnamefont
  {J.}~\bibnamefont {Bozek}}, \bibinfo {author} {\bibfnamefont {L.~F.}\
  \bibnamefont {DiMauro}}, \bibinfo {author} {\bibfnamefont {L.}~\bibnamefont
  {Fang}}, \bibinfo {author} {\bibfnamefont {J.}~\bibnamefont {Frisch}},
  \bibinfo {author} {\bibfnamefont {O.}~\bibnamefont {Gessner}}, \bibinfo
  {author} {\bibfnamefont {M.}~\bibnamefont {Gühr}}, \bibinfo {author}
  {\bibfnamefont {J.}~\bibnamefont {Hajdu}}, \bibinfo {author} {\bibfnamefont
  {M.~P.}\ \bibnamefont {Hertlein}}, \bibinfo {author} {\bibfnamefont
  {M.}~\bibnamefont {Hoener}}, \bibinfo {author} {\bibfnamefont
  {G.}~\bibnamefont {Huang}}, \bibinfo {author} {\bibfnamefont
  {O.}~\bibnamefont {Kornilov}}, \bibinfo {author} {\bibfnamefont {J.~P.}\
  \bibnamefont {Marangos}}, \bibinfo {author} {\bibfnamefont {A.~M.}\
  \bibnamefont {March}}, \bibinfo {author} {\bibfnamefont {B.~K.}\ \bibnamefont
  {McFarland}}, \bibinfo {author} {\bibfnamefont {H.}~\bibnamefont {Merdji}},
  \bibinfo {author} {\bibfnamefont {V.~S.}\ \bibnamefont {Petrovic}}, \bibinfo
  {author} {\bibfnamefont {C.}~\bibnamefont {Raman}}, \bibinfo {author}
  {\bibfnamefont {D.}~\bibnamefont {Ray}}, \bibinfo {author} {\bibfnamefont
  {D.~A.}\ \bibnamefont {Reis}}, \bibinfo {author} {\bibfnamefont
  {M.}~\bibnamefont {Trigo}}, \bibinfo {author} {\bibfnamefont {J.~L.}\
  \bibnamefont {White}}, \bibinfo {author} {\bibfnamefont {W.}~\bibnamefont
  {White}}, \bibinfo {author} {\bibfnamefont {R.}~\bibnamefont {Wilcox}},
  \bibinfo {author} {\bibfnamefont {L.}~\bibnamefont {Young}}, \bibinfo
  {author} {\bibfnamefont {R.~N.}\ \bibnamefont {Coffee}},\ and\ \bibinfo
  {author} {\bibfnamefont {P.~H.}\ \bibnamefont {Bucksbaum}},\ }\bibfield
  {title} {{\selectlanguage {EN}\bibinfo {title} {Time-resolved pump-probe
  experiments at the {LCLS}}},\ }\href {https://doi.org/10.1364/OE.18.017620}
  {\bibfield  {journal} {\bibinfo  {journal} {Optics Express}\ }\textbf
  {\bibinfo {volume} {18}},\ \bibinfo {pages} {17620} (\bibinfo {year}
  {2010})}\BibitemShut {NoStop}%
\bibitem [{\citenamefont {Kheifets}\ \emph {et~al.}(2022)\citenamefont
  {Kheifets}, \citenamefont {Wielian}, \citenamefont {Serov}, \citenamefont
  {Ivanov}, \citenamefont {Wang}, \citenamefont {Marinelli},\ and\
  \citenamefont {Cryan}}]{kheifets_ionization_2022}%
  \BibitemOpen
  \bibfield  {author} {\bibinfo {author} {\bibfnamefont {A.~S.}\ \bibnamefont
  {Kheifets}}, \bibinfo {author} {\bibfnamefont {R.}~\bibnamefont {Wielian}},
  \bibinfo {author} {\bibfnamefont {V.~V.}\ \bibnamefont {Serov}}, \bibinfo
  {author} {\bibfnamefont {I.~A.}\ \bibnamefont {Ivanov}}, \bibinfo {author}
  {\bibfnamefont {A.~L.}\ \bibnamefont {Wang}}, \bibinfo {author}
  {\bibfnamefont {A.}~\bibnamefont {Marinelli}},\ and\ \bibinfo {author}
  {\bibfnamefont {J.~P.}\ \bibnamefont {Cryan}},\ }\bibfield  {title} {\bibinfo
  {title} {Ionization phase retrieval by angular streaking from random shots of
  {XUV} radiation},\ }\href {https://doi.org/10.1103/PhysRevA.106.033106}
  {\bibfield  {journal} {\bibinfo  {journal} {Physical Review A}\ }\textbf
  {\bibinfo {volume} {106}},\ \bibinfo {pages} {033106} (\bibinfo {year}
  {2022})},\ \bibinfo {note} {publisher: American Physical Society}\BibitemShut
  {NoStop}%
\bibitem [{\citenamefont {DeGroot}\ and\ \citenamefont
  {Schervish}(2010)}]{DeGroot_probability_2010}%
  \BibitemOpen
  \bibfield  {author} {\bibinfo {author} {\bibfnamefont {M.~H.}\ \bibnamefont
  {DeGroot}}\ and\ \bibinfo {author} {\bibfnamefont {M.~J.}\ \bibnamefont
  {Schervish}},\ }\href@noop {} {\emph {\bibinfo {title} {Probability and
  Statistics}}},\ \bibinfo {edition} {4th}\ ed.\ (\bibinfo  {publisher}
  {Pearson},\ \bibinfo {address} {Upper Saddle River, NJ},\ \bibinfo {year}
  {2010})\ p.\ \bibinfo {pages} {261}\BibitemShut {NoStop}%
\bibitem [{\citenamefont {Eckart}\ and\ \citenamefont
  {Young}(1936)}]{eckart_approximation_1936}%
  \BibitemOpen
  \bibfield  {author} {\bibinfo {author} {\bibfnamefont {C.}~\bibnamefont
  {Eckart}}\ and\ \bibinfo {author} {\bibfnamefont {G.}~\bibnamefont {Young}},\
  }\bibfield  {title} {\bibinfo {title} {The approximation of one matrix by
  another of lower rank},\ }\href {https://doi.org/10.1007/BF02288367}
  {\bibfield  {journal} {\bibinfo  {journal} {Psychometrika}\ }\textbf
  {\bibinfo {volume} {1}},\ \bibinfo {pages} {211} (\bibinfo {year}
  {1936})}\BibitemShut {NoStop}%
\bibitem [{\citenamefont {Johnson}\ and\ \citenamefont
  {Wichern}(2007)}]{Johnson_Wichern_2007}%
  \BibitemOpen
  \bibfield  {author} {\bibinfo {author} {\bibfnamefont {R.~A.}\ \bibnamefont
  {Johnson}}\ and\ \bibinfo {author} {\bibfnamefont {D.~W.}\ \bibnamefont
  {Wichern}},\ }in\ \href@noop {} {\emph {\bibinfo {booktitle} {Applied
  Multivariate Statistical Analysis}}}\ (\bibinfo  {publisher} {Pearson},\
  \bibinfo {address} {Upper Saddle River, NJ},\ \bibinfo {year} {2007})\
  \bibinfo {edition} {6th}\ ed.,\ pp.\ \bibinfo {pages} {407--410}\BibitemShut
  {NoStop}%
\bibitem [{\citenamefont {Walter}\ \emph {et~al.}(2021)\citenamefont {Walter},
  \citenamefont {Kamalov}, \citenamefont {Gatton}, \citenamefont {Driver},
  \citenamefont {Bhogadi}, \citenamefont {Castagna}, \citenamefont {Cheng},
  \citenamefont {Shi}, \citenamefont {Obaid}, \citenamefont {Cryan},
  \citenamefont {Helml}, \citenamefont {Ilchen},\ and\ \citenamefont
  {Coffee}}]{walter_multi_2021}%
  \BibitemOpen
  \bibfield  {author} {\bibinfo {author} {\bibfnamefont {P.}~\bibnamefont
  {Walter}}, \bibinfo {author} {\bibfnamefont {A.}~\bibnamefont {Kamalov}},
  \bibinfo {author} {\bibfnamefont {A.}~\bibnamefont {Gatton}}, \bibinfo
  {author} {\bibfnamefont {T.}~\bibnamefont {Driver}}, \bibinfo {author}
  {\bibfnamefont {D.}~\bibnamefont {Bhogadi}}, \bibinfo {author} {\bibfnamefont
  {J.-C.}\ \bibnamefont {Castagna}}, \bibinfo {author} {\bibfnamefont
  {X.}~\bibnamefont {Cheng}}, \bibinfo {author} {\bibfnamefont
  {H.}~\bibnamefont {Shi}}, \bibinfo {author} {\bibfnamefont {R.}~\bibnamefont
  {Obaid}}, \bibinfo {author} {\bibfnamefont {J.}~\bibnamefont {Cryan}},
  \bibinfo {author} {\bibfnamefont {W.}~\bibnamefont {Helml}}, \bibinfo
  {author} {\bibfnamefont {M.}~\bibnamefont {Ilchen}},\ and\ \bibinfo {author}
  {\bibfnamefont {R.~N.}\ \bibnamefont {Coffee}},\ }\bibfield  {title}
  {\bibinfo {title} {{Multi-resolution electron spectrometer array for future
  free-electron laser experiments}},\ }\href
  {https://doi.org/10.1107/S1600577521007700} {\bibfield  {journal} {\bibinfo
  {journal} {Journal of Synchrotron Radiation}\ }\textbf {\bibinfo {volume}
  {28}},\ \bibinfo {pages} {1364} (\bibinfo {year} {2021})}\BibitemShut
  {NoStop}%
\bibitem [{\citenamefont {Petkov{\v{s}}ek}\ \emph {et~al.}(1996)\citenamefont
  {Petkov{\v{s}}ek}, \citenamefont {Wilf},\ and\ \citenamefont
  {Zeilberger}}]{petkovsek_A_1996}%
  \BibitemOpen
  \bibfield  {author} {\bibinfo {author} {\bibfnamefont {M.}~\bibnamefont
  {Petkov{\v{s}}ek}}, \bibinfo {author} {\bibfnamefont {H.~S.}\ \bibnamefont
  {Wilf}},\ and\ \bibinfo {author} {\bibfnamefont {D.}~\bibnamefont
  {Zeilberger}},\ }\href@noop {} {\emph {\bibinfo {title} {A=B}}}\ (\bibinfo
  {publisher} {A K Peters},\ \bibinfo {address} {Wellesley, MA},\ \bibinfo
  {year} {1996})\ p.~\bibinfo {pages} {38}\BibitemShut {NoStop}%
\end{thebibliography}%
\end{document}